\documentclass[aps,pra,singlecolumn,preprintnumbers,amsmath,amssymb]{revtex4}
\usepackage{dcolumn}
\usepackage{bm}
\usepackage{amssymb}
\usepackage[german, english]{babel}
\usepackage{graphicx}
\usepackage{color}
\usepackage[letterpaper,total={6.5in,9.5in},top=0.75in,left=1in]{geometry}

\usepackage{array}

\begin{document}
\title{Degenerate quantum gases of strontium}

\author{Simon Stellmer}
 \affiliation{Institut f\"ur Quantenoptik und Quanteninformation (IQOQI),
\"Osterreichische Akademie der Wissenschaften, 6020 Innsbruck, Austria}
\author{Florian Schreck}
\affiliation{Institut f\"ur Quantenoptik und Quanteninformation (IQOQI),
\"Osterreichische Akademie der Wissenschaften, 6020 Innsbruck, Austria}
\author{Thomas C.~Killian}
\affiliation{Rice University, Department of Physics and Astronomy, Houston, Texas 77251, USA}

\date{\today}

\begin{abstract}
Degenerate quantum gases of alkaline-earth-like elements open new opportunities in research areas ranging from molecular physics to the study of strongly correlated systems. These experiments exploit the rich electronic structure of these elements, which is markedly different from the one of other species for which quantum degeneracy has been attained. Specifically, alkaline-earth-like atoms, such as strontium, feature metastable triplet states, narrow intercombination lines, and a non-magnetic, closed-shell ground state. This review covers the creation of quantum degenerate gases of strontium and the first experiments performed with this new system. It focuses on laser-cooling and evaporation schemes, which enable the creation of Bose-Einstein condensates and degenerate Fermi gases of all strontium isotopes, and shows how they are used for the investigation of optical Feshbach resonances, the study of degenerate gases loaded into an optical lattice, as well as the coherent creation of Sr$_2$ molecules.
\end{abstract}

\maketitle

\section{Introduction}
\label{sec:Introduction}

All of the early experiments reaching Bose-Einstein condensation (BEC) \cite{Inguscio1999book} and the Fermi-degenerate regime \cite{Inguscio2008ufg} in ultracold gases were performed with alkali atoms. In recent years, degenerate samples of more complex atoms, such as the alkaline-earth (-like) species ytterbium \cite{Sugawa2013uyg}, calcium \cite{Kraft2009bec}, and strontium became available. These samples bring us closer to the realization of intriguing experiments that are intimately connected to the properties of alkaline-earth elements, ranging from the creation of ultracold open-shell polar molecules to the study of novel, strongly correlated many-body systems. In this Chapter we review the creation of degenerate quantum gases of strontium and the first experiments based on these gases. We will start by introducing the properties of strontium most relevant to quantum gas experiments and some of the possibilities opened up by these properties.

There are four stable isotopes of strontium; three are bosonic and one is fermionic. The bosonic isotopes $^{84}$Sr, $^{86}$Sr, and $^{88}$Sr have zero nuclear spin, just as all other bosonic alkaline-earth (-like) elements. The reason for this zero spin is that bosonic isotopes of atoms with an even number of electrons must have an even-even nucleus, for which the proton and neutron spins pair up such that the total nuclear spin vanishes \cite{Fuller1976nsa}. The absence of a nuclear spin in these isotopes precludes the appearance of hyperfine structure, as well as of Zeeman structure for the $J=0$ states, and thus leads to a simple electronic level scheme. The fermionic isotope $^{87}$Sr carries a nuclear spin of $I=9/2$, which forms the basis of many proposed experiments.

Strontium features two valence electrons. The electronic level structure decomposes into singlet states, in which the spins of the two valence electrons are aligned anti-parallel, and triplet states with parallel spins. Transitions between singlet and triplet states are dipole-forbidden, leading to narrow linewidths and the emergence of metastable triplet states. Some of these intercombination transitions have linewidths on the order of kHz, ideally suited for narrow-line cooling \cite{Katori1999mot}, while others show linewidths well below 1\,Hz and are employed in optical clocks \cite{Derevianko2011poo}.

A key property of atomic gases is the scattering behavior of its constituents. In the limit of low temperatures, we can express the scattering between two atoms by a single parameter:~the scattering length $a$, which is usually stated in units of the Bohr radius $a_0=53\,$pm. The evaporation efficiency, the stability of a quantum gas, and the mean-field energy all depend on the scattering length. Magnetic Feshbach resonances are widely used in quantum gas experiments for interaction tuning \cite{Chin2010fri}, but such resonances are absent in alkaline-earth species due to the $J=0$ nature of the $^1S_0$ ground state. It is thus fortunate that the scattering lengths of the various isotopes are very different; see Tab.~\ref{tab:StrontiumProperties}. The isotope $^{88}$Sr has an extremely small scattering length of $a_{88}=-2.0\,a_0$, making it ideally suited for certain precision measurements due to the almost vanishing mean-field shift. Experiments involving excited-state atoms, such as optical clocks, are suffering from additional shifts arising from interactions involving atoms in the excited state. The $^{84}$Sr isotope, on the other hand, exhibits a moderate scattering length of $a_{84}=123\,a_0$ that allows for efficient evaporation and stable BECs. Optical Feshbach resonances \cite{Fedichev1996ion,Theis2004tts,Ciurylo2005oto,Blatt2011moo,Yan2013ccc}, a means to vary the scattering length by a suitable light field, are discussed in Sec.~\ref{sec:OFR}.

\begin{table}[t]
	\centering
\begin{tabular*}{155mm}{@{\extracolsep{\fill}}ccdccrrrr}\hline\hline \noalign{\smallskip}
&\multicolumn{1}{c}{statistics}&\multicolumn{1}{c}{abundance}&\multicolumn{1}{c}{$I$}& \hspace{3mm} & \multicolumn{1}{c}{$^{84}$Sr} & \multicolumn{1}{c}{$^{86}$Sr} & \multicolumn{1}{c}{$^{87}$Sr} & \multicolumn{1}{c}{$^{88}$Sr}\\
&   &  \multicolumn{1}{c}{(\%)} & & & \multicolumn{4}{c}{($a_0$)} \\   \noalign{\smallskip}\hline\noalign{\smallskip}
    $^{84}$Sr & bosonic   &  0.56 &   0  & & 123     & 32       & $-57$      & 1700      \\
	$^{86}$Sr & bosonic   &  9.86 &   0  & & 32      & 800      & 162        & 97        \\
	$^{87}$Sr & fermionic &  7.00 & 9/2  & & $-57$   & 162      & 96         & 55        \\
	$^{88}$Sr & bosonic   & 82.58 &   0  & & 1700    & 97       & 55         & $-2$      \\ \hline \hline
\end{tabular*}
\caption{Important properties of the four stable strontium isotopes.  The scattering lengths $a$ given in the last four columns are averages of values taken from Refs.~\cite{Martinez2008tpp,Stein2008fts,Stein2010tss} and are given in units of the Bohr radius $a_0=53\,$pm. Only the fermionic $^{87}$Sr isotope has a nuclear spin $I$.}
\label{tab:StrontiumProperties}
\end{table}

The unique combination of properties of alkaline-earth atoms are the long lifetime of the $^3P_J$ states, the associated clock transitions originating from the $^1S_0$ state, and the near-perfect decoupling of electronic and nuclear spin for the $^1S_0$ and $^3P_0$ states of the fermionic isotope. A certain set of elements, namely zinc, cadmium \cite{Brinckmann2007mot}, mercury \cite{Hachisu2008ton}, ytterbium \cite{Kuwamoto1999mot}, and nobelium, share these features with the ``true'' alkaline-earth elements. For simplicity, we will refer to these species as alkaline-earth elements as well.

Proposals demanding some or all of these properties describe the creation of artificial gauge fields \cite{Dalibard2011agp,Gerbier2010gff,Cooper2011ofl,Beri2011zti,Gorecka2011smf}, the implementation of sub-wavelength optical lattices \cite{Yi2008sda}, the processing of quantum information \cite{Stock2008eog}, or the study of many-body systems with dipolar or quadrupolar interaction \cite{Olmos2013lri,Bhongale2013qpo}. The large nuclear spin of the fermionic isotopes is at the heart of many recent proposals to study SU($N$) magnetism \cite{Wu2003ess,Wu2006hsa,Cazalilla2009ugo,Gorshkov2010tos,Xu2010lim,FossFeig2010ptk,FossFeig2010hfi,Hung2011qmo,Manmana2011smi,Hazzard2012htp,Bonnes2012alo,Messio2012edo}, where various phases such as chiral spin liquids \cite{Hermele2009mio}, algebraic spin liquids \cite{Corboz2012soq}, spatial symmetry breaking \cite{Corboz2012ssi} and spontaneous SU($N$) symmetry breaking \cite{Lauchli2006qpo,Toth2010tso,Corboz2011sda} are predicted to occur. Further proposals suggest using alkaline-earth atoms to simulate lattice gauge theories \cite{Banerjee2013aqs}, or to robustly store quantum information and perform quantum information processing \cite{Hayes2007qlv,Daley2008qcw,Gorshkov2009aem,Daley2011sdl}. Quantum gas mixtures of alkaline-earth atoms with alkali atoms can be used as a basis for the production of ground-state open-shell molecules, such as RbSr \cite{Zuchowski2010urm,Guerout2010gso}, which constitute a platform towards the simulation of lattice-spin models \cite{Micheli2006atf,Brennen2007dsl}. Bi-alkaline-earth molecules, such as Sr$_2$, are sensitive and model-independent probes for variations of the electron-to-proton mass ratio \cite{Zelevinsky2008pto,Kotochigova2009pfa}. Narrow optical transitions to $^3P_J$ states are useful, to create molecular condensates through coherent photoassociation \cite{Naidon2008tbt}, and to manipulate the scattering properties through optical Feshbach resonances \cite{Ciurylo2005oto,Blatt2011moo,Yan2013ccc}. Aside from degenerate gases, two-valence-electron atoms have been used for optical clocks \cite{Derevianko2011poo} and other precision experiments \cite{Ferrari2006llb,Poli2011pmo}, and the clock transition has recently been used to explore quantum many-body physics \cite{Martin2013aqm}. Other experiments investigate the coherent transport of light \cite{Bidel2002clt}, as well as the production of ultracold plasmas \cite{Killian2007unp} and Rydberg gases \cite{Millen2010tee,McQuillen2013cou}.

So far, two research groups have reported on the attainment of BECs and degenerate Fermi gases of strontium, and quite naturally, this Chapter is a joint effort of these two teams. We have already discussed the nuclear, electronic, and scattering properties of strontium, and we will show how they combine to form a very powerful platform for quantum gas experiments. In Sec.~\ref{sec:HistoricalOverview} we begin with a historical overview on work performed in various groups around the world. In Secs.~\ref{sec:LaserCooling} to \ref{sec:DFG} we focus on the experimental procedure to generate quantum-degenerate samples of all stable isotopes, both bosonic (Sec.~\ref{sec:BEC}) and fermionic (Sec.~\ref{sec:DFG}). In addition, we present a few experiments carried out with such samples:~the study of optical Feshbach resonances (Sec.~\ref{sec:OFR}), the observation of the Mott-insulator transition (Sec.~\ref{sec:OpticalLattice}), and the creation of Sr$_2$ molecules (Sec.~\ref{sec:Molecules}). The work presented here has already been published by the Rice \cite{Nagel2005psa,Mickelson2005sdo,Martinez2008tpp,Martinez2009bec,Mickelson2010bec,DeSalvo2010dfg,Yan2013ccc} and Innsbruck groups \cite{Stellmer2009bec,Tey2010ddb,Stellmer2010bec,Stellmer2011dam,Stellmer2012cou,Stellmer2013lct,Stellmer2013poq,Stellmer2013dqg} and complements a review on similar work on ytterbium, which appeared in the preceding volume of this series \cite{Sugawa2013uyg}.

\begin{figure}[t]
\includegraphics[width=114mm]{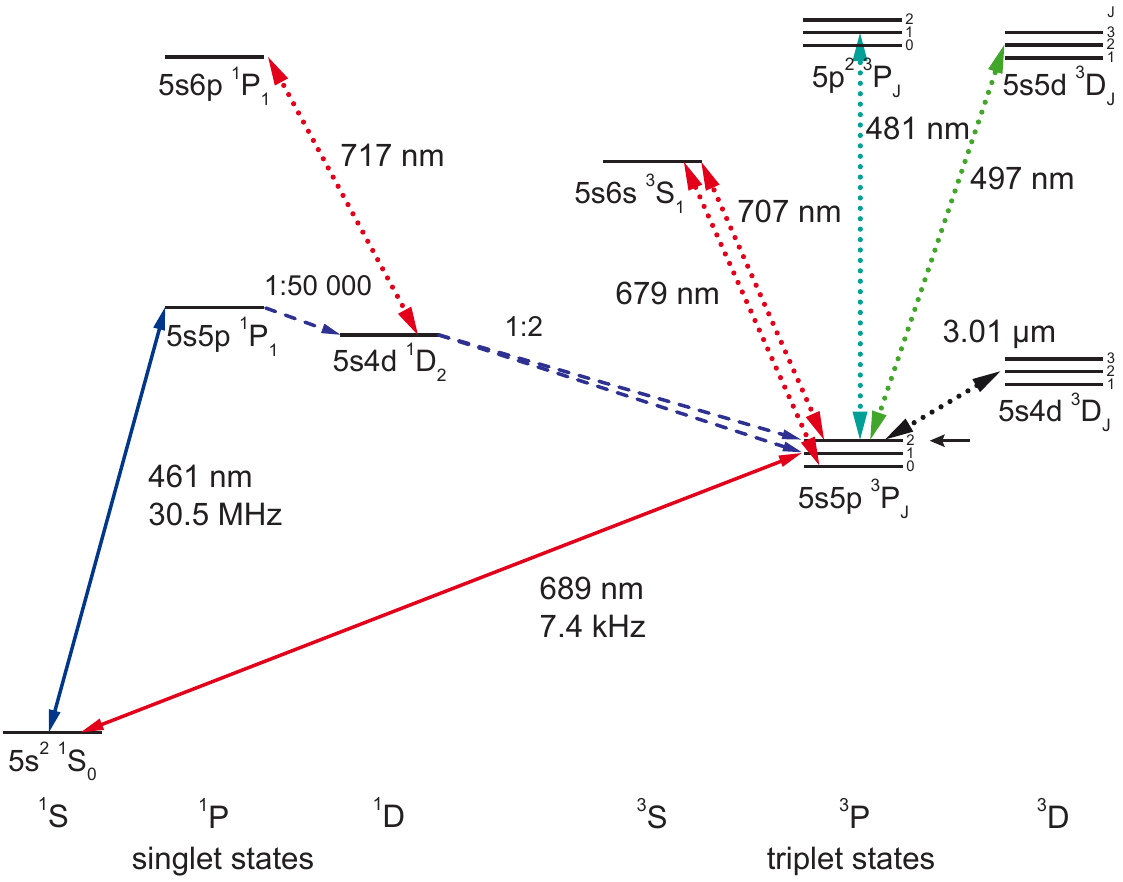}
\caption{Selection of the level scheme of strontium. The cooling (solid arrows) and repump transitions (dotted arrows), dominant decay paths from the $^1P_1$ state (dashed arrows) and related branching ratios are depicted. The $^3P_2$ reservoir state is marked by a small arrow.}
\label{fig_LevelScheme}
\end{figure}

\section{Historical overview}
\label{sec:HistoricalOverview}

\subsection{Laser cooling on the broad transition}

Laser cooling of alkaline-earth atoms was pioneered by the Tokyo groups. The first cooling and trapping of various isotopes of calcium and strontium was reported in the beginning of the 1990s \cite{Kurosu1990lca}. These magneto-optical traps (MOTs) were operated on the blue $^1S_0 - {^1P_1}$ singlet transitions, but the lifetimes were very short compared to typical alkali MOTs. As known from earlier experiments with calcium, the strontium MOT lifetimes are limited to a few 10\,ms due to a weak decay channel from the $^1P_1$ state out of the cooling cycle into the $^1D_2$ state \cite{Lellouch1987mot,Beverini1989mot}; see Fig.~\ref{fig_LevelScheme}. Contrary to early assumptions, atoms do not remain in the $^1D_2$ state \cite{Uhlenberg2000mot}, but decay further into the long-lived $^3P_2$ and the short-lived $^3P_1$ state \cite{Kurosu1990lca,Kurosu1992lca}. Repumping of strontium atoms from the $^1D_2$ state into the $5s6p\,{^1P_1}$ state using light at 717\,nm allowed only for a small increase of the MOT atom number by about a factor of two \cite{Kurosu1992lca,Vogel1999lco,Bidel2002per} due to unfavorably large branching ratios from the $5s6p\,{^1P_1}$ state into long-lived metastable triplet states.

Substantial increase of the MOT atom number came about only when repumping of the $^3P_0$ and $^3P_2$ states was implemented \cite{Vogel1998ews}, using the $^3S_1$ state as an intermediate state to transfer population into the $^3P_1$ state, which decays into the singlet ground state with a comparably short lifetime of $21\,\mu$s. Further studies performed by the Boulder group include the quantification of loss processes from excited-state collisions \cite{Dinneen1999cco}, sub-Doppler cooling of the fermionic isotope \cite{Xu2003sss}, and simultaneous MOTs of two different isotopes \cite{Xu2003cat}.

\subsection{Laser cooling on the narrow transition}

\begin{figure}[t]
\includegraphics[width=165mm]{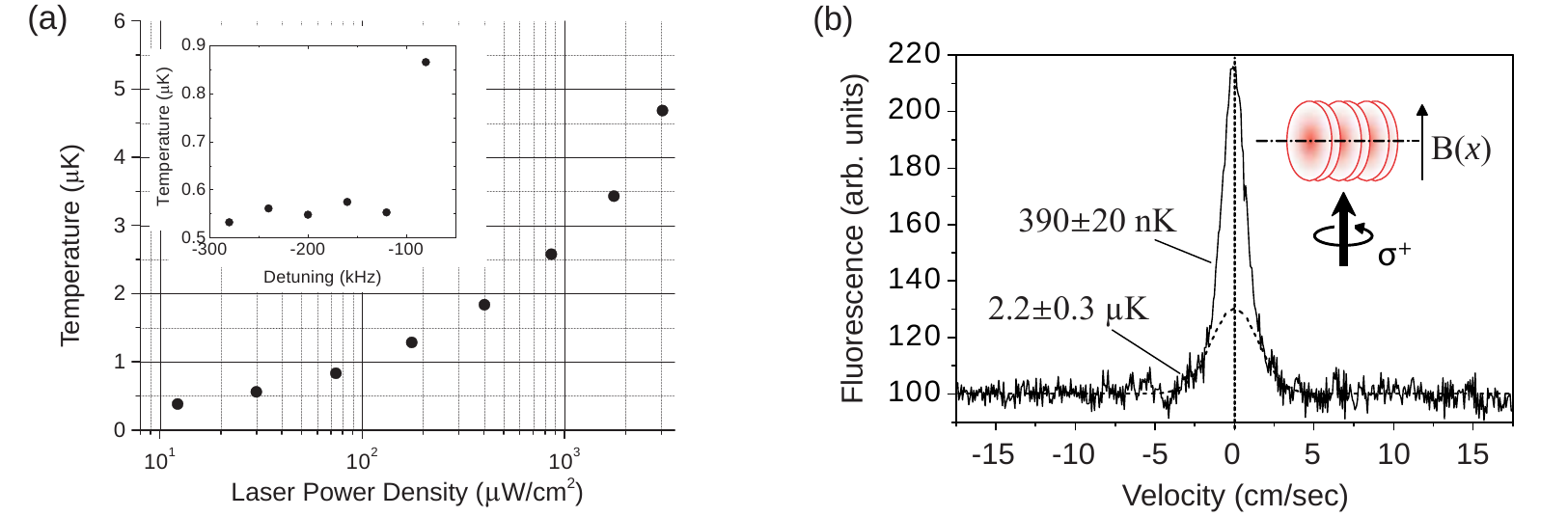}
\caption{Narrow-line cooling in strontium. (a) A MOT of bosonic $^{88}$Sr can reach temperatures of about 400\,nK for very low intensity of the cooling light. (b) The momentum distribution of fermionic $^{87}$Sr atoms released from a 1D optical lattice shows a very narrow feature, corresponding to a large fraction of atoms pumped into a single $m_F$ state and cooled to the recoil limit. Reprinted figures with permission from Refs.~\cite{Katori1999mot} and \cite{Mukaiyama2003rll}. Copyright (1999) and (2003) by the American Physical Society.}
\label{fig:History}
\end{figure}

Alkaline-earth elements, cooled to mK temperatures on the broad transition, constitute an adequate starting point to probe the intercombination lines. This was first done for the $^1S_0 - {^3P_1}$ transition in calcium \cite{Barger1979rop,Beverini1989mot,Kurosu1992oot}. The first MOT operated on the intercombination line of the bosonic $^{88}$Sr isotope was presented by the Tokyo group \cite{Katori1999mot} and showed remarkable features: the attainable temperature reached as low as about 400\,nK, indeed close to the recoil temperature. This group observed the peculiar pancake shape of the atomic cloud and showed that the attainable temperature can be reduced by lowering the MOT light intensity. The temperature was found to be independent of the detuning over a large range; see Fig.~\ref{fig:History}(a). The ``magic'' wavelength for this cooling transition was calculated \cite{Katori1999odo}, and a dipole trap at this wavelength was used to confine atoms at a phase-space density of 0.1, just one order of magnitude away from quantum degeneracy \cite{Ido2000odt}. The behavior of the narrow-line MOT was studied further by the Boulder group \cite{Loftus2004nlc,Loftus2004nlca}.

The narrow-line MOT for fermionic isotopes is more involved due to the appearance of hyperfine structure, and was first described and implemented by the Tokyo group \cite{Mukaiyama2003rll}. This experiment included already the loading into a 1D optical lattice, optical pumping into a single Zeeman substate, and cooling down to the recoil temperature; see Fig.~\ref{fig:History}(b). A value of $T/T_F=2$ was reached in this experiment.

\subsection{Optical clocks}

The potential of optical clocks operated on the ultranarrow $^1S_0 - {^3P_0}$ transition was appreciated very early; see Ref.~\cite{Derevianko2011poo} for a recent review. The narrow linewidth can only be exploited for sufficiently long interrogation times, thus the atoms would need to be trapped. The absence of both charge and sizable magnetic moment suggests an optical trap. The deployment of optical lattices even allows to reach the Lamb-Dicke regime \cite{Katori2002sos}, thus removing the Doppler broadening. These traps, however, are prone to induce a light shift onto the clock transition, thereby shifting and broadening the transition substantially. Spectroscopy of atoms in a lattice of carefully chosen magic wavelength was first proposed \cite{Katori2002sos} and demonstrated \cite{Katori1999odo,Ido2003rfs} by the Tokyo group using the $^1S_0 - {^3P_1}$ cooling transition; see Ref.~\cite{Ye2008qse} for a review.

The same concepts were also applied to the $^1S_0 - {^3P_0}$ clock transition: calculations of the magic wavelength for the bosonic isotope $^{88}$Sr were followed by experimental realizations, \textit{e.g.}~in Paris \cite{Baillard2007aeo}. Optical clocks based on bosonic $^{88}$Sr suffer from interaction shifts due to the interaction between atoms in the $^1S_0$ and $^3P_0$ states \cite{Lisdat2009cld}. This issue can be overcome by advancing from a one-dimensional to a three-dimensional optical lattice \cite{Akatsuka2010tdo}.

As another possible solution, it was proposed to use the fermionic isotope $^{87}$Sr with all atoms prepared in the same $m_F$ state, thus removing collisions of the identical particles through the Pauli exclusion principle. This concept was proposed \cite{Katori2003uoc} and realized \cite{Takamoto2003sot,Takamoto2005aol} by the Tokyo group, and soon after, optical clocks were operated by the Tokyo \cite{Takamoto2005aol}, Boulder \cite{Boyd2006oac,Ludlow2006sso,Ludlow2008slc,Martin2013qma}, Paris \cite{LeTergat2006aol}, and various other groups. Since 2007, strontium lattice clocks constitute the best agreed-upon frequency standard, and have recently almost drawn level with ion clocks \cite{Chou2010fco,Nicholson2012cot,Hinkley2013aac} in terms of the achieved accuracy. A set of measurements of various experiments around the world has been analyzed to calculate limits on possible drifts of fundamental constants \cite{Blatt2008nlo,Blatt2011uca}.

These experiments were performed in 1D optical lattices, where interactions induced by inhomogeneous probing \cite{Campbell2009pib} can be observed despite the fermionic character of the atoms. These experiments opened the door towards the exploration of many-body phenomena in optical clocks \cite{Martin2013aqm}. To overcome the residual influence of interactions, the density of the sample can be reduced by a sufficient increase of the trap volume \cite{Nicholson2012cot}, or a 3D optical lattice can be employed to separate the atoms from one another. Blue-detuned lattices are investigated as well \cite{Takamoto2009pfo}.

\subsection{Struggle to reach quantum degeneracy}

The early experiments reached phase-space densities already very close to quantum degeneracy \cite{Ido2000odt,Poli2005cat}, but it was understood that plain narrow-line cooling in a dipole trap could not yield phase-space densities substantially larger than 0.1 \cite{Ido2000odt,Stellmer2013lct}. This last order of magnitude called for evaporative cooling as an additional cooling stage, which seemed to pose an unsurmountable obstacle at that time. There are two main explanations: First, experiments involving strontium or calcium were primarily aiming for optical clocks, which typically operate at relatively fast cycle times and do not require a sophisticated vacuum. The lifetime of trapped samples in these experiments did not allow for accumulation of low-abundant isotopes or long evaporation times. Second, the scattering properties of the most abundant isotopes $^{40}$Ca, $^{86}$Sr, and $^{88}$Sr are not particularly favorable for evaporation. As a consequence, the first BECs and degenerate Fermi gases in alkali-earth systems were reached with ytterbium \cite{Takasu2003ssb,Fukuhara2007dfg} in 2003 and with calcium \cite{Kraft2009bec} in 2009.

Attempts to reach BEC in the bosonic isotope $^{88}$Sr failed due to the small scattering length of $-2\,a_0$ \cite{Poli2005cat}, which does not allow for efficient thermalization during evaporation. The scattering length of $^{86}$Sr amounts to about $800\,a_0$, leading to strong inelastic losses, which also impede evaporation in a mixture of $^{86}$Sr and $^{88}$Sr \cite{Ferrari2006cos}. At the time of these experiments (2006), the scattering properties of $^{84}$Sr, the third stable bosonic isotope of only 0.56\% abundance, were not yeFt explored.

To circumvent the unfavorable scattering properties, some experiments aimed to increase the phase space density by laser cooling of atoms in the metastable $^3P_2$ state, which has a lifetime of about 500\,s \cite{Xu2003cat,Yasuda2004lmo,Santra2004pom}. This state is naturally populated in the broad-transition MOT, it can be trapped in a magnetic trap, and the magnetic substructure allows for sub-Doppler cooling mechanisms. A variety of cooling transitions could be used, some of which have very low Doppler and recoil limits. So far, all of these attempts were spoiled by large inelastic two-body collisions. These have been quantified in calcium \cite{Hansen2006oom}, ytterbium \cite{Yamaguchi2008ici,Uetake2012sdc}, and strontium \cite{Traverso2009iae}, leaving little hope that laser or evaporative cooling towards quantum degeneracy will be successful in this state.

A number of experiments were instead performed with thermal samples of strontium. Indeed, the $^{88}$Sr isotope possesses remarkable properties: it combines a $J=0$ and $I=0$ ground state with a high natural abundance and a narrow cooling transition, constituting a B-field insensitive and easy-to-cool atomic species. In addition, this particular isotope is almost non-interacting, making it ideally suited for precision measurements besides optical clocks. The Florence group used this isotope to study Bloch oscillations \cite{Ferrari2006llb} and measure the force of gravity \cite{Poli2011pmo}.

\subsection{Photoassociation measurements}

Starting from about 2005, a series of photoassociation (PA) measurements was performed to explore both the ground- and excited molecular potentials. Knowledge of the ground-state energy levels would allow for a precise determination of all scattering lengths, while excited molecular states could be employed for optical Feshbach resonances \cite{Fedichev1996ion,Theis2004tts,Ciurylo2005oto} to tune the scattering length. These PA measurements would therefore elucidate alternative approaches to evaporative cooling.

The first one-color PA measurements were performed near the broad singlet transition at 461\,nm \cite{Nagel2005psa,Mickelson2005sdo,Yasuda2006pso}, quickly followed by measurements near the intercombination line at 689\,nm \cite{Zelevinsky2006nlp}. Precise two-color PA near the intercombination line of ytterbium \cite{Tojo2006hrp} had allowed for a determination of all intra- and interisotope scattering lengths \cite{Kitagawa2008tcp}. This approach was adopted to strontium \cite{Martinez2008tpp,Stellmer2012cou} and allowed for a calculation of all relevant scattering lengths in 2008 \cite{Ciuryloprivate,Martinez2008tpp,Stein2008fts}.

From these calculations, it became immediately clear that $^{84}$Sr would be ideally suited for evaporative cooling:~the scattering length of $123\,a_0$ promises a favorable ratio of elastic to inelastic collisions \cite{Fedichev1996tbr,Bedaque2000tbr}. Provided that the low natural abundance could be overcome using the accumulation scheme introduced by the Florence group \cite{Katori2001lco,Nagel2003mto,Sorrentino2006lca}, evaporative cooling into quantum degeneracy seemed within reach. About one year later, BEC of this isotope was reached by the Innsbruck and Rice groups, and within a few more months, BECs and Fermi gases of all stable isotopes were obtained as well. These experiments will be described in the following sections of this chapter.

\subsection{Proposals for quantum many-body simulations}

In parallel to the experimental advances, an eagerly anticipated stream of theoretical proposals started to swell in 2008. These proposals, some of which were mentioned in the previous chapter, employ the specific properties of alkaline-earth elements, and are often worked out for strontium or ytterbium. They are centered around various flavors of many-body simulations, mostly using the $m_F$ states of $^{87}$Sr, as well as schemes of quantum computation. Experiments and theory have stimulated each other and continue to do so in a very fruitful way.

\section{Two-stage laser cooling}
\label{sec:LaserCooling}

The rich level structure of strontium provides us with a variety of transitions \cite{Sansonetti2010wtp} that could be used for laser cooling; see Fig.~\ref{fig_LevelScheme}. Specifically, these include the broad transition at 461\,nm and the narrow intercombination line at 689\,nm, which have linewidths of 30.5\,MHz and 7.4\,kHz respectively \cite{Kurosu1990lca,Katori1999mot,Mukaiyama2003rll,Sorrentino2006lca,Boyd2007hps,Ludlow2008tso,Stellmer2013dqg}.

A sequence of three cooling stages is employed to bring strontium atoms into the regime of degeneracy. The first stage is a MOT operated on a broad transition, ideally suited to capture atoms from a thermal beam and cool them to mK temperatures. The second stage is a MOT operated on a narrow transition, capable of cooling the atoms to thousand-times lower temperatures at ten thousand-times higher densities. Such a sample is loaded into an optical dipole trap. The third stage, evaporative cooling, leads into quantum degeneracy. While the details of the last cooling stage depend on the respective isotope and the objective of the experiment, the first two cooling stages are rather similar for all experiments and will be described in the following. Further details can be found \textit{e.g.}~in an earlier review \cite{Sorrentino2006lca} and in  Refs.~\cite{Boyd2007hps,Ludlow2008tso,Martinez2010bec,Mickelson2010tae,Stellmer2013dqg}.

\subsection{The blue MOT}

We will now describe the Innsbruck apparatus, to which the Rice experiment is similar. A stream of strontium atoms at about $600\,^{\circ}$C is emitted from an oven and directed into a UHV chamber. The atomic beam has a divergence of order 10\,mrad, which can be reduced by a 2D optical molasses, also known as \emph{transverse cooling}. We use light red-detuned by $-15\,{\rm MHz} \approx - \Gamma/4\pi$ from the $^1S_0 - {^1P_1}$ transition at 461\,nm. This light is split into two beams, propagating orthogonal to each other and to the atomic beam, intersecting with the atoms about 100\,mm downstream from the oven. The interaction region is about 50\,mm long. The beams are elliptically shaped, retro-reflected, and contain about 10\,mW in each axis. Transverse cooling increases the number of atoms in the $^{84}$Sr MOT by a factor of three or even more, depending on the geometric design of the oven.

The broad 30-MHz transition allows for fast Zeeman-slowing and offers a high capture velocity of the MOT. The Zeeman-slower beam contains about 35\,mW of power, it is slightly focussed onto the aperture of the oven and has a waist of about 8\,mm at the position of the MOT.

Atoms in the MOT region are illuminated by three retro-reflected MOT beams, having waists of about 5\,mm. They have intensities of $I_{\rm vert}=0.1\,I_{\rm sat}$ and $I_{\rm hor}=0.25\,I_{\rm sat}$, corresponding to about 1\,mW in the vertical and 4\,mW in the horizontal beams. The saturation intensity $I_{\mathrm{sat}}=\pi h c/3\lambda^3 \tau$ of this transition is $I_{\mathrm{sat}}=40\,{\rm mW/cm}^2$. Here, $\tau=5\,$ns is the lifetime of the $^1P_1$ state. The detuning is $-32\,{\rm MHz} \approx - \Gamma/2\pi$, and the gradient of the quadrupole field is $55\,$G/cm in the vertical direction. The Doppler temperature $T_D=\hbar\Gamma/(2k_B)$ of this ``blue'' MOT is $T_D=720\,\mu$K, much higher than the recoil temperature $T_r=690\,$nK. The recoil temperature is given by $T_r=\hbar^2 k^2/(k_B m)$, where $k=2 \pi /\lambda$ is the wave vector of the light field and $\lambda$ the wavelength. Sub-Doppler cooling requires a magnetic substructure and has indeed been observed by the Boulder group \cite{Xu2003cat} for the fermionic isotope, which has a nonzero nuclear spin $I=9/2$. Repumping of hyperfine states as required for alkali atoms is not required due the lack of hyperfine structure in the $^1S_0$ state.

As we will see later, atoms from the upper MOT level can decay via the $^1D_2$ level into the metastable $^3P_2$ level, which possesses a magnetic moment. Atoms in weak-field-seeking $m_F$ states of this level can be trapped in a magnetic quadrupole field. This decay reduces the lifetime of the blue MOT to a few 10\,ms. We do not optimize the MOT for fluorescence or atom number, but for the loading rate of the \emph{metastable reservoir}, which we define as the container formed by the magnetic trap for $^3P_2$ atoms. The loading rate depends on various experimental parameters, among them the temperature (\textit{i.e.}~the flux) of the oven, the amount of light available at 461\,nm (\textit{i.e.}~the slowing and capture efficiency), the natural abundance of the respective isotope, and the temperature of the blue MOT.

We usually operate the blue MOT until a few $10^7$ to $10^8$ atoms are accumulated in the reservoir. This takes between 50\,ms and 10\,s, depending mainly on the abundance of the isotope.

\subsection{Repumping}

\subsubsection{General considerations}

The electronic structures of calcium, strontium, barium, and radium share a common feature: a $ns(n-1)d\,{^1D_2}$ state appears below the $nsnp\,{^1P_1}$ state. Here, $n$ is the principal quantum number of the valence electrons, ranging from 4 to 7. The nonzero branching ratio between the ${^1P_1}$ and the ${^1D_2}$ states opens a decay channel from the blue MOT cycle. This branching ratio is roughly $1:50\,000$ for strontium and calcium, and roughly $1:300$ for barium and radium. The atoms decay further into the $^3P_{1,2}$ metastable triplet states with a branching ratio of $2:1$. Strontium atoms in the $^3P_1$ state have a lifetime of only $21\,\mu$s and decay back into the $^1S_0$ state. On the other hand, the $^3P_2$ state has a lifetime of 500\,s in the absence of ambient black-body radiation \cite{Yasuda2004lmo}. Atoms in this state have a magnetic moment and, provided they are in a low-field seeking $m_F$ state, can be trapped in the quadrupole field of the MOT.

There is an additional decay channel of the type $^1P_1 \rightarrow {^3D_{1,2}} \rightarrow {^3P_{0,1,2}}$, which is at least two orders of magnitude weaker \cite{Werij1992osa}. Atoms in the very long-lived $^3P_0$ state are not trapped in the quadrupole field and might constitute an additional loss channel.

Atoms in the metastable state can be returned to the ground state either during ($^3P_0$ and $^3P_2$ states) or after (weak-field seeking $^3P_2$ states) the blue MOT stage through optical pumping into the short-lived metastable $^3P_1$ state \cite{Katori2001lco}. Generally, continuous repumping allows for a faster loading rate of the blue MOT, as most of the atoms falling into the $^3P_2$  state appear in non-trapped states and would otherwise be lost. This strategy is followed in optical clock experiments. If however large atom numbers are required, such as in BEC experiments, it is advantageous to accumulate atoms in the metastable reservoir and transfer them into the ground state after the blue MOT has been extinguished. The atom number in the blue MOT is limited by light-assisted collisions \cite{Dinneen1999cco}, which are absent for atoms in the metastable $^3P_2$ state. Losses through inelastic collisions of atoms in the $^3P_2$ state \cite{Traverso2009iae} are negligible due to the low density of about $10^7\,$cm$^{-3}$ in the large reservoir. The ambient black-body radiation reduces the lifetime of the $^3P_2$ state to about 20\,s \cite{Yasuda2004lmo}, which, together with the loading rate of atoms into the reservoir, sets the achievable atom number. For typical experimental parameters, this number is orders of magnitude larger than the blue MOT atom number, allowing for the accumulation of significantly more atoms. This accumulation stage plays a crucial role in generating mixtures and degenerate quantum gases of strontium.

A variety of transitions can be used for repumping. An early experiment tried to close the leakage of atoms into the triplet states by pumping them directly from the $^1D_2$ state into the $5s6p\,{^1P_1}$ state at 717\,nm \cite{Kurosu1992lca}. This approach is inefficient due to a significant branching ratio from the $5s6p\,{^1P_1}$ state into the triplet states. Other experiments use a pair of repump lasers at 679\,nm and 707\,nm to pump both the $^3P_0$ and $^3P_2$ states into the $^3P_1$ state via the $^3S_1$ state \cite{Dinneen1999cco}. This repumping approach rigorously collects atoms from all possible decay paths and facilitates blue MOT lifetimes of many seconds. A third strategy involves any of the $5snd\,{^3D_2}$ states at $3.01\,\mu$m \cite{Mickelson2009ras}, 497\,nm \cite{Poli2005cat}, or 403\,nm \cite{Stellmer2013rso} for $n=4,5,6$, respectively. Repumping via the $5p^2\,{^3P_2}$ state at 481\,nm is also efficient. Current quantum gas experiments repump only the $^3P_2$ state, thus loss through the $^3P_0$ state persists and limits the lifetime of a continuously repumped MOT to about 1\,s. Population of the $^3P_0$ state originates both from decay via the pathway $^1P_1 \rightarrow {^3D_{1,2}} \rightarrow {^3P_{0,1,2}}$ mentioned above, as well as a cascade of transitions from the $^3D_2$ via intermediate states into the $^3P_0$ state. The branching ratios of these indirect pathways are at most a few percent.

The experiments presented in this review follow the accumulation strategy, such that losses into the $^3P_0$ state are below the few-percent level and can be tolerated. The choice of the employed $5snd\,{^3D_2}$ state involves a trade-off between repump efficiency and ease of laser operation: laser systems for the three wavelengths mentioned above tend to become simpler for shorter wavelengths. As one climbs up the ladder of $^3D_2$ states, however, more and more decay channels open up that lead atoms into the dark $^3P_0$ state \cite{Stellmer2013rso}. The experiments presented here employ lasers at either $3.01\,\mu$m, 497\,nm, or 481\,nm.

\subsubsection{Fermions}

Repumping of the bosonic isotopes is straightforward even in isotopic mixtures, since the isotope shifts are at most 100\,MHz. In the case of $^{87}$Sr, efficient repumping is complicated by the hyperfine structure of the states involved. We find that all five hyperfine states $F=5/2$ through $F=13/2$ of the $^3P_2$ level are populated during the blue MOT, however at different relative amounts:~roughly 80\% of the atoms populate the $F=13/2$ and $F=11/2$ states. The hyperfine splittings of the $^3P_2$ and $^3D_2$ states are on the order of GHz. In a typical experimental cycle, repumping is performed only on the $F=11/2 \rightarrow F'=13/2$ and $F=13/2 \rightarrow F'=13/2$ transitions, or the laser is rapidly scanned across all hyperfine transitions.

\subsubsection{Experimental parameters}

The lifetime of metastable atoms in the reservoir is about 30\,s  in our experiment. This value is largely independent on the density of $^3P_2$ atoms in the reservoir. It is likely to be limited by decay along the pathway $^3P_2 \rightarrow 5s4d\,{^3D_1} \rightarrow {^3P_0}$ in presence of blue MOT light, and additional channels  $^3P_2 \rightarrow 5s4d\,{^3D_{1,2}} \rightarrow {^3P_1} \rightarrow {^1S_0}$ when the MOT light is turned off. Here, the first step is an excitation driven by the ambient black-body radiation \cite{Xu2003cat,Yasuda2004lmo}. Collisions with the background gas might limit the lifetime even further. The lifetime is certainly long enough to allow for sequential loading of different isotopes when working with mixtures.

The repumping flash lasts typically 50\,ms and contains a few 100\,$\mu$W of light in a beam collimated to a diameter of 10\,mm, corresponding to roughly $0.05\,I_{\mathrm{sat}}$, if light at a repump transition in the visible range is used. About 10\,mW are used for the repumping at $3.01\,\mu$m.

\subsection{The red MOT}

\begin{figure}[t]
\includegraphics[width=165mm]{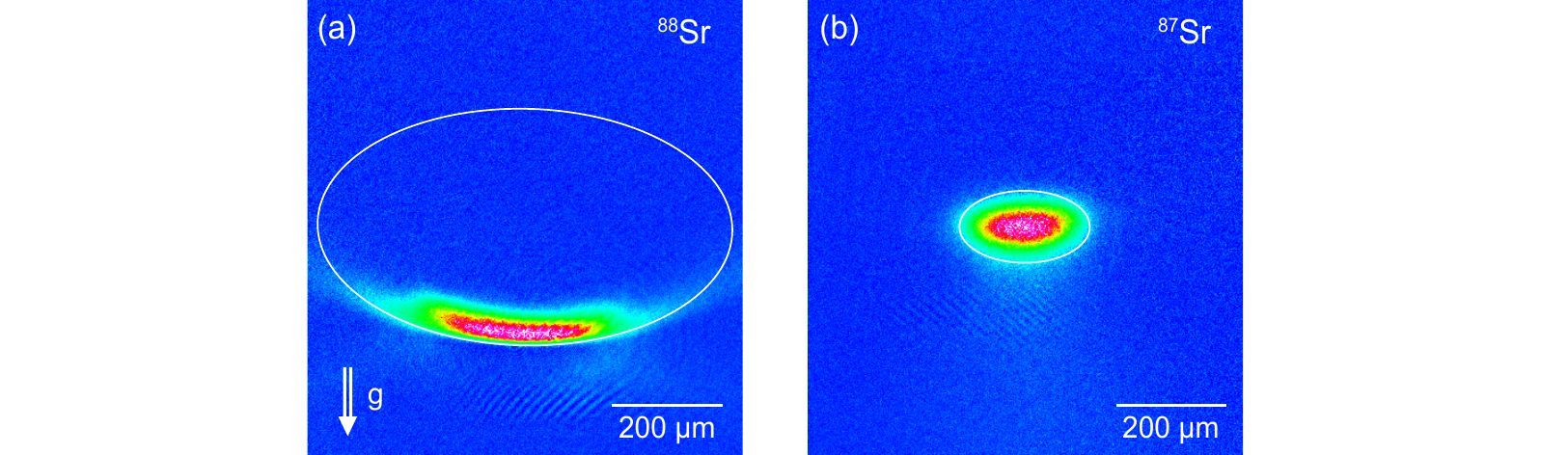}
\caption{Narrow-line MOTs of (a) the bosonic $^{88}$Sr and (b) the fermionic $^{87}$Sr isotopes, shown by \textit{in-situ} absorption images taken along the horizontal direction. In case (a), the MOT beams have a detuning of $-50\,$kHz and a peak intensity of $I_{\rm sat}$. Gravity and laser cooling forces balance each other on the surface of an ellipsoid, which has a vertical radius of $200\,\mu$m, giving rise to a pancake shaped MOT. In the fermionic case (b), we operate at a detuning of about $-20\,$kHz and an intensity equal to $I_{\rm sat}$ for both MOT frequency components. The atoms occupy the volume of an ellipsoid. In both cases, the magnetic field gradient is $\partial B/\partial z=1.15\,$G/cm, the atom number is $1.3(1)\times10^{6}$, and the temperature is $\sim 700\,$nK. The white ellipses are a guide to the eye.}
\label{fig:RedMOTs}
\end{figure}

The availability of narrow intercombination lines in strontium offers the intriguing opportunity to add a second cooling stage after the blue MOT in order to reduce the temperature and increase the density of the ensemble further. The second cooling stage is often referred to as \emph{narrow-line MOT} and frequently named ``red'' MOT, owing to the color of the transition wavelength at 689\,nm.

The linewidth of this transition amounts to $\Gamma=2\pi\times7.4\,$kHz, corresponding to a Doppler temperature of $T_D=179\,$nK; a factor of 4300 smaller than for the blue transition. This impressively low temperature demonstrates the power of narrow-line cooling. For linewidths on the order of kHz, the Doppler temperature might become comparable to the recoil temperature. The recoil temperature is $T_r=460\,$nK for the red transition, where the minimal attainable temperature is $T_r/2$ \cite{Castin1989lod}. Thus, Doppler and recoil limit almost coincide for the red transition in strontium.

\subsubsection{Bosons}

The atoms are repumped from the reservoir at temperatures set by the Doppler temperature of the blue MOT, roughly 1\,mK. A single frequency of the red MOT light would not provide sufficient capture efficiency, and we frequency-broaden the MOT light to match its frequency spectrum to the velocity distribution of the atoms. We use an acousto-optical modulator (AOM) to scan the frequency of the MOT light with a rate of about 20\,kHz, thereby creating a comb of lines extending from roughly $-200$\,kHz to $-5$\,MHz detuning. We have about 2.5\,mW of laser power available on each MOT axis, collimated to a waist of about 3\,mm. The saturation intensity of this transition is $I_{\mathrm{sat}}=3\,\mu\mathrm{W/cm}^2$, yielding a maximum intensity of 2000\,$I_{\mathrm{sat}}$ for our experimental setup. Considering a scan range of 5\,MHz comprising 250 comb lines at a spacing of 20\,kHz, the intensity per comb line is about 10\,$I_{\mathrm{sat}}$. We apply this broad-band red MOT already during the repumping process. The quadrupole field gradient along the vertical direction is ramped to $\partial B/\partial z = 1.15\,$G/cm within about a millisecond once the repumping light is applied. This capture phase lasts 50\,ms and is rather robust:~the lifetime of this MOT exceeds 1\,s at this stage.

It is helpful to visualize the geometric region in which atoms interact with the MOT light:~in the case of narrow-line cooling, the detuning $\Delta$ is much larger than the natural linewidth, $\Delta \gg \Gamma$. The light is only resonant with the atomic transition in regions where the B-field induced Zeeman shift $\Delta \nu = m_J g_J \mu_B B$ balances the detuning. Here, $\mu_B=1.4\,$MHz/G is the Bohr magneton, and the Land\'e g-factor is $g_J=1.5$ for the $^3P_1$ state. This region is the surface of an ellipsoid, where the vertical radius $\zeta$ of this ellipsoid is given by $\zeta=\Delta/(m_J g_J \mu_B \, \partial B/\partial z)$. Typical sizes are about 4\,mm for a gradient of 1\,G/cm and a detuning of 1\,MHz, but only $100\,\mu$m for a detuning of 20\,kHz. The thickness of such a shell is on the order of $10\,\mu$m for a small saturation parameter $s=I/I_{\mathrm{sat}}$. As we apply a frequency comb that stretches from near-zero to about $-5\,$MHz, atoms can get into resonance with the light on 250 narrow, but overlapping shells, filling the entire volume of the ellipsoid. For each shell, $\Delta \gg \Gamma_{\mathrm{sat}} \approx \Gamma$, where $\Gamma$ is the natural and $\Gamma_{\mathrm{sat}}=\Gamma \sqrt{1+s}$ the intensity-broadened linewidth.

In a second phase, we narrow the scan range down to 2\,MHz, where the comb line closest to resonance is 100\,kHz red-detuned to the transition. During this phase of 200\,ms, the total light intensity is reduced to about 100\,$I_{\mathrm{sat}}$ (corresponding to about $I_{\mathrm{sat}}$ per comb line), and the magnetic field gradient remains unchanged. Afterwards, we jump to single-frequency operation with a detuning of $-800$\,kHz and an unchanged intensity of 100\,$I_{\mathrm{sat}}$.

In the third stage, which we call the single-frequency MOT, we shift the frequency very close to resonance while reducing the intensity dramatically to 0.5\,$I_{\mathrm{sat}}$. This stage lasts 200\,ms and is concluded by a 50\,ms wait at the final parameters. It is important to understand that the MOT is driven through very different regimes during this ramp: We begin in the condition $|\Delta| \gg \Gamma_{\mathrm{sat}} \gg \Gamma$. Atoms are in resonance with the light on a single shell, whose thickness is enlarged by the factor $\sqrt{1+s}$ compared to the low-intensity case.  The large intensity ensures that the scattering rate is high enough to keep the atoms in the MOT, and lifetimes are typically 400\,ms. In this regime, the behavior of the atoms can be described semiclassically \cite{Loftus2004nlc}, and the expected temperature is $T = \hbar \Gamma_{\mathrm{sat}}/2k_B$. Note that this temperature is independent of the detuning, and set only by the light intensity. In a simplified picture, the decrease of the detuning provides compression, and the decrease of the intensity provides cooling.

At the end of this stage, the detuning becomes comparable to the linewidth, and $s$ approaches unity: $|\Delta| \sim \Gamma_{\mathrm{sat}} \sim \Gamma$. The behavior of the atoms is determined by single photon recoils, and the system requires a full quantum treatment \cite{Castin1989lod}; the temperature limit approaches $T_r/2$. There is, however, a compromise between atom number and temperature. A temperature of $T_r/2$ is reached only for very low intensity, accompanied by a very low scattering rate. Atoms populate only on a very thin shell (the bottom of the ellipsoid) and interact predominantly with the upward propagating beam; see Fig.~\ref{fig:RedMOTs}(a). At low intensity, an atom is at risk to fall through this shell without absorption of a photon, and be lost. This limits the lifetime at this stage to a few 10\,ms. The attainable temperature is limited by heating due to the re-absorption of photons, and depends on the density and scattering properties of the atoms. We typically achieve temperatures around 800\,nK with a few $10^7$ atoms of $^{84}$Sr, and temperatures as low as 400\,nK for the non-interacting isotope $^{88}$Sr or equivalently $^{84}$Sr at very low densities. Note that the light is still far detuned from the bare atomic transition, such that the atoms occupy the shell of an ellipsoid, about $600\,\mu$m below the quadrupole center, where the diameter of the cloud is typically $500\,\mu$m.

\subsubsection{Fermions}

The bosonic isotopes, for which we have discussed the red MOT dynamics in the previous section, have nuclear spin $I=0$ and therefore only one magnetic substate in the $^1S_0$ ground state. The fact that the magnetic moment $g_J\mu_B$ is zero due to the singlet configuration of the two valence electrons ($J=0$) did not become apparent. This however changes as we consider the fermionic $^{87}$Sr isotope with $I=9/2$ and its ten magnetic states. The magnetic moment is now given by the nuclear moment, which is still orders of magnitude smaller than an electronic magnetic moment. The condition of $J=0$ and $I\neq0$ in the ground state is quite unusual for MOT operation, and is reflected by the fact that the Land\'e $g$-factors of the ground- and excited state differ by about three orders of magnitude.

The experimental realization of a fermionic narrow-line MOT was pioneered by the Tokyo group \cite{Mukaiyama2003rll}. Cooling is performed on the $F=9/2 \rightarrow F'=11/2$ transition, where the large differential $g$-factor leads to a position-dependent restoring force. As a consequence, atoms in certain $m_F$ states at certain locations even experience a force away from the trap center. An effective restoring force for all atoms is obtained by rapid randomization of $m_F$ states. This is achieved by adding a so-called \emph{stirring laser} to the \emph{trapping laser}, operating on the $F=9/2 \rightarrow F'=9/2$ transition.

Just as in the bosonic case, we use the maximum available power on both the trapping and stirring beams to capture the atoms emerging from the metastable reservoir. Conditions for the broadband MOT are identical to the bosonic case described above. Final conditions of the red MOT are a gradient field of 1.15\,G/cm, trapping and stirring beam intensity of a few $I_{\mathrm{sat}}$, and detunings of only a few linewidths. We add a short wait time of 50\,ms to ensure equilibration and attain typical temperatures of 800\,nK with $10^7$ atoms. In contrast to the bosonic case, the fermionic MOT fills the entire volume of an ellipsoid; see Fig.~\ref{fig:RedMOTs}(b).

\subsection{Design and loading of the dipole trap}
\label{sec:DipoleTrapLoading}

Virtually all experiments using a narrow line for cooling towards quantum degeneracy choose a pancake-shaped dipole trap, or at least a trap that is elongated in the horizontal plane \cite{Takasu2003ssb,Kraft2009bec,Stellmer2009bec,Martinez2009bec}.

There are two reasons for this choice. At first, the narrow-line MOT itself is pancake-shaped, and a dipole trap of similar shape provides improved mode-matching. The second reason refers to the evaporation efficiency: During evaporation, atoms will leave the trap predominantly vertically downwards, aided by gravity. The evaporation efficiency benefits from a high vertical trap frequency: once a high-energy atom is produced in a collision, it ought to escape the trap before colliding with another atom. The vertical trap frequency should thus be large compared to the scattering rate: this requirement suggests a pancake-shaped trap. It is fortunate that both the loading of the dipole trap and the evaporation efficiency are optimized with the same trap shape.

The trap can be formed by two intersecting horizontal beams \cite{Martinez2009bec} or by an elliptic horizontal beam, intersecting with a rather large vertical beam that provides additional confinement in the horizontal plane \cite{Stellmer2013poq}. The ellipticity of the horizontal beams can be as extreme as $1:20$, and the ratio between the vertical and the lowest horizontal trap frequency can reach $100:1$.

The dipole trap is turned on from the beginning of the red MOT, and atoms are continuously loaded into the dipole trap once spatial overlap is achieved and the temperature drops below the trap depth. Taking great care to reduce and compensate the light shifts imposed by the dipole trap, we are able to transfer 50\% of the atoms from the single-frequency MOT into the dipole trap while maintaining the temperature of the MOT \cite{Stellmer2013poq,Stellmer2013lct}. Once the atoms are loaded into the dipole trap, the MOT light is kept on for another 100\,ms at an intensity of about $0.5\,I_{\rm sat}$. During this time, the atoms are pushed into the center of the dipole trap by the horizontal MOT beams, thereby increasing the density. For the bosonic case, the quadrupole center is placed about $600\,\mu$m above the horizontal dipole trap beam. The detuning of the cooling light from the Zeeman-shifted and light-shifted $\sigma^+$-resonance position is about $-3\,\Gamma / 2 \pi$. In the fermionic case, the quadrupole center is overlapped with the dipole trap. Working with a mixture of bosonic and fermionic isotopes requires a sequential loading scheme, in which we load the fermions first and then shift the quadrupole center upwards to load the bosons.

\section{Photoassociation of atomic strontium}
\label{sec:Photoassociation}

Photoassociation (PA) spectroscopy is an important tool for determining and manipulating the scattering properties of ultracold atoms and for forming molecules. There has been a significant amount of work in this area with strontium for several reasons. The knowledge of atom-atom interactions gained through PA is critical for designing experiments to reach quantum degeneracy, and this is particularly important in strontium because some of the isotopes have scattering properties that are not ideal for evaporative cooling. The formation of ground-state molecules is now a major theme of ultracold physics research, and strontium offers efficient routes to achieve this through PA. Finally, narrow-line PA  near the $^1S_0 - {^3P_1}$ intercombination transition is different in many ways from traditional PA with broad, electric-dipole-allowed transitions, and it holds promise for optical Feshbach resonances with reduced losses.

\subsection{One-color photoassociation}

For a PA measurement, a sufficiently cold and dense cloud of atoms is illuminated by light detuned from an atomic transition. The frequency of the light is varied, and whenever it comes into resonance with a transition between two free atoms and an excited molecular state, molecules are created. These excited molecules then quickly decay into deeply-bound states, which are invisible on absorption images.

The first PA experiments in strontium involved excitation to molecular states on the $5s^2\,{^1S_0} + {5s5p}\,{^1P_1}$ $^1\Sigma^+_u$ potential using 461\,nm light \cite{Nagel2005psa,Mickelson2005sdo,Yasuda2006pso}. This allowed accurate determination of the $^1S_0 + {^1P_1}$ $C_6$ coefficient and the $^1P_1$ atomic decay rate, $\Gamma/2\pi=(30.24 \pm 0.02)$\,MHz \cite{Yasuda2006pso}. Measurement of the variation of the intensities of the transitions to different molecular levels allowed a preliminary determination of the ground-state $s$-wave scattering lengths \cite{Mickelson2005sdo}.

Of more interest for the study and control of quantum degenerate gases is narrow-line one-color PA to bound states of the $5s^2\,{^1S_0} + {5s5p}\,{^3P_1}$ $0_u$ and $1_u$ molecular potentials to the red of the intercombination-line transition at 689\,nm. The first experiments \cite{Zelevinsky2006nlp} allowed accurate determination of atomic and molecular parameters, especially the $^3P_1$ atomic decay rate, $\Gamma/2\pi=(7.40 \pm 0.07)$\,kHz.

The small decay rate of the molecular states, $\Gamma_{\rm mol}=2\times\Gamma$, has several important implications. Narrow-line PA opens a new regime, also explored in ytterbium \cite{Tojo2006hrp}, in which the transition linewidth is much smaller than the level spacings even for the least-bound molecular levels. Small $\Gamma_{\rm mol}$ also implies a weak dipole-dipole interaction between $^1S_0$ and $^3P_1$ atoms during PA. This gives rise to similar ground and excited molecular potentials and very large Frank-Condon factors for bound-bound transitions.

Large free-bound matrix elements suggest that these transitions can be used to manipulate atomic interactions through an optical Feshbach resonance with reduced inelastic loss \cite{Ciurylo2005oto,Ciurylo2006spa}, which will be described in Sec.~\ref{sec:OFR}. It has also been predicted that strong transitions and long molecular lifetimes can combine to yield atom-molecule Rabi frequencies that exceed decoherence rates to enable coherent single-photon PA \cite{Naidon2008tbt}, which is inaccessible with electric-dipole-allowed transitions. Large bound-bound matrix elements are important for creating ground-state molecules, either through spontaneous decay after one-color PA \cite{Reinaudi2012opo,Kato2012ool} or through two-color PA techniques \cite{Stellmer2012cou}. In fact, near-unity Frank-Condon factors were found for several bound-bound transition \cite{Reinaudi2012opo}, which enables very efficient, state-selective molecule production by spontaneous emission. Driving one or more additional Raman transitions should populate the absolute ground state of the Sr$_2$ system.

Initial experiments with intercombination-line PA were performed with $^{88}$Sr \cite{Zelevinsky2006nlp}, but measurements have been extended to include all the bosonic isotopes \cite{Stellmer2012cou}. Table \ref{tab:1ColorPA} gives binding energies of excited molecular states that have been determined for  $^{84}$Sr, $^{86}$Sr, and $^{88}$Sr, respectively.

\begin{table}[b]
\centering
\begin{tabular*}{155mm}{@{\extracolsep{\fill}}cccccc}\hline\hline \noalign{\smallskip}
&\multicolumn{1}{c}{$^{84}$Sr (Ref.~\cite{Stellmer2012cou})}&\multicolumn{2}{c}{$^{86}$Sr (Ref.~\cite{Borkowski2013fom})}&\multicolumn{2}{c}{$^{88}$Sr  (Ref.~\cite{Zelevinsky2006nlp})} \\
\multicolumn{1}{c}{$\nu$}& \multicolumn{1}{c}{0$_u$} & \multicolumn{1}{c}{0$_u$} & \multicolumn{1}{c}{1$_u$} & \multicolumn{1}{c}{0$_u$} & \multicolumn{1}{c}{1$_u$} \\   \noalign{\smallskip}\hline\noalign{\smallskip}
$-1$ & $-0.32(1)$     & $-1.63(1)$      & $-159.98(5)$    &  $-0.435(37)$      &   $-353.236(35)$         \\
$-2$ & $-23.01(1)$         & $-44.25(1)$     &                 &  $-23.932(33)$      &  $-2683.722(32)$         \\
$-3$ & $-228.38(1)$        & $-348.74(1)$    &                 &  $-222.161(35)$      &  $-8200.163(39)$        \\
$-4$ & $-1288.29(1)$       &                   &               &  $-1084.093(33)$      &           \\
$-5$ &                     &                   &               &  $-3463.280(33)$      &               \\
$-5$ &                     &                   &               &  $-8429.650(42)$      &           \\ \hline \hline
\end{tabular*}
\caption{Binding energies in MHz of the $\ell=1$  states of the highest vibrational levels of the 0$_u$ and 1$_u$ potentials, where $\ell$ is the total angular momentum quantum number. The levels are labeled by $\nu$, starting from above with $\nu=-1$.}
\label{tab:1ColorPA}
\end{table}

\subsection{Two-color photoassociation}

\begin{figure}
  \includegraphics[width=165mm]{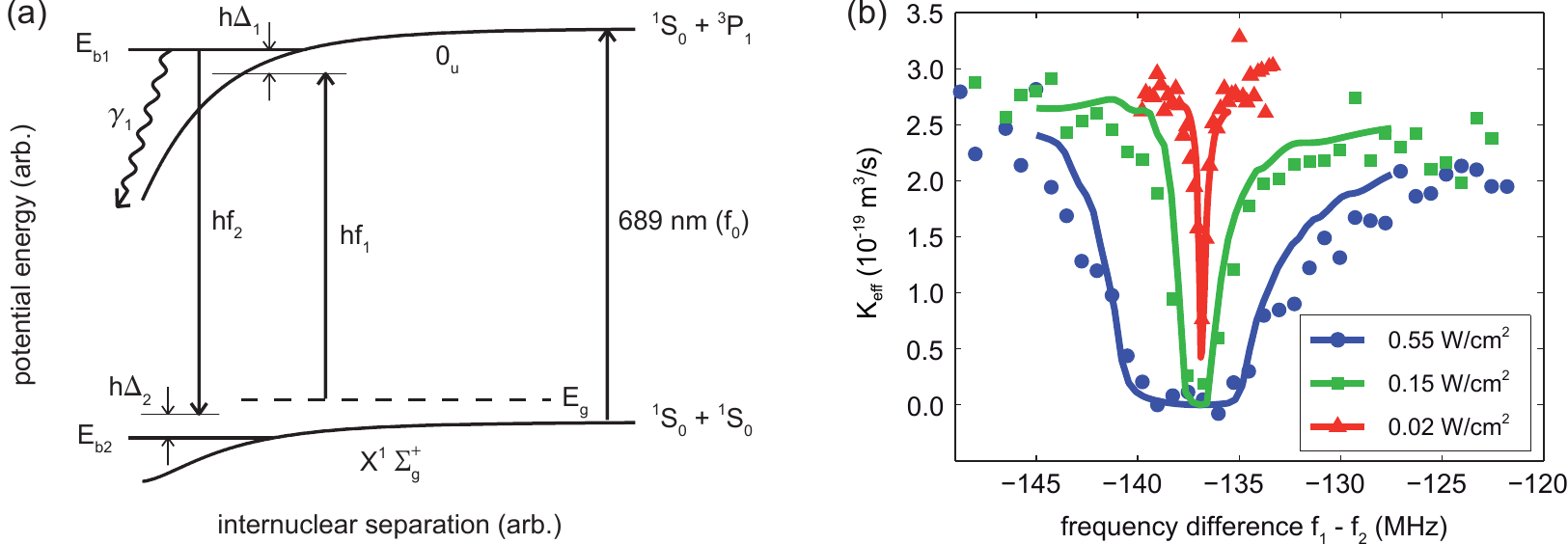}
  \caption{(a) Two-color PA spectroscopy diagram. $E_g$ is the kinetic energy of the colliding atom pair. $E_{b_1}$ is the energy of the bound state of the excited molecular potential that is near resonance with the free-bound laser. $E_{b_2} <0$ is the unperturbed energy of the bound state of the ground molecular potential. The photon of energy $h f_1$  is detuned from $E_{b1}$ by $h\Delta_1$, while the photon of energy $h f_2$ is detuned from $E_{b2}$ by $h\Delta_2$. The decay rate of $b_1$ is $\Gamma_1$. (b) Inelastic-collision event rate constant $K_{\rm eff}$ versus frequency difference between free-bound and bound-bound lasers for spectroscopy of the $\ell=0$, $v=62$ level of the $X{^1\Sigma_g^+}$ potential. The frequency of the free-bound laser is fixed close to the one-photon PA resonance and its intensity is 0.05\,W/cm$^2$. The bound-bound laser frequency is scanned, and its intensity is indicated in the legend. On two-photon resonance, PA loss is suppressed due to quantum interference. The solid lines are model fits yielding the binding energy $E_{b2}/h=-136.7(2)$\,MHz. The figure is taken from Ref.~\cite{Martinez2008tpp}.}
\label{PASDiagram}
\end{figure}

In two-color PA, two laser fields couple colliding atoms  to a weakly bound state of the ground molecular potential via a near-resonant intermediate state. In strontium, this was first performed in a thermal gas of $^{88}$Sr with the goal of measuring the  binding energies of weakly bound levels of the ground-state $X$$^1\Sigma^+_g$ potential \cite{Martinez2008tpp}; see Fig.~\ref{PASDiagram}(a). It was also used in a stimulated Raman adiabatic passage (STIRAP) process \cite{Vitanov2001lip} to coherently produce molecules in the ground electronic state from quantum degenerate $^{84}$Sr in an optical lattice \cite{Stellmer2012cou}. The STIRAP experiment will be described in Sec.~\ref{sec:Molecules}.

For experiments with a thermal gas of $^{88}$Sr, we closely follow the description in Ref.~\cite{Martinez2008tpp}. Atoms are held in an optical dipole trap, with a temperature of several $\mu$K and peak densities on the order of $10^{14}$\,cm$^{-3}$.  A dark resonance is used to determine the binding energy of  molecular levels of the ground-state potential. The frequency of the free-bound laser is held fixed close to the one-color resonance, $\Delta_1\simeq0$, while the bound-bound laser detuning $\Delta_2$ is scanned. When $\Delta_2-\Delta_1=0$, the system is in two-color resonance from state $g$ to $b_2$, and one-color photoassociative loss is suppressed due to quantum interference. At this point, $f_1-f_2=(E_{b2}-E_g)/h$, so the spectrum allows accurate determination of  $E_{b_2}$. Averaging over $E_g$ is necessary in order to properly account for thermal shifts of the
resonance.

At the low temperatures of atoms in the dipole trap, only $s$-wave collisions occur so only $\ell=1$ intermediate levels and $\ell=0$ and 2 final states are populated. Figure \ref{PASDiagram}(b) shows a series of spectra taken at various bound-bound intensities for $b_2$ equal to the $\ell=0$, $v=62$ state; denoted as $\nu=-1$ in Tab.~\ref{tab:2ColorPA}. The detuning of the free-bound laser frequency $f_1$ from the free-bound resonance, which depends on the collision energy $E_g$ and AC Stark shift from the dipole trap, causes slight asymmetry in the lines and broadening, but this can be accounted for by a simple model \cite{Martinez2008tpp}. We also measured the binding energy of the $\ell=2$, $v=62$ state; see Tab.~\ref{tab:2ColorPA}.

Knowledge of the binding energies in $^{88}$Sr allowed accurate determination of the $s$-wave scattering lengths for all isotopic collision possibilities.  This relied upon  a relativistic many-body calculation  of the dispersion coefficients for the long-range behavior of the ground-state molecular potential \cite{Porsev2006}. PA measurements were later combined with Fourier-transform spectroscopy of molecular levels of the $X$$^1\Sigma_g$ potential to yield further improvements and the most accurate determination of the ground molecular potential and scattering lengths \cite{Stein2010tss}.

\begin{figure}
 \includegraphics[width=165mm]{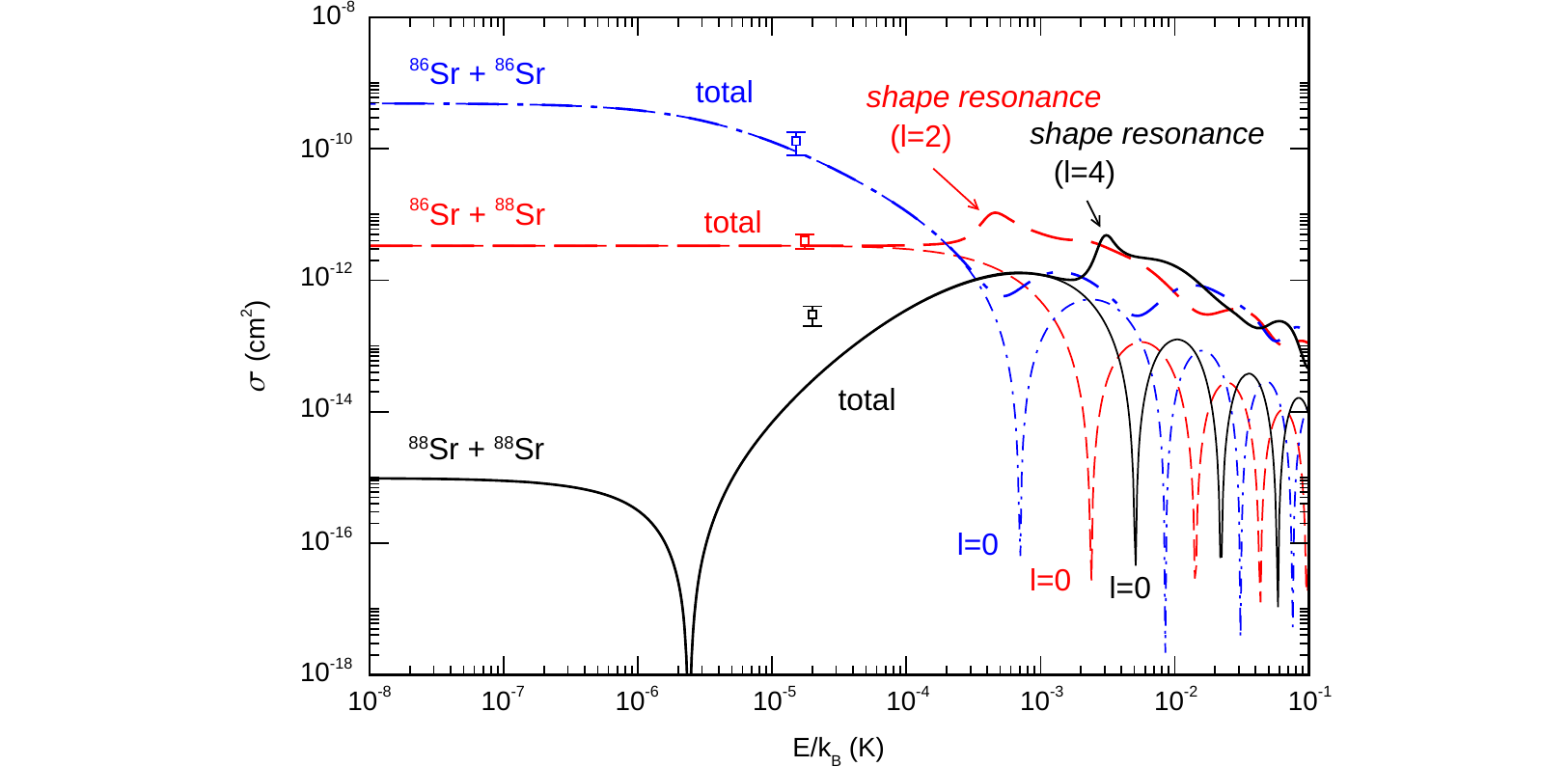}
 \caption{Dependence of elastic-scattering cross sections $\sigma$ on collision energy E in Kelvin for selected strontium isotopes. The thick lines are cross sections including partial waves up to $\ell=4$. Shape resonances are indicated. Thin lines indicate cross section contributions from $\ell=0$ only. The data symbols are cross section measurements from thermalization experiments \cite{Ferrari2006cos}, and the respective collision energies are set to $E=k_B T$, where $T$ is the sample temperature. The figure is taken from Ref.~\cite{Martinez2008tpp}.}
\label{figure:BosonicCrossSecVsEnergy}
\end{figure}

According to the Wigner threshold law, the elastic cross-section for collisions between neutral particles approaches a constant as the collision energy goes to zero. Most experiments with ultracold atoms reach this limit, and the cross-section is well described by an energy-independent $\ell =0$ partial wave for distinguishable particles or indistinguishable bosons. However, this is not the case when there is a low-energy scattering resonance or when the scattering length is very small. Figure \ref{figure:BosonicCrossSecVsEnergy} demonstrates that $^{88}$Sr$-{^{88}}$Sr and $^{86}$Sr$-{^{86}}$Sr collision cross-sections vary significantly with collision energy, even at energies below $1\,\mu$K.

Table \ref{tab:2ColorPA} lists all the binding energies that have been determined for  molecular Sr$_2$ in the ground electronic state.

\begin{table}[t]
\centering
\begin{tabular*}{130mm}{@{\extracolsep{\fill}}ccccc}\hline\hline \noalign{\smallskip}
&\multicolumn{2}{c}{$^{84}$Sr (Ref.~\cite{Stellmer2012cou})}&\multicolumn{2}{c}{$^{88}$Sr  (Ref.\ \cite{Martinez2008tpp})} \\
\multicolumn{1}{c}{$\nu$}& \multicolumn{1}{c}{$\ell=0$} & \multicolumn{1}{c}{$\ell=2$} & \multicolumn{1}{c}{$\ell=0$} & \multicolumn{1}{c}{$\ell=2$} \\ \noalign{\smallskip}\hline\noalign{\smallskip}
$-1$ & $-13.7162(2)$ & -                       &  $-136.7(2)$      &   $-66.6(2)$        \\
$-2$ & $-644.7372(2)$ & $-519.6177(5)$ &                           &              \\ \hline \hline
\end{tabular*}
\caption{Binding energies in MHz of the $\ell=0,2$  states of the highest vibrational levels of the  $X{^1\Sigma_g^+}$ potentials, where $\ell$ is the rotational angular momentum quantum number. The levels are labeled by $\nu$, starting from above with $\nu=-1$.}
\label{tab:2ColorPA}
\end{table}

\section{Bose-Einstein condensation of strontium}
\label{sec:BEC}

\subsection{Bose-Einstein condensation of $^{84}$Sr}

The early experiments towards quantum degeneracy in strontium were focused on the three relatively abundant isotopes $^{86}$Sr (9.9\%), $^{87}$Sr (7.0\%), and $^{88}$Sr (82.6\%), the first and the last one being bosonic. The necessary phase-space density for BEC or Fermi degeneracy could not be achieved in spite of considerable efforts \cite{Katori2001lco,Ferrari2006cos}. For the two bosonic isotopes the scattering properties turned out to be unfavorable for evaporative cooling \cite{Ferrari2006cos}. The scattering length of $^{88}$Sr is close to zero, such that elastic collisions are almost absent. In contrast, the scattering length of $^{86}$Sr is very large, leading to detrimental three-body recombination losses. Magnetic Feshbach resonances are absent in the bosonic alkaline-earth systems, and optical Feshbach resonances are accompanied by strong losses on the timescales required for evaporation.

\subsubsection{First attainment of BEC in strontium}

The first BECs of strontium were attained in 2009 using the isotope $^{84}$Sr. This isotope has a natural abundance of only 0.56\% and, apparently for this reason, had received little attention up to that time. The low abundance does not represent a serious disadvantage for BEC experiments, as it can be overcome by the accumulation scheme described in Sec.~\ref{sec:LaserCooling}. Because of the favorable scattering length of $+123\,a_0$ \cite{Ciuryloprivate,Stein2008fts,Martinez2008tpp}, there is no need of Feshbach tuning. Ironically, this low-abundance isotope turned out to be the prime candidate among all alkaline-earth isotopes to obtain large BECs and might also allow for sympathetic cooling of other isotopes and elements.

In the following, we will state the experimental procedure of the early Innsbruck experiment \cite{Stellmer2009bec}. The laser cooling stages were already described in Sec.~\ref{sec:LaserCooling}. To prepare the evaporative cooling stage, the atoms are transferred into a crossed-beam dipole trap, which is derived from a 16-W laser source operating at 1030\,nm in a single longitudinal mode. Our trapping geometry follows the basic concept successfully applied in experiments on ytterbium and calcium BEC \cite{Takasu2003ssb,Fukuhara2007bec,Fukuhara2009aof,Kraft2009bec}. The trap consists of a horizontal and a near-vertical beam with waists of $32\,\mu$m and $80\,\mu$m, respectively, thus creating a cigar-shaped geometry. Initially the horizontal beam has a power of 3\,W, which corresponds to a potential depth of $110\,\mu$K and oscillation frequencies of 1\,kHz radially and a few Hz axially. The vertical beam contains 6.6\,W, which corresponds to a potential depth of $37\,\mu$K and a radial trap frequency of 250\,Hz. Axially, the vertical beam does not provide any confinement against gravity. In the crossing region the resulting potential represents a nearly cylindrical trap. In addition the horizontal beam provides an outer trapping region of much larger volume, which is advantageous for the trap loading.

The dipole trap is switched on at the beginning of the red MOT compression phase. After switching off the red MOT, we observe $2.5\,\times 10^6$ atoms in the dipole trap with about $10^6$ of them residing in the crossing region. At this point we measure a temperature of $\sim10\,\mu$K, which corresponds to roughly one tenth of the potential depth. We then apply forced evaporative cooling by exponentially reducing the power of both beams with a $1/e$ time constant of $\sim3\,$s. The evaporation process starts under excellent conditions, with a peak number density of $1.2 \times 10^{14}$\,cm$^{-3}$, a peak phase-space density of $\sim2 \times 10^{-2}$, and an elastic collision rate of about 3500\,s$^{-1}$. During the evaporation process the density stays roughly constant and the elastic collision rate decreases to $\sim700\,$s$^{-1}$ before condensation. The evaporation efficiency is very large as we gain at least three orders of magnitude in phase-space density for a loss of atoms by a factor of ten.

\begin{figure}
\centering
\includegraphics[width=165mm]{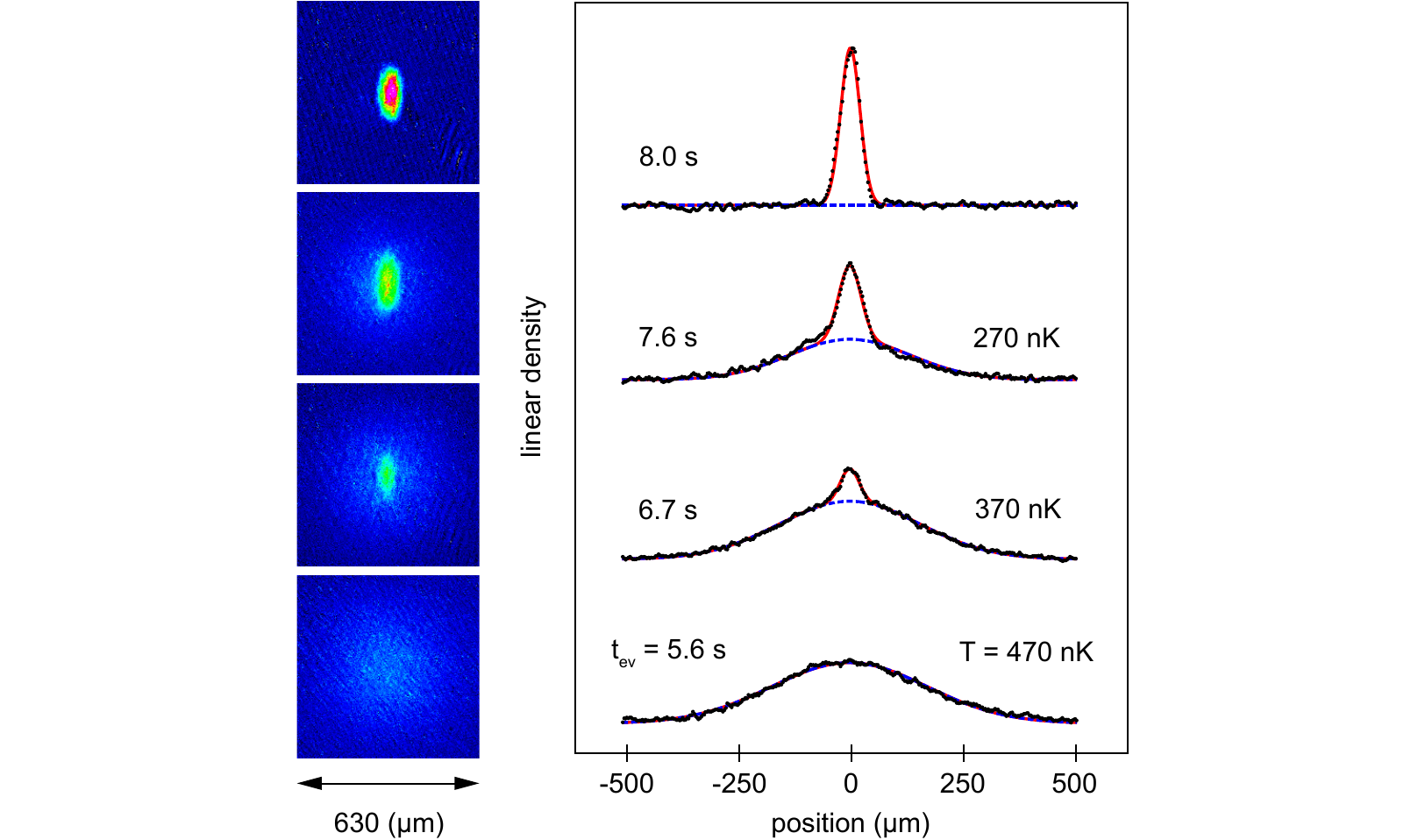}
\caption{Absorption images and integrated density profiles showing the BEC phase transition for different times $t_{\rm ev}$ of the evaporative cooling ramp. The images are taken along the vertical direction 25\,ms after release from the trap. The solid line represents a fit with a bimodal distribution, while the dashed line shows the Gaussian-shaped thermal part, from which the given temperature values are derived. The figure is taken from Ref.~\cite{Stellmer2009bec}.}
\label{fig:BEC1}
\end{figure}

The phase transition from a thermal cloud to BEC becomes evident in the appearance of a textbooklike bimodal distribution, as clearly visible in time-of-flight absorption images and the corresponding linear density profiles shown in Fig.~\ref{fig:BEC1}. At higher temperatures the distribution is thermal, exhibiting a Gaussian shape. Cooling below the critical temperature $T_c$ leads to the appearance of an additional, narrower and denser, elliptically shaped component, representing the BEC. The phase transition occurs after 6.3\,s of forced evaporation, when the power of the horizontal beam is 190\,mW and the one of the vertical beam is 410\,mW. At this point, with the effect of gravitational sag taken into account, the trap depth is $2.8\,\mu$K. The oscillation frequencies are 59\,Hz in the horizontal axial direction, 260\,Hz in the horizontal radial direction, and 245\,Hz in the vertical direction.

For the critical temperature we obtain $T_c = 420$\,nK by analyzing profiles as displayed in Fig.~\ref{fig:BEC1}. This agrees within 20\%, \textit{i.e.}~well within the experimental uncertainties, with a calculation of $T_c$ based on the number of $3.8 \times 10^5$ atoms and the trap frequencies at the transition point. Further evaporation leads to an increase of the condensate fraction and we obtain a nearly pure BEC without discernable thermal fraction after a total ramp time of 8\,s. The pure BEC that we can routinely produce in this way contains $1.5\times10^5$ atoms and its lifetime exceeds 10\,s.

\begin{figure}[b]
\centering
\includegraphics[width=165mm]{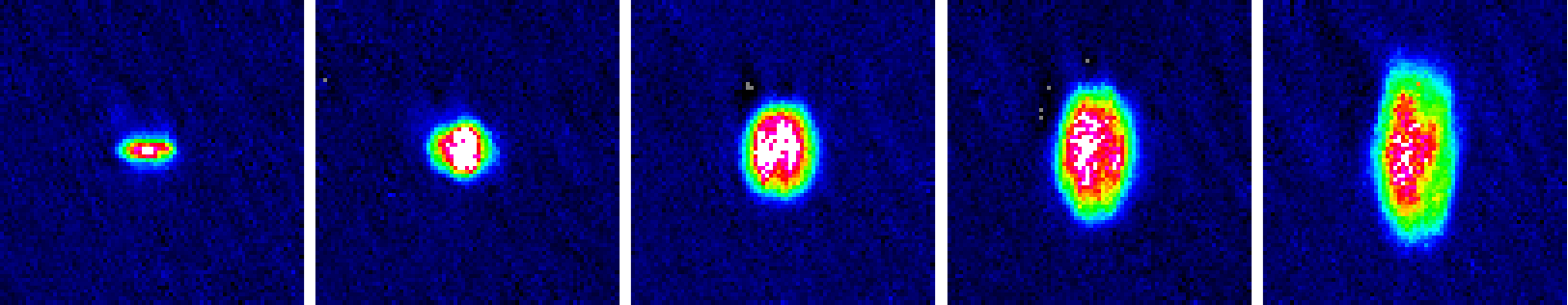}
\caption{Inversion of the aspect ratio during the expansion of a pure BEC. The images (field of view $250\,\mu{\rm m}\times 250\,\mu{\rm m}$) are taken along the vertical direction. The first image is an \textit{in-situ} image recorded at the time of release. The further images are taken 5\,ms, 10\,ms, 15\,ms, and 20\,ms after release. The figure is taken from Ref.~\cite{Stellmer2009bec}.}
\label{fig:BEC2}
\end{figure}

The expansion of the pure condensate after release from the trap clearly shows another hallmark of BEC. Figure~\ref{fig:BEC2} demonstrates the well-known inversion of the aspect ratio \cite{Anderson1995oob,Inguscio1999book}, which results from the hydrodynamic behavior of a BEC and the fact that the mean field energy is released predominantly in the more tightly confined directions.
Our images show that the cloud changes from an initial prolate shape with an aspect ratio of at least 2.6 (limited by the resolution of the \textit{in-situ} images) to an oblate shape with aspect ratio 0.5 after 20\,ms of free expansion. From the observed expansion we determine a chemical potential of $\mu/k_B\approx150$\,nK for the conditions of Fig.~\ref{fig:BEC2}, where the trap was recompressed to the setting at which the phase transition occurs in the evaporation ramp. Within the experimental uncertainties, this agrees with the calculated value of $\mu/k_B\approx180$\,nK.\\

The corresponding Rice experiment reaching BEC in $^{84}$Sr \cite{Martinez2009bec} resembles the Innsbruck experiment very closely. Nearly all experimental parameters of the sequence are almost identical to the one described above. The main difference lies in the fact that two near-horizontal dipole trap beams of $100\,\mu$m waist are used. The large trap allows for the loading of more atoms at a lower temperature, however at a lower density and smaller collision rate. To improve evaporative cooling, the trap is re-compressed after loading of atoms from the red MOT. The evaporation time of 4.5\,s is slightly shorter than in the Innsbruck experiment and leads to pure BECs of typically $3\times10^5$ atoms.

\subsubsection{BECs of large atom number}

The first BECs of $^{84}$Sr contained a few $10^5$ atoms, but were far from being optimized. A careful optimization of various parameters of the experimental sequence, most of all the transfer into the dipole trap and its geometric shape, allowed us to increase the BEC atom number into the $10^7$ range \cite{Stellmer2013poq}. To the best of our knowledge, these BECs are the largest ones ever created by evaporative cooling in an optical dipole trap. This experiment represents the current state of the art and will be described in the following.

To overcome the low natural abundance of $^{84}$Sr, we accumulate atoms in the metastable reservoir for 40\,s. This time is slightly longer than the lifetime of the gas in the reservoir, and further loading does not increase the atom number significantly. The atoms are returned into the ground state, cooled and compressed by the red MOT, and transferred into the dipole trap. For this experiment, we use only the horizontal dipole trap beam, which has an initial depth of $k_B \times 12\,\mu$K and provides initial trapping frequencies of $f_x=45\,$Hz and $f_y=6\,$Hz in the horizontal and $f_z=650\,$Hz in the vertical directions. This beam has an aspect ratio of $1:15$, with waists of about $\omega_x=120\,\mu$m and $\omega_z=18\,\mu$m. After ramping the red MOT light off over 100\,ms, the gas is allowed to thermalize in the dipole trap for 250\,ms. At this point, about $4 \times 10^7$ atoms reside in the dipole trap at a temperature of 1.5\,$\mu$K. The peak density of the gas is $7\times 10^{13}\,{\rm cm}^{-3}$, the average elastic collision rate is $650\,{\rm s}^{-1}$, and the peak phase-space density is 0.3. The power of the dipole trap is reduced exponentially from its initial value of 2.4\,W to 425\,mW within 10\,s.

After 7\,s of evaporation a BEC is detected. At this time, $2.5\times10^7$ atoms remain in the trap at a temperature of about 400\,nK. The evaporation efficiency is high with four orders of magnitude gain in phase-space-density for a factor ten of atoms lost. After 10\,s of evaporation, we obtain an almost pure BEC of $1.1(1)\times 10^7$ atoms. The trap oscillation frequencies at this time are $f_x=20\,$Hz, $f_y=2.5\,$Hz, and $f_z=260\,$Hz. The BEC has a peak density of $2.2\times 10^{14}\,{\rm cm}^{-3}$ and the shape of an elongated pancake with Thomas-Fermi radii of about $R_x=40\,\mu$m, $R_y=300\,\mu$m, and $R_z=3\,\mu$m. The lifetime of the BEC is 15\,s, likely limited by three-body loss.

An increase of the BEC atom number towards the range of $10^8$ should be achievable by simple improvements. A larger volume of the dipole trap, facilitated by an increased ellipticity of the horizontal dipole trap beam, would allow us to support more atoms without a change to the density. The increase in atom number would be accomplished by an increased atomic flux of the oven, while the single-frequency red MOT would be operated at a larger detuning to avoid loss by light-assisted collisions. The larger detuning increases the size of the MOT, such that the peak density does not increase despite a larger atom number.

\subsubsection{Short cycle times}
\label{sec:fast_cycle}

In the previous section, we reported on experiments optimized for a large number of atoms in the BEC. We can also optimize our experimental sequence for a short cycle time. Nearly all experiments profit from the higher data rate made possible by a shorter cycle time. Precision measurement devices, such as atom interferometers, do require high repetition rates or a favorable ratio of probe time versus cycle time and might profit from the coherence of a BEC. Quantum gas experiments taking place in an environment of poor vacuum quality also require a short production time. Most quantum gas experiments have cycle times of a few ten seconds. Experiments that have been optimized for speed while using an all-optical approach achieve cycle times of 3\,s for degenerate bosonic gases \cite{Kinoshita2005aob,Kraft2009bec}. Cycle times down to 1\,s can be reached by using magnetic trapping near the surface of a microchip \cite{Hannover}.

Making use of the very high phase-space density achieved already in the red MOT, as well as the excellent scattering properties of $^{84}$Sr, we are able to reduce the cycle time to 2\,s \cite{Stellmer2013poq}. At the beginning of the cycle, we operate the blue MOT for 800\,ms to load the metastable reservoir. A short flash of repump light returns the metastable atoms into the ground state, where they are trapped, compressed, and cooled to about 1.2\,$\mu$K by the red MOT. Close to $4\times10^6$ atoms are loaded into a dipole trap, which is formed by the horizontal sheet and a vertical beam of 25\,$\mu$m $1/e^2$-radius in the plane of the horizontal dipole trap. The atomic cloud is not only populating the cross of the dipole trap, but extends $\sim1\,$mm along the horizontal dipole trap. Forced evaporation reduces the trap depth over 550\,ms with an exponential time constant of about 250\,ms.

During evaporation, a large fraction of the atoms in the horizontal beam migrate into the crossing region. The phase transition occurs after about 270\,ms of evaporation, and after 480\,ms, the thermal fraction within the crossing region cannot be discerned, indicating an essentially pure BEC in this region. Further evaporation does not increase the BEC atom number, but efficiently removes the thermal atoms residing in the horizontal beam. The BEC is formed by about $10^5$ atoms at the end of evaporation.

The read-out of the charged-coupled device (CCD) chip used for imaging can be performed during the consecutive experimental cycle and is therefore not included in the 2\,s period. The cycle time could be improved substantially if the reservoir loading time was reduced, \textit{e.g.}~by increasing the oven flux. It seems that cycle times approaching 1\,s are within reach.

\subsubsection{Laser cooling to quantum degeneracy}
\label{sec:oBEC}

The remarkable conjunction of supreme laser cooling performance and the excellent scattering properties allow us to reach a phase space density of about 0.1 directly after loading into the dipole trap; just one order of magnitude shy of quantum degeneracy \cite{Ido2000odt}. It is now a challenging and amusing task to bridge this last order of magnitude and create a BEC without the cooling stage of evaporation.

Reaching a high phase space density not only requires a low temperature and a high density, but also a mechanism to suppress the (re-)absorption of cooling light photons, which counteracts the advancement towards BEC by constituting an effective repulsion and leading to heating and loss.

Here, we present an experiment that overcomes these challenges and creates a BEC of strontium by laser cooling \cite{Stellmer2013lct}. Our scheme essentially relies on the combination of three techniques, favored by the properties of this element, and does not rely on evaporative cooling. The narrow 7.4-kHz cooling transition enables simple Doppler cooling down to temperatures of 350\,nK \cite{Ido2000odt}. Using this transition, we prepare a laser cooled sample of $10^7$ atoms of $^{84}$Sr in a large ``reservoir'' dipole trap. To avoid the detrimental effects of laser cooling photons, we render atoms transparent for these photons in a small spatial region within the laser cooled cloud. Transparency is induced by a light shift on the optically excited state of the laser cooling transition. In the region of transparency, we are able to increase the density of the gas, by accumulating atoms in an additional, small ``dimple'' dipole trap \cite{StamperKurn1998rfo,Weber2003bec}. Atoms in the dimple thermalize with the reservoir of laser cooled atoms by elastic collisions and form a BEC. A striking feature of our technique is that the BEC is created within a sample that is being continuously laser cooled.

\begin{figure}[b]
\centering
\includegraphics[width=165mm]{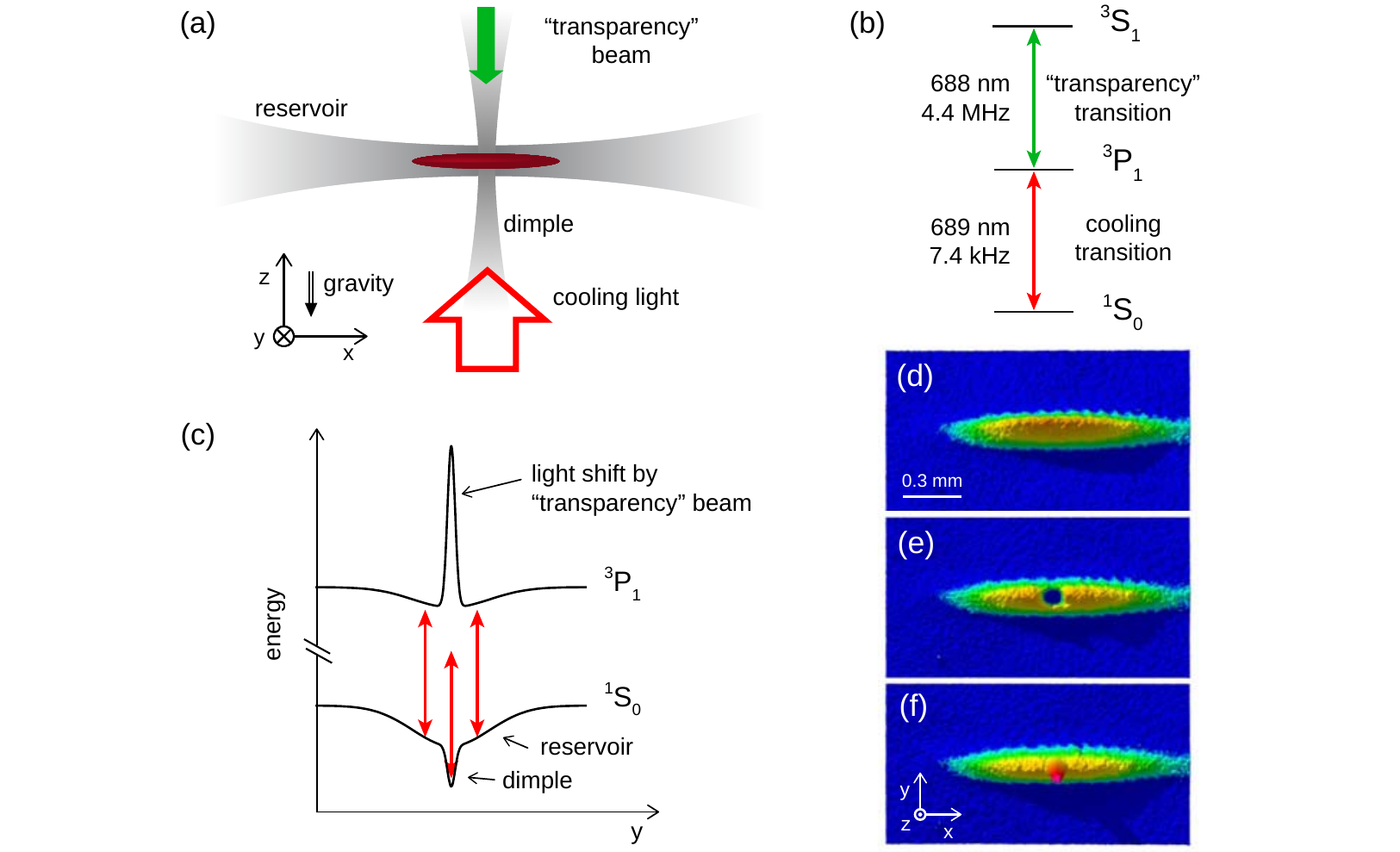}
\caption{Scheme to reach quantum degeneracy by laser cooling. (a) A cloud of atoms is confined in a deep reservoir dipole trap and exposed to a single laser cooling beam (red arrow). Atoms are rendered transparent by a ``transparency'' laser beam (green arrow) and accumulate in a dimple dipole trap
by elastic collisions. (b) Level scheme showing the laser cooling transition and the transparency transition. (c) Potential experienced by $^1S_0$ ground-state atoms and atoms excited to the $^3P_1$ state. The transparency laser induces a light shift on the $^3P_1$ state, which tunes the atoms out of resonance with laser cooling photons. (d) to (f) Absorption images of the atomic cloud recorded using the laser cooling transition. The images show the cloud from above and demonstrate the effect of the transparency laser (e) and the dimple (f). (d) is a reference image without these two laser beams. The figure is taken from Ref.~\cite{Stellmer2013lct}.}
\label{fig:BEC3}
\end{figure}

The details of our scheme are shown in Fig.~\ref{fig:BEC3}. The pre-cooling stages of two sequential MOTs and the dipole trap loading are identical to the protocol described in Sec.~\ref{sec:LaserCooling}. The trap consists of a 1065-nm laser beam propagating horizontally. The beam profile is strongly elliptic, with a beam waist of $300\,\mu$m in the transverse horizontal direction and $17\,\mu$m along the field of gravity. The depth of the reservoir trap is kept constant at $k_B \times 9\,\mu$K. After preparation of the sample, another laser cooling stage is performed on the narrow $^1S_0 - {^3P_1}$ intercombination line, using a single laser beam propagating vertically upwards. The detuning of the laser cooling beam from resonance is about $-2\,\Gamma$ and the peak intensity is $0.15\,\mu{\rm W/cm}^2=0.05\,I_{\rm sat}$. These parameters result in a photon scattering rate of $\sim 70\,$s$^{-1}$. At this point, the ultracold gas contains $9\times 10^6$ atoms at a temperature of 900\,nK.

To render the atoms transparent to cooling light in a central region of the laser cooled cloud, we induce a light shift on the $^3P_1$ state, using a ``transparency'' laser beam 15\,GHz blue-detuned to the $^3P_1 - {^3S_1}$ transition. This beam propagates downwards under a small angle of $15^{\circ}$ to vertical, it has a waist of 26\,$\mu$m in the plane of the reservoir trap and a peak intensity of $0.7\,$kW/cm$^2$. It upshifts the $^3P_1$ state by more than 10\,MHz and also influences the nearest molecular level tied to the $^3P_1$ state significantly. Related schemes of light-shift engineering were used to image the density distribution of atoms \cite{Thomas1995ppm,Brantut2008lst}, to improve spectroscopy \cite{Kaplan2002soi}, or to enhance loading of dipole traps \cite{Griffin2006ssl,Clement2009aor}. To demonstrate the effect of the transparency laser beam, we take absorption images of the cloud on the laser cooling transition. Figure~\ref{fig:BEC3}(d) shows a reference image without the transparency beam. In presence of this laser beam, atoms in the central part of the cloud are transparent for the probe beam, as can be seen in Fig.~\ref{fig:BEC3}(e).

To increase the density of the cloud, a dimple trap is added to the system. It consists of a 1065-nm laser beam propagating upwards under a small angle of $22^{\circ}$ to vertical and crossing the laser cooled cloud in the region of transparency. In the plane of the reservoir trap, the dimple beam has a waist of 22\,$\mu$m. The dimple is ramped to a depth of $k_B \times 2.6\,\mu$K, where it has trap oscillation frequencies of 250\,Hz in the horizontal plane. Confinement in the vertical direction is only provided by the reservoir trap and results in a vertical trap oscillation frequency of 600\,Hz. Figure~\ref{fig:BEC3}(f) shows a demonstration of the dimple trap in absence of the transparency beam:~the density in the region of the dimple increases substantially. However, with the dimple alone no BEC is formed because of photon reabsorption.

\begin{figure}
\centering
\includegraphics[width=165mm]{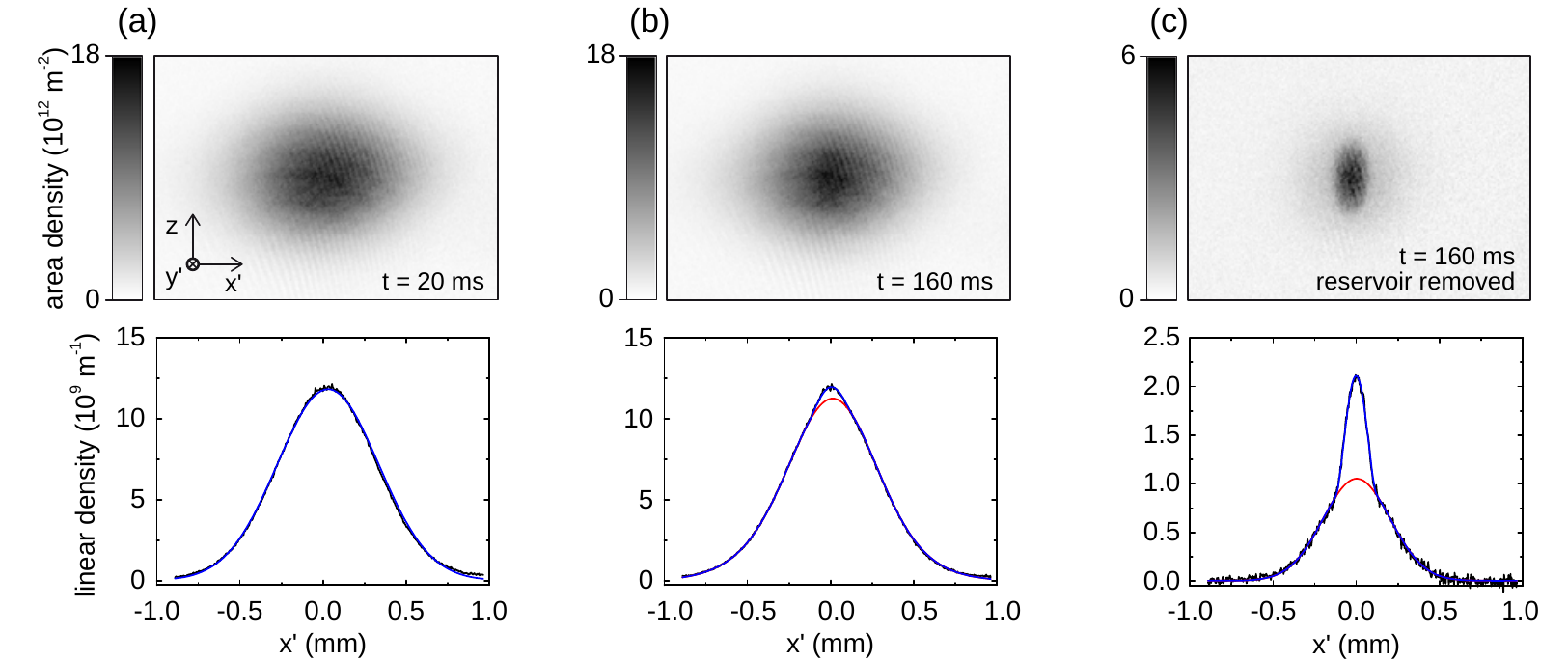}
\caption{Creation of a BEC by laser cooling. Shown are time-of-flight absorption images and integrated density profiles of the atomic cloud for different times $t$ after the transparency laser has been switched on, recorded after 24\,ms of free expansion. The images are taken in the horizontal direction, at an angle of $45^\circ$ with respect to the horizontal dipole trap beam, and the field of view of the absorption images is $2\,{\rm mm} \times 1.4\,{\rm mm}$. (a) and (b) The appearance of an elliptic core at $t=160\,$ms indicates the creation of a BEC. (c) Same as in (b), but to increase the visibility of the BEC, atoms in the reservoir trap were removed before the image was taken. The fits (blue lines) consist of Gaussian distributions to describe the thermal background and an integrated Thomas-Fermi distribution describing the BEC. The red lines show the component of the fit corresponding to the thermal background. The figure is taken from Ref.~\cite{Stellmer2013lct}.}
\label{fig:BEC4}
\end{figure}

The combination of the transparency laser beam and the dimple trap leads to BEC. Starting from the laser cooled cloud held in the reservoir trap, we switch on the transparency laser beam and ramp the dimple trap within 10\,ms to a depth of $k_B \times 2.6\,\mu$K. The potentials of the $^1S_0$ and $^3P_1$ states in this situation are shown in Fig.~\ref{fig:BEC3}(c). About $10^6$ atoms accumulate in the dimple without being disturbed by photon scattering, and elastic collisions thermalize atoms in the dimple with the laser cooled reservoir during the next $\sim100\,$ms. The temperature of the reservoir gas is hereby not increased, since the energy transferred to it is dissipated by laser cooling. Figure~\ref{fig:BEC4}(a) shows the momentum distribution 20\,ms after switching on the transparency beam, which is well described by a thermal distribution. By contrast, we observe that 140\,ms later, an additional, central elliptical feature has developed; see Fig.~\ref{fig:BEC4}(b). This is the hallmark of the BEC, which appears about 60\,ms after ramping up the dimple. Its atom number saturates at $1.1\times 10^5$ after 150\,ms. The atom number in the reservoir decreases slightly, initially because of migration into the dimple and on longer timescales because of light assisted loss processes in the laser cooled cloud. We carefully check that evaporation of atoms out of the dimple region is negligible even for the highest temperatures of the gas.

Although clearly present, the BEC is not very well visible in Fig.~\ref{fig:BEC4}(b), because it is shrouded by $8\times 10^6$ thermal atoms originating from the reservoir. To show the BEC with higher contrast, we have developed a background reduction technique. We remove the reservoir atoms by an intense flash of light on the $^1S_0-{^3P_1}$ transition applied for 10\,ms. Atoms in the region of transparency remain unaffected by this flash. Only $5\times10^5$ thermal atoms in the dimple remain and the BEC stands out clearly; see Fig.~\ref{fig:BEC4}(c). This background reduction technique is used only for demonstration purposes, but not for measuring atom numbers or temperatures.

The ability to reach the quantum degenerate regime by laser cooling has many exciting prospects. This method can be applied to any element possessing a laser cooling transition with a linewidth in the kHz range and suitable collision properties. The technique can also cool fermions to quantum degeneracy and it can be extended to sympathetic cooling in mixtures of isotopes or elements. Another tantalizing prospect enabled by variations of our techniques is the realization of a continuous atom laser, which converts a thermal beam into a laser-like beam of atoms.

\subsection{Bose-Einstein condensation of $^{86}$Sr}

Some isotopes of alkaline-earth atoms feature large positive scattering lengths, such as $^{40}$Ca, $^{42}$Ca, $^{44}$Ca \cite{Dammalapati2011sco}, and $^{86}$Sr. While scattering between atoms provides thermalization during evaporation, there is a downside of a very large scattering length $a$: Inelastic three-body losses have an upper limit proportional to $a^4$ \cite{Fedichev1996tbr,Bedaque2000tbr}, and can reduce the evaporation efficiency drastically. Magnetic Feshbach resonances, a widely used means to tune the scattering length in ultracold samples, are absent in the alkaline-earth species, and a different strategy to reach degeneracy despite the large scattering length is needed.

Quantum degeneracy in $^{86}$Sr has been reached \cite{Stellmer2010bec,Stellmer2013poq} despite the large scattering length of about 800\,$a_0$ \cite{Martinez2008tpp}. In this experiment, the crucial innovation is to perform evaporation at a comparatively low density in a dipole trap of large volume. Two-body collisions, vital for thermalization, scale proportional with the density $n$, while detrimental three-body collisions scale as $n^2$. At small enough densities, evaporation can be efficient even for large scattering lengths.

The dipole trap has an oblate shape with initial trap frequencies of about $f_x=30\,$Hz, $f_y=3\,$Hz, and $f_z=500\,$Hz. Using a 500-ms reservoir loading stage, we load $9\times10^5$ atoms at a temperature of about 1\,$\mu$K into the dipole trap. The initial density is about $10^{12}\,{\rm cm}^{-3}$ and the average elastic collision rate 200\,s$^{-1}$. The large vertical trap frequency allows us to perform evaporation very quickly, in just 800\,ms, which helps to avoid strong atom loss from three-body collisions. The onset of BEC is observed after 600\,ms of evaporation at a temperature of about 70\,nK with $3.5\times10^5$ atoms present. Further evaporation results in almost pure BECs of 25\,000 atoms. The cycle time of this experiment is again short, just 2.1\,s. Such a BEC with a large scattering length might constitute a good starting point for studies of optical Feshbach resonances.

\subsection{Bose-Einstein condensation of $^{88}$Sr}

The most abundant strontium isotope, $^{88}$Sr, presents significant challenges to reaching quantum degeneracy because of the small and negative $s$-wave scattering length, $a_{88}=-2.0$\,$a_0$. Fortunately, because of the good $^{87}$Sr$-{^{88}}$Sr interspecies scattering length ($a_{88-87}=55$\,$a_0$), $^{87}$Sr can serve as an effective refrigerant for $^{88}$Sr for dual-species evaporative cooling. Use of an equal mixture of the 10 distinguishable ground states for $^{87}$Sr arising from the nuclear spin $I=9/2$ diminishes any limitation on fermion-fermion collisions due to Pauli blocking. Essentially pure condensate cans be created with up to 10\,000 atoms, limited by the critical number for condensate collapse due to attractive interactions \cite{Ruprecht1995tds,Houbiers1996sob}. This is adequate for many experiments that benefit from working with a BEC of a nearly ideal gas. Here we will describe results from the Rice group \cite{Mickelson2010bec}; similar results were reported by the Innsbruck group \cite{Stellmer2013poq}. Sympathetic cooling of $^{88}$Sr with $^{86}$Sr can also produce quantum degenerate $^{88}$Sr \cite{Stellmer2013poq}, although this is less efficient.

We closely follow the presentation in Ref.~\cite{Mickelson2010bec}. $^{88}$Sr atoms are accumulated in the metastable state reservoir for 3\,s, followed by 30\,s of loading for $^{87}$Sr. $^3P_2$ atoms are returned to the ground state with 60\,ms of excitation on the $^3P_2 - {^3D_2}$ transition at 3.01\,$\mu$m. We typically recapture approximately $1.1\times 10^7$ $^{88}$Sr and $3\times 10^7$ $^{87}$Sr in the blue MOT at temperatures of a few mK.

The 461\,nm light is then extinguished and 689\,nm light is applied to drive the $^1S_0-^3P_1$ transitions and create intercombination-line MOTs
for each isotope \cite{Katori1999mot,Mukaiyama2003rll,Mickelson2010tae}. After 400\,ms of $^1S_0 - {^3P_1}$ laser cooling, an optical dipole trap consisting of two crossed beams is overlapped for 100\,ms with the intercombination-line MOT with 3.9\,W per beam and waists of approximately $w=90\,\mu$m in the trapping region. The dipole trap is formed by a single beam derived from a 20\,W multimode, 1.06\,$\mu$m fiber laser that is recycled through the chamber in close to the horizontal plane.

After extinction of the 689\,nm light, the sample is compressed by ramping the dipole trap power to 7.5\,W in 30\,ms, resulting in a trap depth of $25\,\mu$K. Typically the atom number, temperature, and peak density at this point for both $^{88}$Sr and $^{87}$Sr are  $3\times 10^6$, 7\,$\mu$K, and $2.5 \times 10^{13}$\,cm$^{-3}$. The peak phase space density for $^{88}$Sr is 0.01.

We decrease the laser power  according to $P=P_0/(1+t/\tau)^{\beta}+P_{\rm offset}$, with time denoted by $t$, $\beta=1.4$, and $\tau=1.5$\,s. This trajectory without $P_{\rm offset}$ was designed \cite{OHara2001slf} to yield efficient evaporation when gravity can be neglected. Gravity is a significant effect in this trap, and to avoid decreasing the potential depth too quickly at the end of the evaporation, we set $P_{\rm offset}=0.7$\,W, which corresponds to the power at which gravity causes the trap depth to be close to zero. The $^{87}$Sr and $^{88}$Sr remain in equilibrium with each other during the evaporation, and we observe an increase of $^{88}$Sr phase space density by a factor of 100 for a loss of one order of magnitude in the number of atoms. $^{87}$Sr atoms are lost at a slightly faster rate, as expected because essentially every collision involves an $^{87}$Sr atom.

A Maxwell-Boltzmann distribution fits the momentum distribution well at 5\,s of evaporation. At 6\,s, however, a Boltzmann distribution fit to the high velocity wings underestimates the number of atoms at low velocity. A Bose-Einstein distribution matches the distribution well. This sample  is close to the critical temperature for condensation and has a fit fugacity of $1.0$. Further evaporation to 7.5\,s produces  a narrow peak  at low velocity, which is a clear signature of the presence of a BEC. A pure condensate is observed near the end of the evaporation trajectory, which takes 9\,s.

At the transition temperature, $2\times10^5$ $^{87}$Sr  atoms remain at a temperature of 200\,nK. This corresponds to ${T}/{T_F}=0.9$ for an unpolarized sample, which is non-degenerate and above the point at which Pauli blocking significantly impedes evaporation efficiency \cite{Demarco1999oof}.

$^{88}$Sr has a negative scattering length, so one expects a collapse of the condensate when the system reaches a critical number of condensed atoms given by \cite{Ruprecht1995tds}
\begin{equation}\label{Equation: Maximum BEC number}
    N_{cr}\approx 0.575\frac{a_{ho}}{\left| a_{88} \right|}
\end{equation}
for a spherically symmetric trap. Here $a_{ho}=[\hbar/(m\omega)]^{1/2}$ is the harmonic oscillator length, where $m$ is the atom mass, $\hbar$
is the reduced Planck constant, and $\omega$ is the trap oscillation frequency. Our initial studies \cite{Mickelson2010bec} showed significant fluctuation in condensate number, bounded by the critical number $N_{cr}\approx 10\,000$ for the trap. Subsequent optimization showed that reducing the initial number of $^{88}$Sr atoms in the dipole trap to be about half that of $^{87}$Sr yields much less variation in condensate number. We reliably create condensates with  about 90\% of the critical number with a standard deviation of about 10\% \cite{Stellmer2013poq,Yan2013ccc}. The ability to make reproducible condensates is critical for experiments with $^{88}$Sr, such as investigation of an optical Feshbach resonance \cite{Yan2013ccc}.

\subsection{Bose-Bose mixtures}

Mixtures of two Bose-degenerate gases of different isotopes or elements allow the study of interesting phenomena, such as the miscibility and phase separation of two quantum fluids \cite{Hall1998doc,Riboli2002tot,Jezek2002ide}. The many bosonic isotopes of alkaline-earth elements in principle allow the creation of many different Bose-Bose mixtures. Unfortunately, for many of these mixtures, the interaction properties are unfavorable to create large and stable BECs. To avoid rapid decay, the absolute value of the two intra- and the interspecies scattering length must not be too large, but it must be large enough for efficient thermalization. The intraspecies scattering lengths should not be strongly negative to permit the formation of detectably large BECs \cite{Fukuhara2009aof}. The scattering length of alkaline-earth-like atoms can only be tuned by optical Feshbach resonances, which introduce losses \cite{Ciurylo2005oto,Enomoto2008ofr,Blatt2011moo}. These limitations reduce the number of possible binary mixtures considerably. In particular, all combination of bosonic calcium isotopes seem unfavorable, since all intraspecies scattering lengths of the most abundant calcium isotopes are quite large \cite{Kraft2009bec,Dammalapati2011sco}. In ytterbium, two out of five bosonic isotopes have large negative scattering lengths \cite{Kitagawa2008tcp}, excluding many possible combinations of isotopes. One remaining combination, $^{170}$Yb~+~$^{174}$Yb, has a large and negative interspecies scattering length. One of the two remaining combinations, $^{168}$Yb~+~$^{174}$Yb, has been brought to double degeneracy very recently, with 9000 atoms in the BEC of each species \cite{Sugawa2011bec}. The interspecies scattering length between these two isotopes is $2.4 (3.4)\,a_0$ and provides only minuscule interaction between the two. The three bosonic isotopes of strontium give rise to three different two-isotope combinations; see Tab.~\ref{tab:StrontiumProperties}. Of these the mixtures, $^{84}$Sr~+~$^{88}$Sr suffers from a large interspecies scattering length.

We will now present double-degenerate Bose-Bose mixtures of the combinations $^{84}$Sr~+~$^{86}$Sr and $^{86}$Sr~+~$^{88}$Sr, which have interspecies scattering lengths of $32\,a_0$ and $97\,a_0$, respectively. The experimental realization is straightforward: We consecutively load the two isotopes into the reservoir, repump them simultaneously on their respective $^3P_2-{^3D_2}$ transitions, and operate two red MOTs simultaneously. The mixture is loaded into the dipole trap and subsequently evaporated to form two BECs. Imaging is performed on the blue $^1S_0-{^1P_1}$ transition, and we image only one isotope per experimental run. The frequency shift between the isotopes is only about 4.5 linewidths. To avoid a contribution of the unwanted isotope to the absorption image, we remove the unwanted species by an 8-ms pulse of resonant light on the very isotope selective $^1S_0-{^3P_1}$ intercombination transition. To avoid a momentum distribution change of the imaged species by interspecies collisions, the pulse of light is applied after 17\,ms of free expansion, when the density of the sample has decreased sufficiently.

We will discuss the $^{86}$Sr~+~$^{88}$Sr combination first: $2.3\times 10^6$ ($3.3\times 10^6$) atoms of $^{86}$Sr ($^{88}$Sr) are loaded into the dipole trap, consisting of a horizontal beam and a weak vertical beam for additional axial confinement. The initial temperatures of the two species are quite different: 950\,nK for $^{86}$Sr and 720\,nK for $^{88}$Sr, which reflects the different intraspecies scattering behavior. The interspecies scattering length is around $100\,a_0$, and the two species clearly thermalize to reach equilibrium after 1\,s of evaporation. As the trap depth is lowered further, we observe the onset of BEC in $^{86}$Sr ($^{88}$Sr) after 2.0\,s (2.3\,s). At the end of our evaporation ramp, which lasts 2.4\,s, we obtain 10\,000 (3000) atoms of $^{86}$Sr ($^{88}$Sr) in the condensate fraction. Further evaporation does not increase the BEC atom numbers.

In a second experiment, we investigate the $^{84}$Sr~+~$^{86}$Sr mixture with an interspecies scattering length of $32\,a_0$. Starting out with $10\times 10^6$ ($1.5\times 10^6$) atoms of $^{84}$Sr ($^{86}$Sr) in the dipole trap, we perform forced evaporation over 2\,s, and the two species remain in perfect thermal equilibrium throughout this time. The phase transition of $^{84}$Sr is observed already after 1.3\,s, with about $2.5 \times 10^6$ atoms present at a temperature of 200\,nK. After 1.9\,s, the BEC is essentially pure and contains up to $2\times 10^6$ atoms. The atom number of $^{86}$Sr is kept considerably lower to avoid three-body loss. The phase transition occurs later: after 1.7\,s, with $4 \times 10^5$ atoms at a temperature of 130\,nK. Till the end of evaporation, the BEC component grows to 8000 atoms but remains accompanied by a large thermal fraction.

We have here presented two binary Bose-Bose mixtures of alkaline-earth atoms with appreciable interaction between the two species. These mixtures enjoy the property that isotope-selective optical traps can be operated close to one of the intercombination lines. This might allow for an individual addressing of the isotopes by a dipole trap operated close to these transitions \cite{Yi2008sda}, reminiscent of the case of rubidium in its hyperfine states $F=1$ and $F=2$ \cite{Mandel2003cto} or nuclear substates in ytterbium and strontium \cite{Taie2010roa,Stellmer2011dam}.

\section{Spin state control in $^{87}$Sr}
\label{sec:OSG}

Fermionic $^{87}$Sr has a nuclear spin of $I=9/2$. This large nuclear spin has many applications in quantum simulation and computation, for which preparation, manipulation, and detection of the spin state are requirements. For an ultracold $^{87}$Sr cloud, we show two complementary methods to characterize the spin-state mixture: optical Stern-Gerlach state separation and state-selective absorption imaging. We use these methods to optimize the preparation of a variety of spin-state mixtures by optical pumping and to measure an upper bound of the $^{87}$Sr spin relaxation rate.

\subsection{Optical Stern-Gerlach separation}
\label{sec:detection}

 Several alkaline-earth spin-state detection schemes have been demonstrated. The number of atoms in the highest $m_F$ state can be determined by selectively cooling \cite{Mukaiyama2003rll} or levitating \cite{Tey2010ddb} atoms in this state. The number of atoms in an arbitrary $m_F$ state was determined using state-selective shelving of atoms in a metastable state \cite{Boyd2007nse}. Recording the full $m_F$-state distribution with this method is possible, but needs one experimental run per state. Determination of the $m_F$-state distribution in only two experimental runs was shown for quantum-degenerate ytterbium gases, using optical Stern-Gerlach (OSG) separation \cite{Taie2010roa}.

\begin{figure}[t]
\centering
\includegraphics[width=165mm]{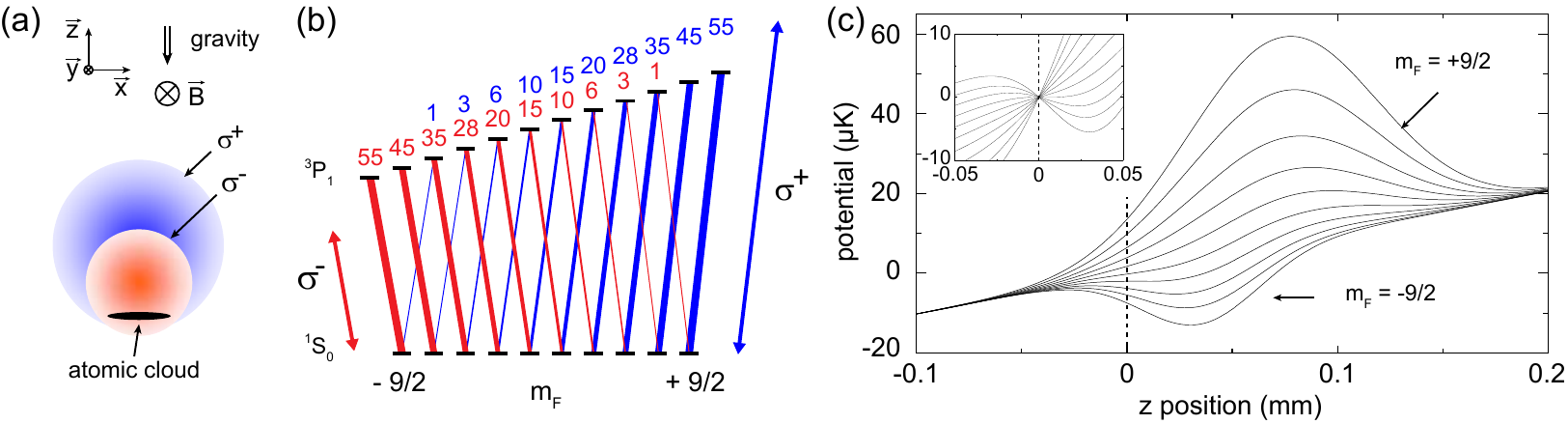}
\caption{Principle of OSG separation. (a) $\sigma^{+}$- and $\sigma^{-}$-polarized laser beams propagating in the $y$-direction create dipole forces on an atomic cloud that is located on the slopes of the Gaussian beams. (b) The laser beams are tuned close to the $^1S_0\,(F=9/2) - {^3P_1}\,(F'=11/2)$ intercombination line, creating attractive ($\sigma^{-}$ beam) or repulsive ($\sigma^{+}$ beam) dipole potentials. Each $m_F$ state experiences a different potential because of the varying line strength of the respective transition. (c) The potentials resulting from dipole potentials and the gravitational potential. The dashed line marks the initial position of the atoms. The inset shows the relevant region of the potentials, offset shifted to coincide at the position of the atoms, which clearly shows the different gradient on each $m_F$ state. The figure is adapted from Ref.~\cite{Stellmer2011dam}.}
\label{fig:OSG1}
\end{figure}

The Stern-Gerlach technique separates atoms in different internal states by applying a state-dependent force and letting the atomic density distribution evolve under this force \cite{Stern1922den}. The implementation of this technique for alkali atoms is simple. Their single valence electron provides them with a $m_F$-state dependent magnetic moment that, for easily achievable magnetic field gradients, results in $m_F$-state dependent forces sufficient for state separation \cite{StamperKurn1998oco}. By contrast, atoms with two valence electrons possess only a weak, nuclear magnetic moment in the electronic ground state, which would require the application of impractically steep magnetic field gradients. An alternative is OSG separation, where a state dependent dipole force is used. OSG separation was first shown for a beam of metastable helium \cite{Sleator1992edo}, where orthogonal dressed states of the atoms were separated by a resonant laser field gradient. The case of interest here, OSG $m_F$-state separation, has been realized as well for a quantum degenerate gas of ytterbium, by using $m_F$-state dependent dipole forces \cite{Taie2010roa}.

We first explain the basic operation principle of strontium OSG separation before discussing our experimental implementation. The experimental situation is shown in Fig.~\ref{fig:OSG1}(a). An ultracold cloud of $^{87}$Sr atoms in a mixture of $m_F$ states is released from an optical dipole trap. The $m_F$-state dependent force is the dipole force of two laser beams propagating in the plane of the pancake-shaped cloud, one polarized $\sigma^{+}$, the other $\sigma^{-}$. The diameter of these OSG laser beams is on the order of the diameter of the cloud in the $x$-direction. The beams are displaced vertically by about half a beam radius to produce a force in the $z$-direction on the atoms. To create a $m_F$-state dependent force, the OSG beams are tuned close to the $^1S_0\,(F=9/2) - {^3P_1}\,(F'=11/2)$ intercombination line, so that this line gives the dominant contribution to the dipole force. A guiding magnetic field is applied in the direction of the laser beams such that the beams couple only to $\sigma^{+}$ or $\sigma^{-}$ transitions, respectively. The line strength of these transitions varies greatly with the $m_F$ state \cite{Metcalf1999book}, see Fig.~\ref{fig:OSG1}(b), resulting in different forces on the states. For $^{173}$Yb, this variation, together with a beneficial summation of dipole forces from transitions to different $^3P_1$ hyperfine states, is sufficient to separate four of the six $m_F$ states using just one OSG beam \cite{Taie2010roa}. The remaining two $m_F$ states could be analyzed by repeating the experiment with opposite circular polarization of the OSG beam.

Strontium, which has nearly twice as many nuclear spin states, requires an improved OSG technique to separate the states. The improvement consists of applying two OSG beams with opposite circular polarization at the same time. The $\sigma^{+}$-polarized beam produces dipole forces mainly on the positive $m_F$ states, the $\sigma^{-}$ beam mainly on the negative $m_F$ states. By positioning the beams in the appropriate way (see below), the forces point in opposite directions and all $m_F$ states can be separated in a single experimental run. A second improvement is to enhance the difference in the dipole forces on neighboring $m_F$ states by tuning already strong transitions closer to the OSG beam frequency using a magnetic field, which splits the excited state $m_{F'}$ states in energy. For our settings, the difference in forces on neighboring high $|m_F|$ states is enhanced by up to 25\%, which helps to separate those states. This enhancement scheme requires the $\sigma^{+}$-polarized OSG beam to be tuned to the blue of the resonance, whereas the $\sigma^{-}$ beam has to be tuned to the red of the resonance; see Fig.~\ref{fig:OSG1}(b). Both beams are centered above the atomic cloud so that the repulsive blue detuned beam produces a force pointing downwards, whereas the attractive red detuned beam produces a force pointing upwards.

\subsection{Experimental demonstration}
\label{sec:OSGExperimentalDemonstration}

We demonstrate OSG separation of a cloud of $4.5 \times 10^4$ $^{87}$Sr atoms in a mixture of $m_F$ states. To prepare the cloud, Zeeman slowed $^{87}$Sr atoms are laser cooled in two stages, first in a blue magneto-optical trap (MOT) on the broad-linewidth $^1S_0 - {^1P_1}$ transition, then in a red MOT on the narrow-linewidth $^1S_0 - {^3P_1}$ transition. Next, the atoms are transferred to a pancake-shaped optical dipole trap with strong confinement in the vertical direction. The sample is evaporatively cooled over 7\,s. At the end of evaporation the trap oscillation frequencies are $f_x=19\,$Hz, $f_y=11$\,Hz, and $f_z=85$\,Hz, where the coordinate system is defined in Fig.~\ref{fig:OSG1}(a). The collision rate at this stage is only $1\,$s$^{-1}$, which is insufficient for complete thermalization. Since atoms are evaporated mainly downwards, along the $z$-direction, the sample is not in cross-dimensional thermal equilibrium, having a temperature of 25\,nK in the $z$-direction and twice that value in the $xy$-plane. The sample is non-degenerate and the $1/e$-widths of the Gaussian density distribution are $w_x=55\,\mu$m, $w_y=85\,\mu$m, and $w_z=7\,\mu$m.

The OSG beams propagate along the $y$-direction. The power of the $\sigma^{+}$ ($\sigma^{-}$) beam is 4\,mW (0.5\,mW), the waist is $\sim80\,\mu$m ($\sim60\,\mu$m), and the beam center is displaced $\sim70\,\mu$m ($\sim40\,\mu$m) above the cloud. Both beams create dipole forces of similar magnitude since the reduced power of the $\sigma^{-}$ beam compared to the $\sigma^+$ beam is partially compensated by its decreased waist. At zero magnetic field, the $\sigma^{\pm}$ beam is detuned $\pm100\,$MHz from resonance. To increase the difference in dipole potential on neighboring $m_F$ states, a magnetic field of 16\,G is applied parallel to the OSG beams, which splits neighboring $^3P_1(F'=11/2$) $m_{F'}$ states by 6.1\,MHz. With this field applied, the $\sigma^{\pm}$ beam has a detuning of $\pm 66.4$\,MHz to the $^1S_0\,(F=9/2$, $m_F=\pm9/2) - {^3P_1}\,(F'=11/2$, $m_{F'}=\pm11/2$) transition and a detuning of $\pm121.4$\,MHz to the $^1S_0\,(F=9/2$, $m_F=\mp9/2) - {^3P_1}\,(F'=11/2$, $m_{F'}=\mp7/2$) transition; see Fig.~\ref{fig:OSG1}(b).

\begin{figure}[t]
\centering
\includegraphics[width=165mm]{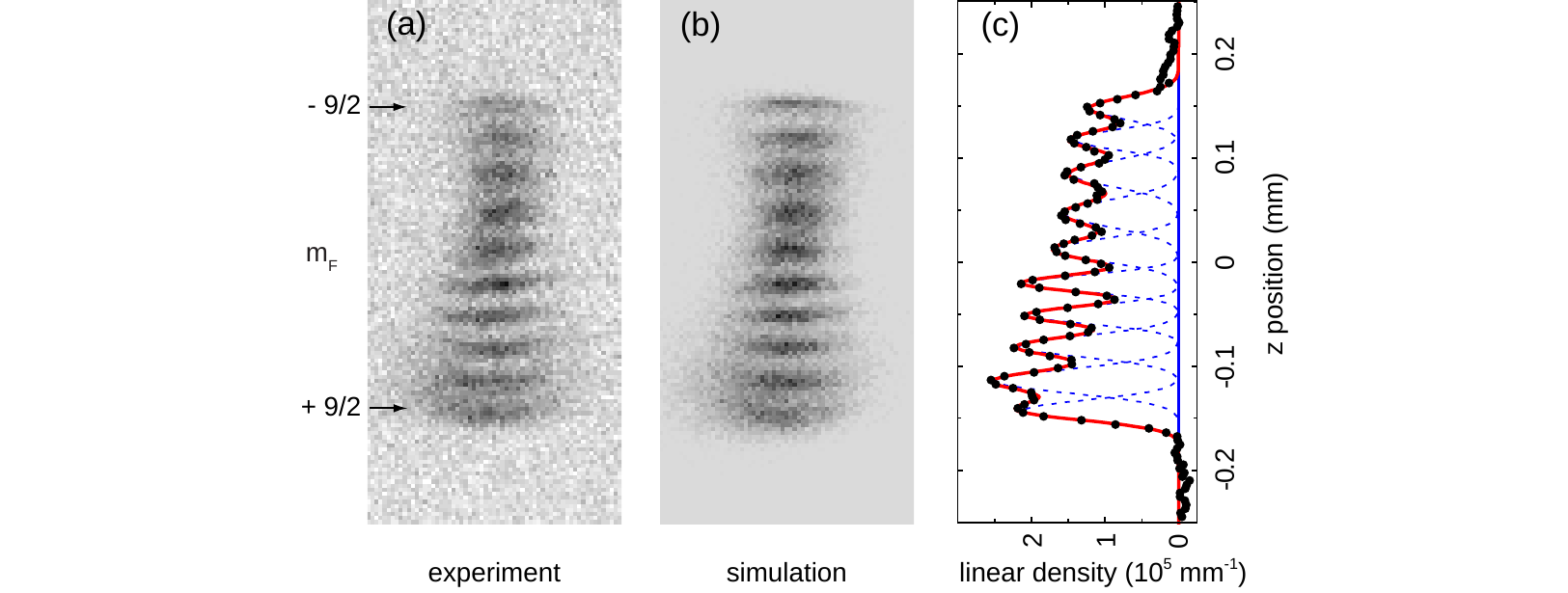}
\caption{OSG separation of the ten $^{87}$Sr nuclear spin states. The images show the atomic density distribution after OSG separation integrated over the $(\mathbf{\hat{x}}+\mathbf{\hat{y}})$-direction as obtained in (a) the experiment and (b) the simulation. (c) The density distribution of the experiment integrated along the $x$- and $y$-directions is shown together with a fit consisting of ten Gaussian distributions. The figure is adapted from Ref.~\cite{Stellmer2011dam}.}
\label{fig:OSG2}
\end{figure}

OSG separation is started by simultaneously releasing the atoms from the dipole trap and switching on the OSG beams. The atoms are accelerated for 1.6\,ms by the OSG beams. Then the beams are switched off to avoid oscillations of atoms in the dipole trap formed by the red detuned OSG beam. The atoms freely expand for another 2.3\,ms before an absorption image on the $^1S_0 - {^1P_1}$ transition is taken. The result is shown in Fig.~\ref{fig:OSG2}(a). All ten $m_F$ states are clearly distinguishable from each other.

To quantify the separation of the states, we fit ten Gaussian distributions to the density distribution integrated along the $x$- and $y$-directions, see Fig.~\ref{fig:OSG2}(c). We obtain a separation of adjacent states very similar to the $1/e$-widths of the distributions, which are between 24 and 36\,$\mu$m. From the Gaussian fits we also obtain an estimation of the atom number in each state. The $m_F$-state dependence of the line strength of the blue imaging transition, as well as optical pumping processes during imaging, need to be taken into account to accurately determine the atom number in each spin state. Detailed simulations of classical atom trajectories describing the OSG separation process can be found in Ref.~\cite{Stellmer2011dam}. The simulations show very good agreement with the experiment, which can be appreciated by a comparison of Figs.~\ref{fig:OSG2}(a) and (b).

OSG separation works only well for very cold samples. If the temperature is too high, the sample expands too fast and the individual $m_F$-state distributions cannot be distinguished. For a density minimum to exist between two neighboring $m_F$-state distributions of Gaussian shape, the $1/e$-widths have to be smaller than $\sqrt{2}$ times the distance between the maxima of the distributions. For our smallest separation of 24\,$\mu$m, this condition corresponds to samples with a temperature below 100\,nK, which can only be obtained by evaporative cooling.

\subsection{Spin-state dependent absorption imaging}
\label{sec:AbsorptionImaging}

\begin{figure}
\centering
\includegraphics[width=165mm]{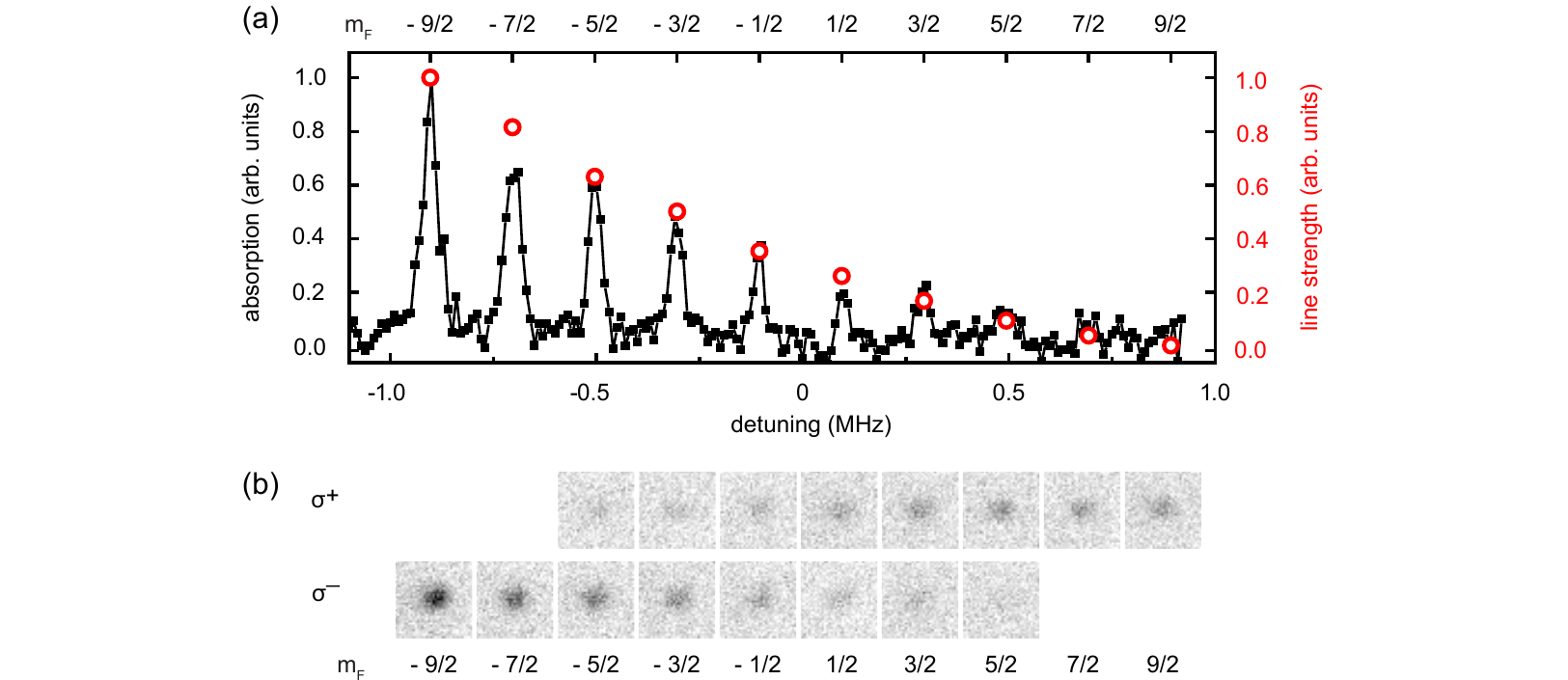}
\caption{$m_F$-state resolved absorption imaging on the $^1S_0\,(F=9/2) - {^3P_1}\,(F'=11/2$) intercombination line. (a) Spectrum of a $^{87}$Sr sample with nearly homogeneous $m_F$-state distribution. The spectrum was obtained using $\sigma^{-}$-polarized light and shifting transitions corresponding to different $m_F$ states in frequency by applying a magnetic field of 0.5\,G. The circles give the line strengths of the transitions. (b) Absorption images taken on the maxima of absorption of each $m_F$ state using $\sigma^{+}$ or $\sigma^{-}$ polarized light. The figure is taken from Ref.~\cite{Stellmer2011dam}.}
\label{fig:OSG3}
\end{figure}

We also demonstrate a complementary method of $m_F$-state detection:~$m_F$-state dependent absorption imaging. This method is often used for alkali atoms employing a broad linewidth transition \cite{Matthews1998dro}. For strontium, $m_F$-state resolved imaging on the broad $^1S_0 - {^1P_1}$ transition is not possible since the magnetic field splitting of the exited state $m_{F'}$ states is smaller than the linewidth of the transition \cite{Boyd2007nse}. But $m_F$-state dependent imaging can be realized using the narrow $^1S_0\,(F=9/2) - {^3P_1}\,(F'=11/2$) intercombination line. To achieve state selectivity, we apply a magnetic field of 0.5\,G, which splits neighboring $m_{F'}$ states by 200\,kHz, which is 27 times more than the linewidth of the imaging transition. The advantages of this method compared to OSG separation is its applicability to samples that have not been evaporatively cooled, spatially resolved imaging, and a near perfect suppression of signal from undesired $m_F$ states. A disadvantage of this method is that it delivers a reduced signal compared to imaging on the $^1S_0 - {^1P_1}$ transition, as done after OSG separation. The reduction comes from the narrower linewidth, optical pumping to dark states during imaging, and weak line strengths for some $m_F$ states.

To demonstrate absorption imaging on the intercombination line, we use a sample of $10^6$ atoms at a temperature of 500\,nK in a trap with oscillation frequencies of $f_x=45\,$Hz, $f_y=40\,$Hz, and $f_z=220\,$Hz, obtained after 1.4\,s of evaporation. Figure~\ref{fig:OSG3} shows a spectroscopy scan and absorption images taken on the maxima of the absorption signal of this sample. The absorption is strongly $m_F$-state dependent and to obtain the best signal, the polarization of the absorption imaging light has to be adapted to the $m_F$ state of interest: $\sigma^{+}$($\sigma^{-}$) for high (low) $m_F$ states and $\pi$ for low $|m_F|$ states. For our absorption imaging conditions (an intensity of 15\,$\mu$W/cm$^2$, which is five times the saturation intensity, and an exposure time of 40\,$\mu$s), even atoms in $m_F$ states corresponding to the strongest transition will on average scatter less than one photon. Therefore, for a sample with homogeneous $m_F$-state distribution, the maximum absorption is expected to be nearly proportional to the $m_F$-state dependent line strength of the transition, which we confirm using a simulation of the absorption imaging process. This proportionality is observed in the experimental data, indicating that the sample used has a nearly homogeneous $m_F$-state distribution. The observed Lorentzian linewidth of the absorption lines is $\sim40\,$kHz. We expect a linewidth of $\sim30\,$kHz arising from power and interaction-time broadening. Doppler broadening and collisional broadening will contribute to the linewidth as well \cite{Ido2005psa}.

\subsection{Preparation of spin-state mixtures}
\label{sec:OpticalPumping}

\begin{figure}
\centering
\includegraphics[width=165mm]{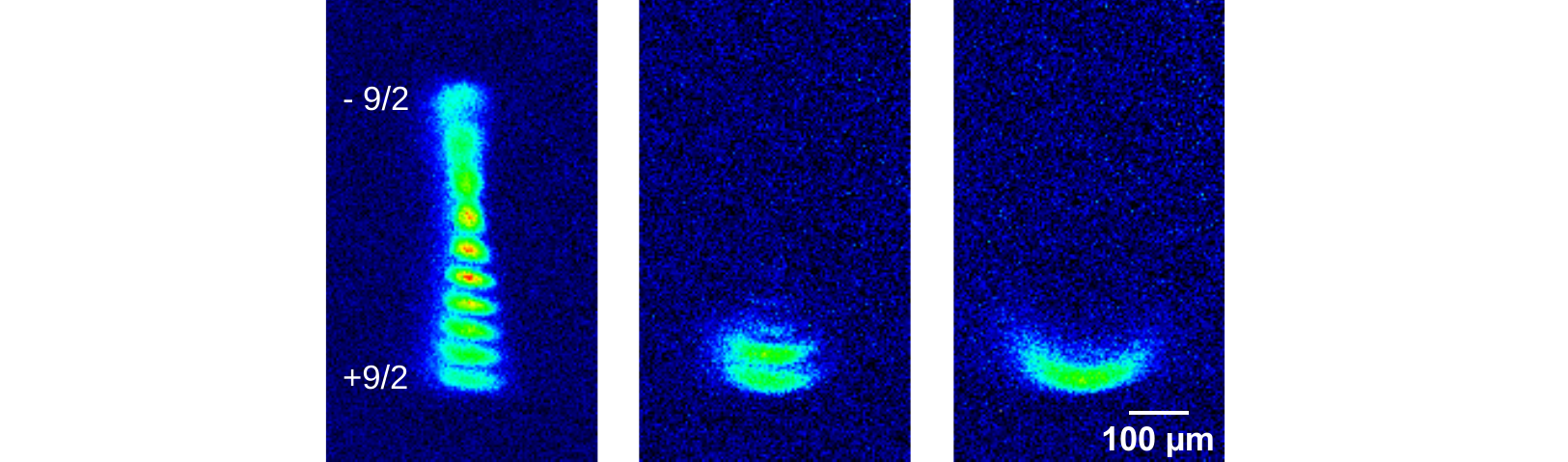}
\caption{Detection of spin-state distribution using the optical Stern-Gerlach technique. Samples of $^{87}$Sr in a ten-state mixture or optically pumped into two or one spin states are shown. The figure is taken from Ref.~\cite{Stellmer2013poq}.}
\label{fig:OSG4}
\end{figure}

For applications of $^{87}$Sr to quantum simulation and computation, the $m_F$-state mixture needs to be controlled. We produce a variety of different mixtures by optical pumping, making use of OSG separation to quickly optimize the optical pumping scheme and quantify the result. Optical pumping is performed on the $^1S_0\,(F=9/2)-{^3P_1}\,(F'=9/2)$ intercombination line, before evaporative cooling. A field of 3\,G splits neighboring excited state $m_{F'}$ states by 255\,kHz. This splitting is well beyond the linewidth of the transition of 7.4\,kHz, allowing transfer of atoms from specific $m_F$ states to neighboring states using $\sigma^{\pm}$- or $\pi$-polarized light, the choice depending on the desired state mixture. Sequences of pulses on different $m_F$ states can create a wide variety of state mixtures, of which three examples are shown in Fig.~\ref{fig:OSG4}. The fidelity of state preparation can reach 99.9\%, as confirmed by state-dependent absorption imaging.

\subsection{Determination of an upper bound of the spin-relaxation rate}
\label{sec:SpinRelaxation}

A low nuclear spin-relaxation rate is an essential requirement to use $^{87}$Sr for quantum simulation and computation \cite{Cazalilla2009ugo,Gorshkov2010tos}. The rate is expected to be small since the nuclear spin does not couple to the electronic degrees of freedom in the ground state. Here, we use our nuclear spin state preparation and detection techniques to determine an upper bound for this spin relaxation rate. We start with a sample of $1.5\times 10^6$ atoms with near uniform $m_F$-state distribution and a temperature of $T=1.5\,\mu$K, confined in a trap with oscillation frequencies $f_x=68\,$Hz, $f_y=67\,$Hz, and $f_z=360\,$Hz, obtained after transferring the atoms from the MOT to the dipole trap and adiabatic compression of the trap. We optically pump all atoms from the $m_F=5/2$ state to neighboring states and look out for the reappearance of atoms in this state by spin relaxation during 10\,s of hold. The atom number in the $m_F=5/2$ state and, as a reference, the $m_F=7/2$ state are determined from absorption images. During 10\,s of hold at a magnetic field of either 5\,G or 500\,G the number of $m_F=5/2$ atoms remains below our detection threshold of about $10^4$ atoms, indicating a low spin-relaxation rate. To obtain a conservative upper bound for the spin-relaxation rate, we assume that the dominant process leading to the creation of $m_F=5/2$-state atoms are collisions of $m_F=7/2$- with $m_F=3/2$-state atoms, forming two $m_F=5/2$-state atoms. Since the second order Zeeman effect is negligible, no energy is released in such a collision and the resulting $m_F=5/2$-state atoms will remain trapped. The number of atoms created in the $m_F=5/2$ state by spin relaxation after a hold time $t$ is $N_{5/2}=2 N_{\rm state} g_{\rm sr} \overline{n} t$, where $N_{\rm state}=1.5 \times 10^5$ is the atom number in each populated state, $g_{\rm sr}$ the spin-relaxation rate constant, $\overline{n}=7.5 \times 10^{11}$\,cm$^{-3}$ the mean density and the factor 2 takes into account that two atoms are produced in the $m_F=5/2$ state per collision. From our measurement we know that $N_{5/2}<10^4$, from which we obtain an upper bound of $5 \times 10^{-15}\,$cm$^3$s$^{-1}$ for the spin-relaxation rate constant. This bound for the rate constant corresponds for our sample to a spin relaxation rate which is 2000 times smaller than the elastic scattering rate. This value can be converted into a deviation of less than $5 \times 10^{-4}$ from an assumed SU($N$) symmetry \cite{Bonnes2012alo}. The rate constant could be even orders of magnitude smaller than the already low upper bound we obtained \cite{JuliennePrivateComm}.

\section{Degenerate Fermi gases of $^{87}$Sr}
\label{sec:DFG}

Ground-breaking experiments with ultracold Fermi gases \cite{Inguscio2008ufg,Giorgini2008tou} have opened possibilities to study fascinating phenomena, as the BEC-BCS crossover, with a high degree of control. Most experiments have been performed with the two alkali fermions $^{40}$K and $^6$Li. Fermions with two valence electrons, like $^{43}$Ca, $^{87}$Sr, $^{171}$Yb, and $^{173}$Yb, have a much richer internal state structure, which is at the heart of recent proposals for quantum computation and simulation; see Sec.~\ref{sec:Introduction}. Unlike bosonic isotopes of these elements, the fermions have a nuclear spin, which decouples from the electronic state in the $^1S_0$ ground state and the $^3P_0$ metastable state. This gives rise to a SU$(N)$ spin symmetry, where $N$ is the number of nuclear spin states, which is ten for $^{87}$Sr. A wealth of recent proposals suggest employing such atoms as a platform for the simulation of SU($N$) magnetism \cite{Wu2003ess,Wu2006hsa,Cazalilla2009ugo,Hermele2009mio,Gorshkov2010tos,Xu2010lim,FossFeig2010ptk,FossFeig2010hfi,Hung2011qmo,Manmana2011smi,Hazzard2012htp,Bonnes2012alo}, for the generation of non-Abelian artificial gauge fields \cite{Dalibard2011agp,Gerbier2010gff}, to simulate lattice gauge theories \cite{Banerjee2013aqs}, or for quantum computation schemes \cite{Hayes2007qlv,Daley2008qcw,Gorshkov2009aem,Daley2011sdl}.

Elements with a large nuclear spin are especially well suited for some of these proposals. They allow to encode several qubits in one atom \cite{Gorshkov2009aem}, and could lead to exotic quantum phases, as chiral spin liquids, in the context of SU($N$) magnetism \cite{Hermele2009mio}. Furthermore, it has been shown that the temperature of a lattice gas is lower for a mixture containing a large number of nuclear spin states after loading the lattice from a bulk sample \cite{Hazzard2012htp,Taie2012asm,Bonnes2012alo}. Pomeranchuk cooling \cite{Richardson1997tpe}, which benefits from a large number of spins, has recently been observed in ytterbium \cite{Taie2012asm}. The largest nuclear spin of any alkaline-earth-like atom is 9/2, and it occurs in the nuclei of $^{87}$Sr and of two radioactive nobelium isotopes. This fact makes $^{87}$Sr with its ten spin states an exceptional candidate for the studies mentioned.

The study of SU($N$) magnetism in a lattice requires the temperature of the sample to be below the super-exchange scale, $t^2/U$, where $t$ is the tunnel matrix element and $U$ the on-site interaction energy \cite{Bonnes2012alo}. A high degree of degeneracy in the bulk would constitute a good starting point for subsequent loading of the lattice.

\subsection{A degenerate Fermi gas of ten spin states}

The first degenerate Fermi gases of strontium were produced in a mixture of all ten spin states \cite{DeSalvo2010dfg}. The procedure of laser cooling the fermionic isotope is very similar to the bosonic case, but complicated by the hyperfine structure. In this experiment, $3\times10^6$ atoms are loaded into an optical dipole trap consisting of two horizontal beams of $90\,\mu$m waist, powers of 3.9\,W per beam, and a wavelength of 1064\,nm.

\begin{figure}[t]
\centering
\includegraphics[width=165mm]{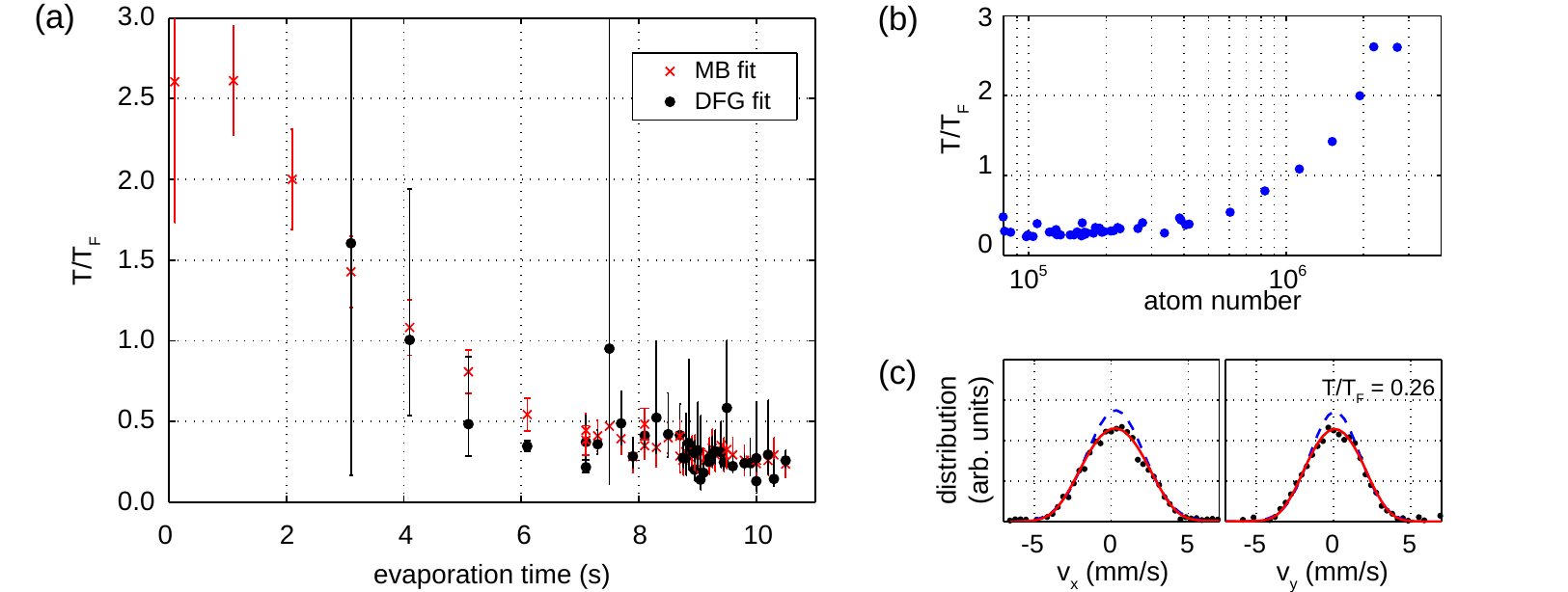}
\caption{Evaporation of $^{87}$Sr. (a) $T/T_F$ is obtained from either a fit of the velocity distribution to a single Fermi- Dirac distribution (DFG) or a determination of number and temperature from fits to a Maxwell-Boltzmann distribution (MB) and knowledge of the trap oscillation frequencies. (b) Variation of $T/T_F$ with atom number. $T/T_F$ is calculated using the latter of the two methods. (c) Velocity distributions along axes perpendicular
to the imaging beam. At the end of evaporation, classical particle statistics (dashed blue line) overestimates the population at low velocities,
while Fermi-Dirac statistics for $T/T_F = 0.26$ (solid red line) accurately fit the data. The figure is adapted from Ref.~\cite{DeSalvo2010dfg}.}
\label{fig:DFG1}
\end{figure}

The atoms are in a roughly even distribution of spin states. The trap depth is increased by about a factor of two to increase the rate of collisions required for thermalization. Forced evaporation increases the phase space density significantly and allows us to enter the quantum degenerate regime; see Fig.~\ref{fig:DFG1}. The evaporation efficiency decreases in the degenerate regime, and we can reach $T/T_F = 0.25(5)$ with about $10^4$ atoms per spin state. The degree of degeneracy is quantified by fitting a Fermi-Dirac distribution to the two-dimensional momentum distribution of time-of-flight absorption images \cite{DeMarco2001qbo}.

\subsection{A degenerate Fermi gas of a single spin state}

A subsequent experiment featured already some degree of control over the spin state composition \cite{Tey2010ddb}, yet not all of the techniques presented in Sec.~\ref{sec:OSG} were available at that time. The goal of this experiment is to generate a degenerate Fermi gas of only one single, fully controlled spin state.

Just after dipole trap loading, optical pumping is performed on the $^1S_0\,(F=9/2) - {^3P_1}\,(F'=9/2)$ transition using circularly polarized light. A small guiding field is applied, and the laser frequency is swept across all transitions. The $m_F=+9/2$ state is a dark state for this transition, and atoms accumulate in this state. The performance of optical pumping is evaluated using a levitation technique: The atoms are released from the dipole trap and subjected to an upward propagating, circularly polarized beam on the $^1S_0\,(F=9/2) - {^3P_1}\,(F'=11/2)$ transition. A magnetic field of 3\,G splits adjacent $m_{F'}$ states by $\sim150$ linewidths, and the levitation light addresses only the cycling $m_F=9/2 \rightarrow m_{F'}=11/2$ transition. In this way, atoms in the $m_F=+9/2$ state can be levitated for about 100\,ms, while atoms in all other states drop down with gravity. Horizontal imaging on the blue transition allows us to access the number of levitated and non-levitated atoms, and we find that more than 95\% of the atoms are in the $m_F=+9/2$ state after optical pumping.

As identical fermions do not collide at low temperatures, we add $^{84}$Sr atoms to the dipole trap. Forced evaporation commences with $2.2\times10^6$ atoms of $^{84}$Sr and $0.7\times10^6$ atoms of $^{87}$Sr, both at a temperature of $1.7\,\mu$K. At the end of evaporation, we observe a Fermi degeneracy of $T/T_F = 0.30(5)$ with $4\times10^4$ atoms, together with a pure BEC of  $1\times10^5$ atoms. This mixture constitutes the first double-degenerate Bose-Fermi mixture of strontium.

As in all of these experiments, we determine the atom number, temperature $T$, and Fermi temperature $T_F$ by fitting two-dimensional Fermi-Dirac distributions to absorption pictures \cite{DeMarco2001qbo}. Two methods are used to determine $T/T_F$. Either $T$ is determined by the fit and $T_F$ is calculated from the atom number $N_{\rm at}$ and average trap frequency as $T_F=\hbar \bar\omega (6N_{\rm at})^{1/3}/k_B$. Alternatively, $T/T_F$ is calculated directly from the fugacity, which is a fit parameter. The momentum distribution of a gas at small $T/T_F$ strongly deviates from a Gaussian shape, which we show in azimuthally integrated profiles; see Fig.~\ref{fig:DFG2}(a).

\subsection{Degenerate Fermi gases of arbitrary spin composition}

The proposed experiments of quantum simulation require a deeper degeneracy than presented before, as well as full control over the entire spin state composition. A third set of experiments was performed to meet these criteria \cite{Stellmer2013poq}. Here, about $5\times10^6$\,atoms of $^{87}$Sr are loaded into the optical dipole trap, where we measure a temperature of $1.2\,\mu$K. At this time, atoms are in a roughly even mixture of all spins. To prepare the desired spin mixture, we perform optical pumping on the $^1S_0\,(F=9/2) - {^3P_1}\,(F'=9/2)$ transition at a small guiding field of 3\,G, which splits adjacent $m_{F'}$-states by 260\,kHz, corresponding to 35 linewidths. As discussed in Sec.~\ref{sec:OSG}, we can prepare any combination and relative population of the ten spin states. The optical pumping is optimized using the optical Stern-Gerlach technique, and quantified using state-selective absorption imaging on the intercombination line. We can reduce the population of undesired spin states to below 0.1\%, where this value is limited by our detection threshold of 3000 atoms.

After the spin preparation, which does not heat the sample, we perform evaporative cooling. Evaporation proceeds in two stages. The first stage lasts 16\,s, during which the power of the horizontal dipole trap beam is reduced by a factor of five. At this point, the gas enters the degenerate regime with a typical temperature of around $0.3\,T_F$ and evaporative cooling becomes less efficient because of Pauli blocking \cite{Demarco2001pbo}. We compensate for this effect by a reduced ramp speed during the second evaporation stage. The power of the horizontal beam is only reduced by a factor 1.5 during 10\,s. Slightly different final trap depths are used for different numbers of populated spin states. Trap frequencies at the end of evaporation are $f_x\sim30\,$Hz, $f_y\sim30\,$Hz, and $f_z\sim200\,$Hz.

\begin{figure}[t]
\centering
\includegraphics[width=165mm]{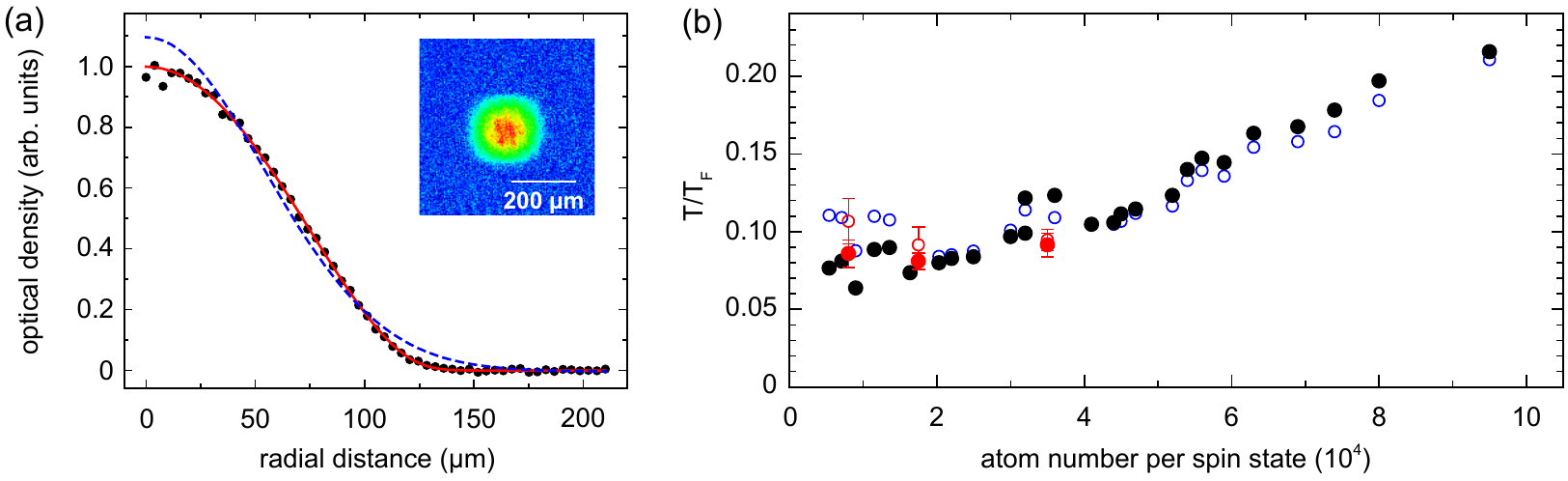}
\caption{Deeply degenerate Fermi gases of $^{87}$Sr in a balanced mixture of ten nuclear spin states. Panel (a) shows the azimuthally averaged density distribution of a degenerate Fermi gas at $T/T_F = 0.08(1)$ after 25.4\,ms of free expansion (black circles). The measurement is well described by a Fermi-Dirac distribution (red line) but not by a Gaussian (dashed blue line). The corresponding absorption image is shown in the inset. Panel (b) shows the value of $T/T_F$ in dependence of atom number per spin state, derived from the fugacity (open circles) or calculated from the temperature, the atom number, and the trap oscillation frequencies (solid circles). Some data points have been taken multiple times (red circles with error bars) to determine the statistical uncertainty. The figure is adapted from Ref.~\cite{Stellmer2013poq}.}
\label{fig:DFG2}
\end{figure}

We will limit the presentation of our data to the cases of $N=10$, $N=2$, and $N=1$. For the ten-state mixture, the fraction of Pauli-forbidden collisions is small, and evaporative cooling performs well to yield a stack of ten spatially overlapping Fermi seas. With about 30\,000 atoms per spin state, we obtain $T/T_F=0.10(1)$ at $T_F=160\,$nK; see Fig.~\ref{fig:DFG2}(b). The errors given here are statistical errors of multiple experimental realizations, and we estimate systematic errors to be of similar magnitude.

For the two-state mixture, we pump all atoms into a balanced population of $m_F=+9/2$ and $m_F=+7/2$ states. The total atom number, initial temperature, and evaporation trajectory are identical to the previous case of a ten-state mixture, but there is a crucial difference: Only half of all collisions possible for distinguishable particles are Pauli-allowed in the binary mixture, leading to a decreased thermalization rate. The reduced evaporation efficiency immediately manifests itself in the degree of degeneracy reached: Despite the higher atom number per spin state, we can reach only $T/T_F=0.20(1)$ with $1.0\times 10^5$\, atom per spin state and $T_F=230\,$nK. The lowest value reached is $T/T_F=0.17(1)$ with 60\,000 atoms remaining. Further evaporation does not reduce $T/T_F$.

In the case of a single-spin sample of $2.5\times 10^6$ fermionic atoms, we add $6.5\times 10^6$ atoms of $^{84}$Sr to the dipole trap to facilitate thermalization. Evaporation is performed in one single exponential ramp over 8\,s and results in a pure BEC of $7\times 10^5$\,atoms. Towards the end of evaporation, the fermionic cloud contains $10^5$\,atoms. The sample is not well thermalized in the axial direction of the trap, which we take into account by fitting the vertical and horizontal directions independently to obtain $T/T_F=0.12$ and $T/T_F=0.23$, respectively.

\subsection{Bose-Fermi mixtures}

In the previous section, we discussed sympathetic cooling of a single $^{87}$Sr spin state with the bosonic $^{84}$Sr isotope. Here, we will combine each of the three bosonic strontium isotopes with a ten-spin-state mixture of $^{87}$Sr \cite{Stellmer2013poq}. The distinguishing property of the three mixtures are the different inter- and intraspecies scattering lengths. The interspecies scattering length of $^{84}$Sr, $^{86}$Sr, and $^{88}$Sr with $^{87}$Sr are $-57\,a_0$, $162\,a_0$, and $55\,a_0$, respectively, and all allow efficient interspecies thermalization.

In a first experiment, we prepare a mixture of $^{84}$Sr~+~$^{87}$Sr. About $6\times10^6$ fermions at $1.15\,\mu$K and $2.1\times10^6$ bosons at $1.1\,\mu$K are loaded into the dipole trap. The first of two exponential evaporation ramps takes 12\,s and yields an essentially pure BEC of $^{84}$Sr. At the same time, we obtain $7\times10^4$ fermions per spin state at a temperature of about 100\,nK, but still well outside the degenerate regime. For further cooling, we add a very slow second evaporation ramp of 7.5\,s duration. Such a slow ramp is required because Pauli blocking decreases the scattering rate between the fermions, and superfluidity of the BEC decreases the scattering rate between fermions and bosons. The degree of Fermi degeneracy increases substantially to $T/T_F=0.15(1)$ with 15\,000 atoms in each spin component at the end of evaporation. The BEC atom number amounts to $2\times10^5\,$atoms. This experiment reaches an eleven-fold degeneracy of distinguishable particles, possibly the largest number of overlapping degenerate gases ever reported.

In a second experiment, we use the bosonic isotope $^{86}$Sr. The bosonic intraspecies and the interspecies scattering lengths are much larger than in the previous case, which we account for by decreasing the density of the sample. We load less atoms, and we keep the average trap frequency low by reducing the horizontal confinement. We maintain the concept of two sequential evaporation ramps of different time constants and evaporate slightly deeper than in the previous case, but reduce the total evaporation time to 3.4\,s. This isotopic combination performs worse than the previous one, yielding a bosonic sample with only 15\% condensate fraction. The horizontal trap frequencies, required to be small to keep three-body loss of $^{86}$Sr low, does not ensure thermalization of the fermionic sample in this direction. We obtain $T/T_F=0.15(5)$ for the vertical direction. The BEC contains 5000 atoms, and each Fermi sea contains 10\,000 atoms.

As a last experiment, we use the $^{88}$Sr isotope as the boson. Starting out with $1.2\times10^6$ bosons and $6.5\times10^6$ fermions both at $1.2\,\mu$K, we reduce the trap depth in two ramps of 12\,s and 8\,s. The atom number of the $^{88}$Sr BEC is limited by the negative scattering length, and evaporation to a low trap depth is required to remove the thermal fraction. We finally obtain a pure BEC of 4000 atoms immersed in ten Fermi seas, each comprising 10\,000 atoms at $T/T_F=0.11(1)$.

\section{Optical Feshbach resonances}
\label{sec:OFR}

The ability to tune interactions in ultracold atomic gases  is central to some of the most important experiments in the field, such as the exploration of many-body physics \cite{Bloch2008mbp} and the creation of quantum degenerate molecules \cite{Jochim2003bec,Greiner2003eoa}. Magnetic Feshbach resonances \cite{Chin2010fri} are commonly used for this purpose, but they are not present in atoms with non-degenerate ground states, such as alkaline-earth atoms. However, these systems do possess optical Feshbach resonances (OFRs). In OFRs, a laser tuned near a photoassociative resonance tunes interatomic interactions by coupling a colliding atom pair to a bound molecular level of an excited-state potential \cite{Fedichev1996ion}. OFRs offer several important new opportunities. For example, they can modulate scattering lengths on much smaller spatial and temporal scales than possible with magnetic Feshbach resonances, and they offer the ability to modify interactions between chosen pairs of species in mixtures without affecting other components. Theoretical proposals have highlighted the potential of optical Feshbach resonances for study of  nonlinear matter-wave phenomena \cite{Saito2003dsb,RodasVerde2005cse,Kartashov2011sin} and creation of novel quantum fluids \cite{Fisher1989bla,Qi2011bsa,Chien2011svi}.

Inelastic losses are a significant concern whenever a near-resonant laser field is applied to ultracold atoms. One of the most intriguing appeals of working with alkaline-earth atoms is the possibility of utilizing an OFR induced by a laser tuned near a weakly allowed intercombination-line transition, such as the $^1S_0 - {^3P_1}$ line. This was predicted \cite{Ciurylo2005oto,Ciurylo2006spa} to result in significantly less induced losses than in experiments with electric-dipole allowed transitions \cite{Fatemi2000ooo,Theis2004tts,Thalhammer2005iao}. Measurements in a thermal gas of $^{88}$Sr and a full coupled-channels calculation \cite{Blatt2011moo} showed that the performance is not as favorable as originally hoped. For example, modifying interactions to improve evaporative cooling efficiency appears infeasible. Nonetheless, OFRs are still promising for experiments that can proceed on faster timescales. The high spatial resolution possible with an OFR was demonstrated by modulating the mean field energy in a $^{174}$Yb BEC with an OFR-laser standing wave, which modified the atomic diffraction pattern \cite{Yamazaki2010ssm}. In addition, a $p$-wave OFR was observed in fermionic $^{171}$Yb \cite{Yamazaki2013ooa}.

Here we describe the use of an OFR to control collapse and expansion of an $^{88}$Sr BEC, and we closely follow the discussion in Ref.~\cite{Yan2013ccc}. This experiment benefits from the initially weak interactions in $^{88}$Sr, which allows convenient modification of the scattering length either positive or more negative. Large relative change in scattering length $a_{\textrm{opt}}/a_{{\textrm{bg}}}=\pm10$ is demonstrated, with the loss-rate constant $K_{\textrm{in}} \sim 10^{-12}$ cm$^3$/s comparable to what is seen in combined optical-magnetic Feshbach resonances \cite{Bauer2009com,Bauer2009coa}. Here, $a_{{\textrm{bg}}}$ is the background scattering length in the absence of the OFR. Controlling condensate collapse and expansion requires application of near-resonant light on the ms timescale of hydrodynamic phenomena.

According to the isolated resonance model \cite{Ciurylo2005oto,Ciurylo2006spa}, a laser of wavelength $\lambda$ detuned by $\Delta$ from  a photoassociative transition to an excited molecular state $|n\rangle$ modifies the atomic scattering length according to $a=a_{{\textrm{bg}}}+a_{\textrm{opt}}$ and induces two-body inelastic collisional losses described by the loss rate constant $K_{\textrm{in}}$, where

\begin{eqnarray}\label{OFRFormulas}
a_{\textrm{opt}}&=& \frac{\ell_{\textrm{opt}}\Gamma_{\textrm{mol}}\Delta}{\Delta^2+\frac{(\eta\Gamma_{\textrm{mol}})^2}{4}};
\nonumber\\
K_{\textrm{in}}&=&
\frac{2\pi\hbar}{\mu} \frac{\ell_{\textrm{opt}}\eta\Gamma_{\textrm{mol}}^2}{\Delta^2+\frac{(\eta\Gamma_{\textrm{mol}}+\Gamma_{\textrm{stim}})^2}{4}}.
\end{eqnarray}

$K_{\textrm{in}}$ is defined such that it contributes to the evolution of density $n$ as $\dot{n}=-K_{\textrm{in}} n^{2}$ for a BEC. The optical length $\ell_{\textrm{opt}}$, which characterizes the strength of the OFR, is proportional to laser intensity and the Franck-Condon factor for the free-bound photoassociative transition. $\Gamma_{\textrm{mol}} =2\pi \times 15$\,kHz is the natural linewidth of the excited molecular level in strontium, and $\Gamma_{\textrm{stim}}$ is the laser-stimulated linewidth of the transition \cite{Ciurylo2005oto,Ciurylo2006spa}, which can be neglected for our conditions.

We find the isolated-resonance-model expressions (Eq.\,\ref{OFRFormulas}) useful for describing our measurements with the modification that the total loss rate constant is given by $K_{\textrm{total}}=K_{\textrm{in}}+K_\textrm{b}$, where the background loss is described phenomenologically in our regime as  $K_\textrm{b}=K_0[\Gamma_{\mathrm{mol}}/(2\delta)]^2$, where $\delta$ is the detuning from atomic resonance. The parameter $\eta > 1$  in Eq.\,\ref{OFRFormulas} accounts for enhanced molecular losses, as observed in previous OFR experiments \cite{Theis2004tts,Blatt2011moo}.

To characterize the effect of the OFR on scattering length and loss, we measure the expansion of an $^{88}$Sr BEC after release from the optical dipole trap  with time-of-flight absorption imaging using the $^1S_0 - {^1P_1}$ transition.  We create condensates with about 7000 atoms and a peak density of $n_0=1\times 10^{15}\,\mathrm{cm}^{-3}$. About 10\% of the trapped atoms are in the condensate and this represents about 95\% of the critical number for collapse with the background scattering length of $^{88}$Sr for our dipole trap, which is close to spherically symmetric.  The 689\,nm OFR laser beam is tuned near the photoassociative transition to the second least bound vibrational level on the $^1S_0 + {^3P_1}$ molecular potential, which has the binding energy of $h\times 24$\,MHz \cite{Zelevinsky2006nlp}.

The OFR laser, with a beam waist of $725\,\mu$m, is applied to the condensate 20\,$\mu$s before extinguishing the dipole trap and left on for the first few milliseconds of the expansion. The exposure time in the dipole trap is short, such that its potential and the background scattering length determine the initial density distribution of the condensate, while interaction energy determined by $a=a_{{\textrm{bg}}}+a_{\textrm{opt}}$ affects the expansion dynamics.

\begin{figure}
\includegraphics[width=165mm]{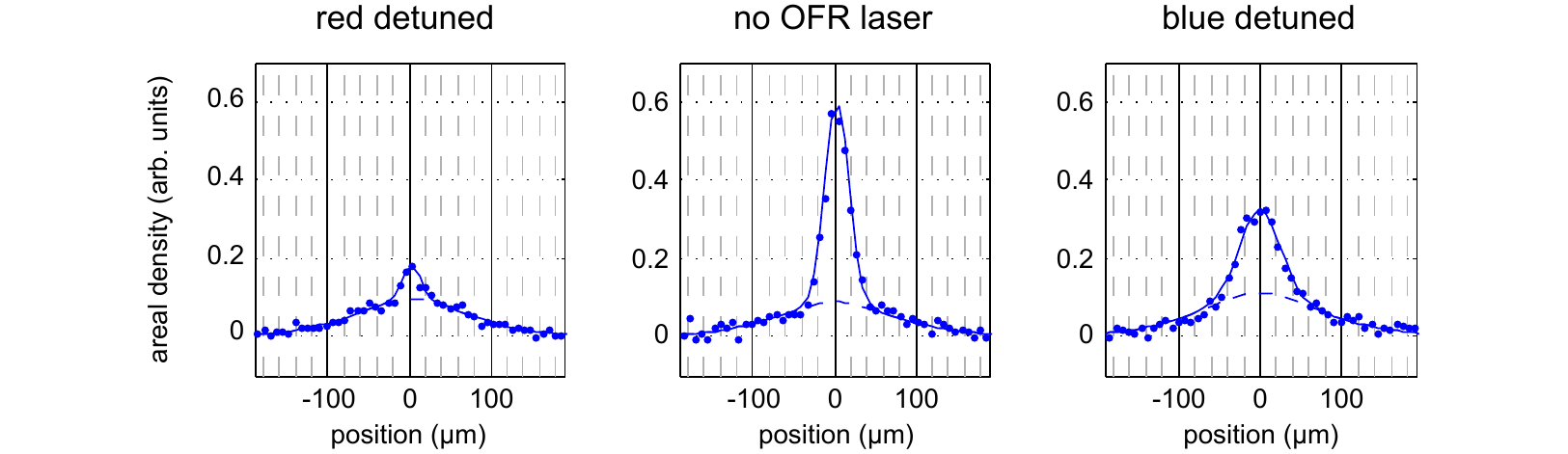}
\caption{Line profiles through absorption images showing OFR-induced variation of BEC expansion. Data correspond to no OFR laser and an OFR laser blue and red detuned by 0.5\,MHz  with respect to the $-24\,$MHz PA line \cite{Zelevinsky2006nlp} applied for $\tau=1.2$\,ms. Expansion times are 35\,ms. Fits are a Bose distribution for the thermal atoms (- -) and a Gaussian density distribution for the BEC. The figure is taken from Ref.~\cite{Yan2013ccc}.}
\label{feshbachonoff}
\end{figure}

Figure \ref{feshbachonoff} shows 1D slices through absorption images of atoms after a 35\,ms time of flight with and without application of the OFR laser. We fit the data with a Bose distribution for the thermal atoms plus a narrow Gaussian for the BEC, $n(r)=\frac{N_{0}}{2\pi\sigma^2} \exp\left[-\frac{r^2}{2 \sigma^2}\right]$, to determine the number of atoms in the BEC $N_{0}$ and BEC size $\sigma$. A blue detuning of the OFR laser from the PA resonance increases $a$, leading to more interaction energy and larger expansion velocity and BEC size. Red detuning produces the opposite behavior that leads to condensate collapse and significant atom loss for strong enough attractive interactions.

\begin{figure}
\includegraphics[width=130mm]{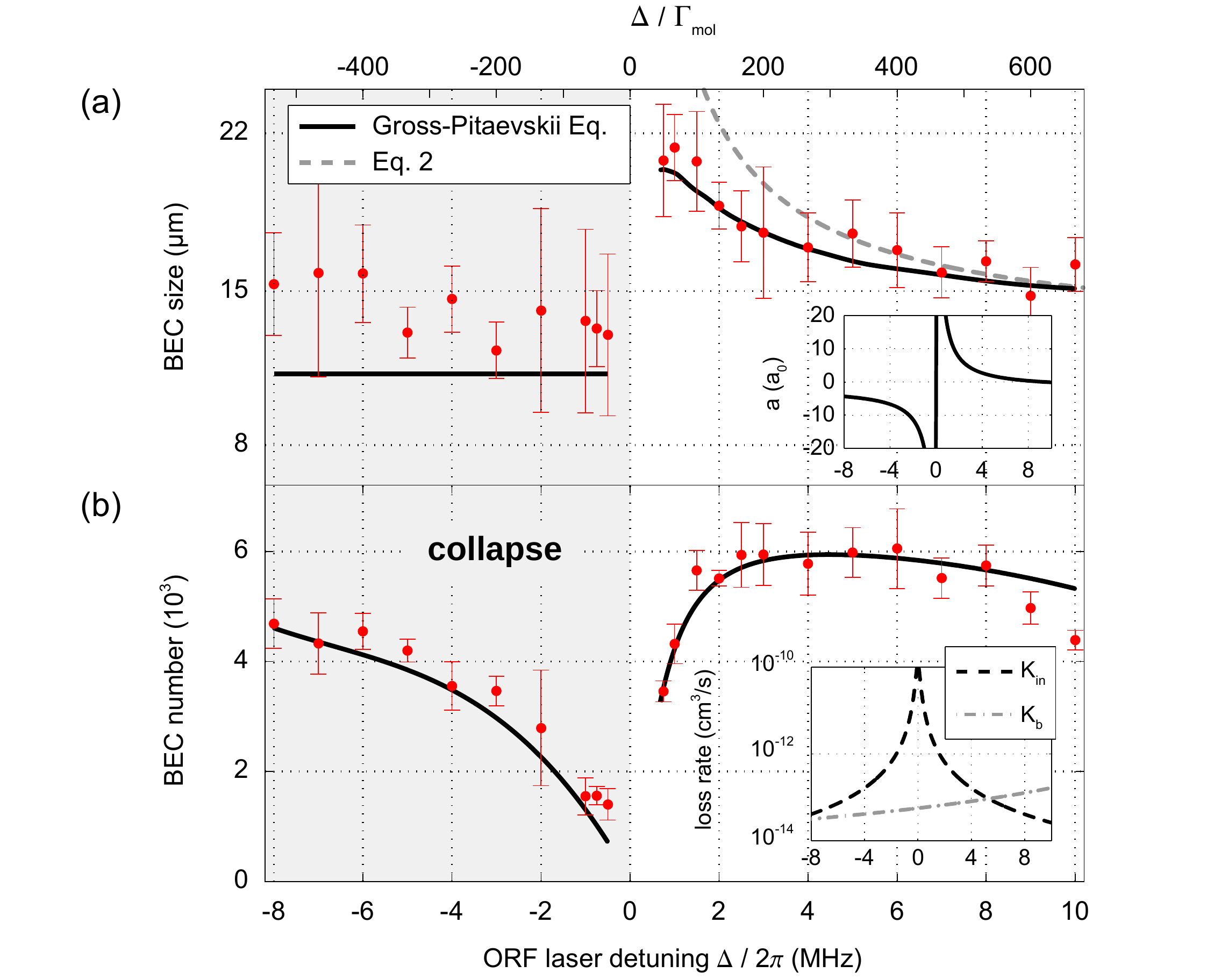}
 \caption{The BEC size (a) and number (b) versus the detuning with respect to the $-24\,$MHz PA resonance for an intensity of 0.057\,W/cm$^2$. The OFR beam is applied for 4.0\,ms, and the data are recorded after 35\,ms of expansion. The insets give the total scattering length $a$ and the loss rate constants. The figure is adapted from Ref.~\cite{Yan2013ccc}.}
\label{BECsizeNumVsOFRdetuning}
\end{figure}

The dependence of the BEC size and number on detuning from the $-24\,$MHz PA line is shown in Fig.~\ref{BECsizeNumVsOFRdetuning} for a fixed laser intensity and interaction time. Note that the number of atoms initially increases with blue detuning from PA resonance as the loss $K_{\textrm{in}}$ from the OFR decreases. The number then slowly decreases because the background loss $K_\textrm{b}$ increases approaching atomic resonance. The BEC size data predicted by  simple conservation of energy neglecting atom loss is also shown in Fig.\ \ref{BECsizeNumVsOFRdetuning}(a), which highlights that atom loss is significant  at smaller detunings. A red-detuned OFR laser makes the scattering length more negative, which triggers a collapse of the condensate. This is evident as large loss in the plot of condensate number remaining after expansion; see Fig.~\ref{BECsizeNumVsOFRdetuning}(b). The asymmetry of loss with respect to detuning shows that the loss reflects condensate dynamics \cite{Kagan1998cab,Gerton2000doo,Donley2001doc}, not direct photoassociative loss.

We quantitatively describe the evolution of the condensate with the time-dependent non-linear Gross-Pitaevskii equation, including the effects of $a_{{\textrm{opt}}}$, $K_{\textrm{total}}$, and single atom light scattering, and neglecting effects of thermal atoms. The parameters $\ell_{{\textrm{opt}}}/I$, $\eta$, and $K_0$ are varied to fit the data. The rate of atomic light scattering  varies from 12 to 17\,s$^{-1}$, and is included in the simulation assuming every scattering event results in the loss of one atom. Fig.~\ref{BECsizeNumVsOFRdetuning} also shows the results of this model. The fit optical length is $\ell_{{\textrm{opt}}}/I=(2.2\pm 1.0)\times 10^4 \,a_0/$(W/cm$^2$), and the fit parameter $K_0=(5.8\pm1.3) \times 10^{-7}$\,cm$^3$/s. Loss from the OFR is described by $\ell_{\textrm{opt}}$ and $\eta=19.5^{+8}_{-3}$, and there is strong anti-correlation between $\ell_{\textrm{opt}}$ and $\eta$. The uncertainty is dominated by systematic uncertainty in the trap oscillation frequency and imaging resolution. These results are in good agreement with the measured value  $\ell_{{\textrm{opt}}}/I=1.58\times10^4\,a_0/$(W/cm$^2$) and disagree slightly with $\ell_{{\textrm{opt}}}/I=8.3\times 10^3\,a_0/$(W/cm$^2$) calculated directly from knowledge of the molecular potentials \cite{Blatt2011moo}. A typical total scattering length is $a=20\,a_0$ for $\Delta=2\pi\times1$\,MHz\,$\simeq\,67$\,$\Gamma_{\textrm{mol}}$; see the inset of Fig.~\ref{BECsizeNumVsOFRdetuning}(a).

Experiments with a thermal strontium gas \cite{Blatt2011moo} found  losses that were described by $\eta=2.7$. These measurements probed  the core of the photoassociative transition ($|\Delta|<50\,\Gamma_{\textrm{mol}}$).  We see a similar resonance width in a BEC when we significantly reduce the laser intensity and interaction time and take a photoassociative loss spectrum  of this core region.  Our use of the OFR probes the distant wings ($50\,\Gamma_{\textrm{mol}}\,<\,\Delta\,<\,667\,\Gamma_{\textrm{mol}}$), and a fit of the loss using the single resonance model requires an even larger value of $\eta$. The additional loss is not well understood, but we interpret the varying $\eta$ values as meaning that the full spectrum of photoassociative loss, including the far wings, is not well described by a Lorentzian.

There are several ways to obtain a larger OFR effect or increased sample lifetime, which holds promise to bring many possible experiments involving optical Feshbach resonances and quantum fluids into reach. Because of the attractive interactions, the  peak density of the $^{88}$Sr condensate in these experiments is extremely high. Lower densities, such as the densities commensurate with single-atom-per-site loading of an optical lattice, would reduce the loss. Improvements could also be made by working at larger detuning from PA resonance and larger laser intensities. Working with a more deeply bound excited molecular state such as the photoassociative line at $-1.08$\,GHz \cite{Zelevinsky2006nlp} may offer advantages in this direction, such as greater suppression of atomic light scattering and reduced background two-body loss.

\section{Strontium atoms in a 3D optical lattice}
\label{sec:OpticalLattice}

Many experiments targeted at the study of magnetism, topological phases, and related topics take place on a lattice geometry. The sites of an optical lattice also form perfect test tubes for the preparation of rovibronic ground-state molecules. Over the last decade, optical lattices have matured into a versatile tool for the study of solid-state phenomena with ultracold atoms \cite{Bloch2008mbp}. The first lattice experiments with alkaline-earth atoms were performed with both bosonic \cite{Fukuhara2009mio} and fermionic \cite{Taie2010roa} atoms of ytterbium.

In the following, we will describe first experiments performed with ultracold strontium in a lattice.  A simple cubic lattice structure is formed by three mutually orthogonal, retroreflected laser beams derived from a solid-state laser at 532\,nm. The beams have waists of about $100\,\mu$m and create an attractive potential for the atoms. The calibration of the lattice depth relies on measuring the energy gap $\Delta\,E$ between the zeroth and second band, and relating this energy to the lattice depth via a simple band structure calculation \cite{Ovchinnikov1999doe,Denschlag2002abe}. The lattice depth is usually expressed in units of the recoil energy $E_{\rm rec}=\hbar^2 k_L^2/2m$, where $k_L=2\pi/\lambda_L$ is the lattice wave vector.

The number of atoms residing on either singly, doubly, or triply occupied sites is accessed in the following way. When loading atoms into the lattice, we observe strong atom loss beyond a certain value of the initial atom number. We attribute this loss to the formation of triply occupied sites in the central region, which rapidly eliminate themselves through inelastic three-body recombination on a timescale of 10\,ms. After this loss, no lattice site harbors more than two atoms. The number of doubly occupied sites is measured through atom loss induced by light tuned to a PA resonance. All atoms surviving the lattice loading and subsequent exposure to PA light must be located on singly occupied sites.

\subsection{Superfluid-to-Mott insulator transition in $^{84}$Sr}

\begin{figure}[t]
\centering
\includegraphics[width=165mm]{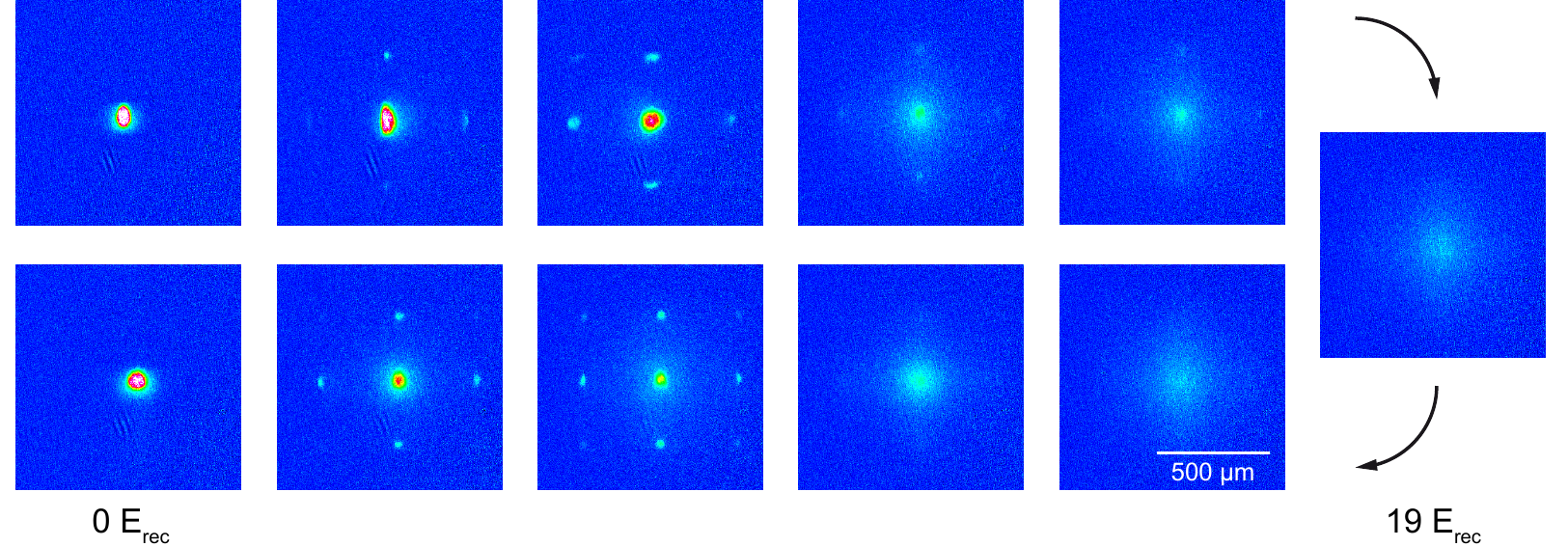}
\caption{The MI transition in $^{84}$Sr. Starting from the top left, the lattice depth is adiabatically increased, and we observe the appearance of superfluid diffraction peaks. For deep lattices (right), the visibility of diffraction peaks vanishes, and the atoms are in the insulating state. The rebirth of a BEC upon lowering of the lattice (bottom left) shows that the heating of the lattice is indeed small. The time of flight is 24\,ms.}
\label{fig:Lattice1}
\end{figure}

We will now present the famous superfluid-to-Mott insulator transition \cite{Greiner2002qpt} using strontium as the atomic species. We start out with a pure $^{84}$Sr BEC in a crossed dipole trap. The lattice depth is adiabatically increased from zero to $19\,E_{\rm rec}$, held there for a short time, and decreased to zero again. Snapshots of the momentum distribution are taken at various times throughout the cycle and shown in Fig.~\ref{fig:Lattice1}. For small lattice depths, the BEC remains superfluid, and the periodic potential is a small perturbation. Particles seek to minimize their kinetic energy by spreading out over the lattice, establishing a fixed phase relation across the sample. We observe the appearance of diffraction peaks, caused by the constructive interference of atoms from all lattice sites; equivalent to the diffraction of light from a periodic structure. As the lattice depth increases, atoms begin to localize on individual lattice sites, and the phase relation dwindles away: the visibility of the diffraction pattern drops. For the very deep lattice, the atom number on each lattice site is a well-defined Fock state, and the atomic wavepacket is tightly confined to a specific lattice site; all phase coherence between lattice sites is lost. This is called the Mott-insulator (MI) regime, detected by the complete disappearance of interference peaks upon release from the trap. The external potential leads to a ``wedding-cake'' structure with plateaus of uniform filling $n$ and superfluid layers in between.

As the lattice depth is then reduced to zero, the sample becomes superfluid again, and phase coherence across the sample is regained:~the original BEC re-appears, provided the ramping was adiabatic and insignificant heating was applied to the system.

The images shown in Fig.~\ref{fig:Lattice1} are strong evidence, but not sufficient proof of the MI transition. The Mott state is characterized by large regions of lattice sites with equal (\textit{i.e.}~at least unity) filling, but the corresponding information is not contained in these images. We can, however, deduce the existence of a dense central region in an indirect way. We observe the formation of triply occupied sites by strong atom loss, and the occupation of doubly occupied sites by PA measurements. The detection of such multiply occupied sites affirms the existence of large $n=1$ and $n=2$ Mott shells. These measurements are backed by a calculation of the wedding cake structure in dependence of atom number using a simple model \cite{DeMarco2005sas,Sherson2010sar}.

\subsection{Fermions on a lattice}

An ultracold gas of fermions loaded into an optical lattice allows for a realization of the famous Fermi-Hubbard model \cite{Bloch2008mbp,Esslinger2010fhp}. Ground-breaking experiments with alkali atoms have been performed, measuring basic properties like the incompressibility of a fermionic lattice gas \cite{Joerdens2008ami,Schneider2008mai} and establishing fundamental techniques such as band spectroscopy \cite{Heinze2011mbs}. Many of the proposals for alkaline-earth systems outlined in Sec.~\ref{sec:Introduction} utilize fermions on a lattice as the underlying structure. First experiments with alkaline-earth atoms have been performed with ytterbium \cite{Taie2010roa} and have recently demonstrated Pomeranchuk cooling, an important step towards the envisioned low entropy phases in SU($N$) lattice systems \cite{Taie2012asm}.

We will now present the loading of degenerate fermionic $^{87}$Sr atoms into a lattice. The maximum number of spin states in strontium is larger than in ytterbium, which extends the SU($N$) systems up to $N=10$ and might allow to reach lower temperatures \cite{Bonnes2012alo}.

\begin{figure}[t]
\centering
\includegraphics[width=165mm]{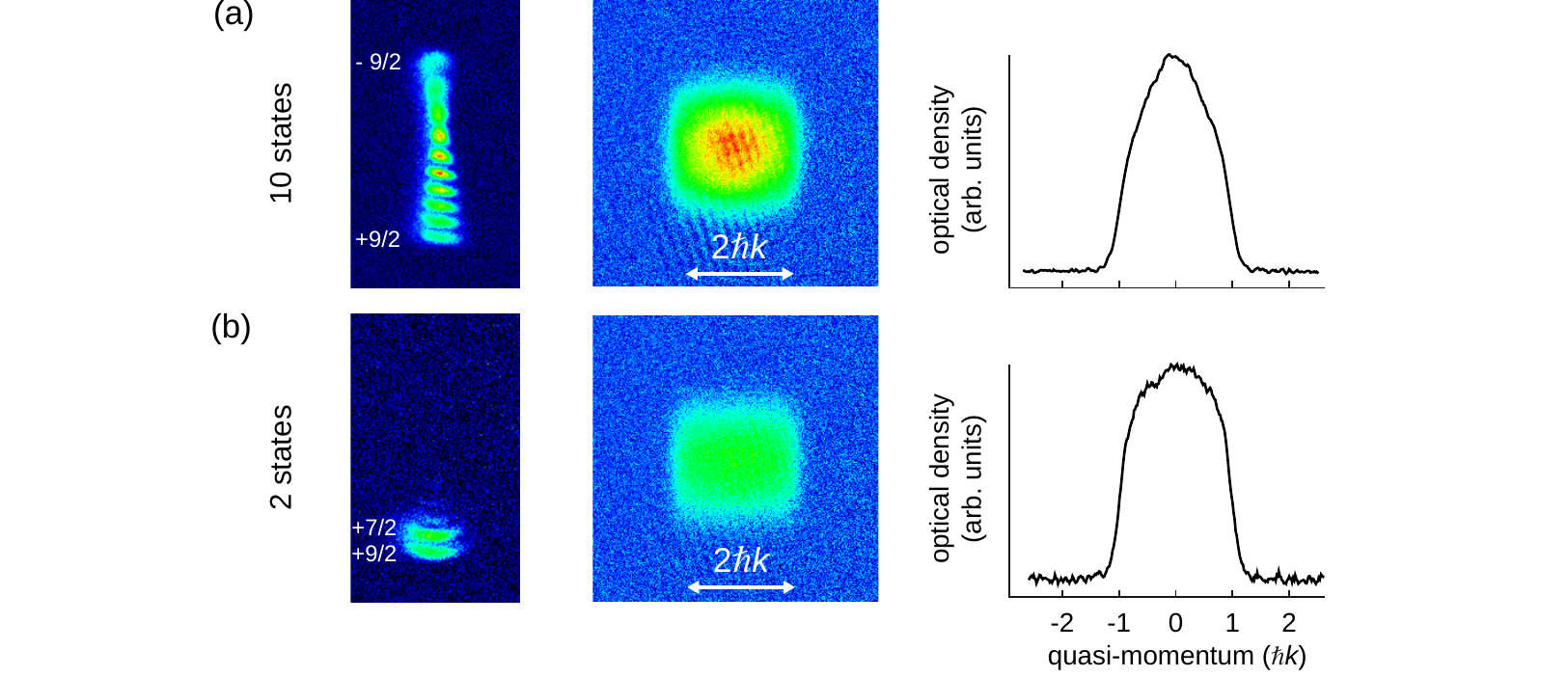}
\caption{Fermions in a lattice with (a) ten spin states and (b) only two spin states. The degree of degeneracy is about $T/T_F=0.15$ in both cases, the number of occupied spin states is visualized by the OSG technique. The band mapping technique shows the population of the lowest Bloch band. There are about twice as many atoms per spin state in case (b) than in case (a), giving rise to a more uniform filling of the first Brillouin zone. These images are taken after 24\,ms of free expansion, they have the same color scale and are averaged over five experimental runs.}
\label{fig:Lattice2}
\end{figure}

In a first experiment, we evaporate a mixture of $N=10$ spin states to a degeneracy of $T/T_F=0.14(2)$ and load the sample of 28\,000 atoms per spin state into a lattice with a depth of about $16\,E_{\rm rec}$. We perform band mapping \cite{Kohl2005fai} and observe a partially-filled first Brillouin zone; see Fig.~\ref{fig:Lattice2}(a). The occupation of higher momenta is not caused by thermal atoms, but a beautiful signature of the finite Fermi momentum even for $T\rightarrow 0$. The loading into and out of the lattice is not entirely adiabatic, and we observe a decrease in degeneracy to $T/T_F=0.30(5)$ as the lattice is ramped down again. This heating is comparable to previous experiments with alkali atoms \cite{Joerdens2008ami,McKay2011cis}, and the sample remains in the degenerate regime throughout the sequence.

In a second experiment, we optically pump the atoms into a mixture of only $N=2$ spin states before evaporation. All other experimental parameters, such as the final trap depth, the lattice depth, and the timing are identical to the previous case. Evaporation proceeds a little less efficient, and we obtain 62\,000 atoms per spin state at $T/T_F=0.17(3)$. The larger atom number per spin state leads to an increased filling of the first Brillouin zone, as nicely seen by a comparison of Figs.~\ref{fig:Lattice2}(a) and (b). The heating during the sequence seems to be smaller than in the case of $N=10$, and we measure a degeneracy of $T/T_F=0.20(2)$ after the lattice ramp-down.

\subsection{A Mott-insulator with impurities}

Optical lattices are ideally suited to create perfectly periodic potentials that mimic the crystal structure of solids. Real solid state materials, however, feature various types of defects such as dopants and dislocations, and some of their most important properties are determined by the concentration of such impurities. While standing waves of light fail to capture these defects, impurities can be introduced by admixture of a second atomic species into the system. Such a setting allows for the study of disorder-related phenomena and of two-species mixtures over a wide range of relative concentrations, which is particularly interesting for Bose-Fermi mixtures \cite{Lewenstein2004abf,Illuminati2004hta,Roth2004qpo,Buchler2003svp}.

This system was first studied with ultracold atoms using a mixture of potassium and rubidium \cite{Ospelkaus2006lob}. In this experiment, the visibility of the diffraction peaks was monitored across the MI transition for different amounts of fermionic impurities, and it was found that the MI transition shifts towards smaller lattice depths with increasing impurity concentration, which was attributed to the appearance of a localized phase. Here, we adapt the potassium-rubidium experiment to mixtures of alkaline-earth atoms \cite{Sugawa2011iaf}.

\begin{figure}[t]
\centering
\includegraphics[width=165mm]{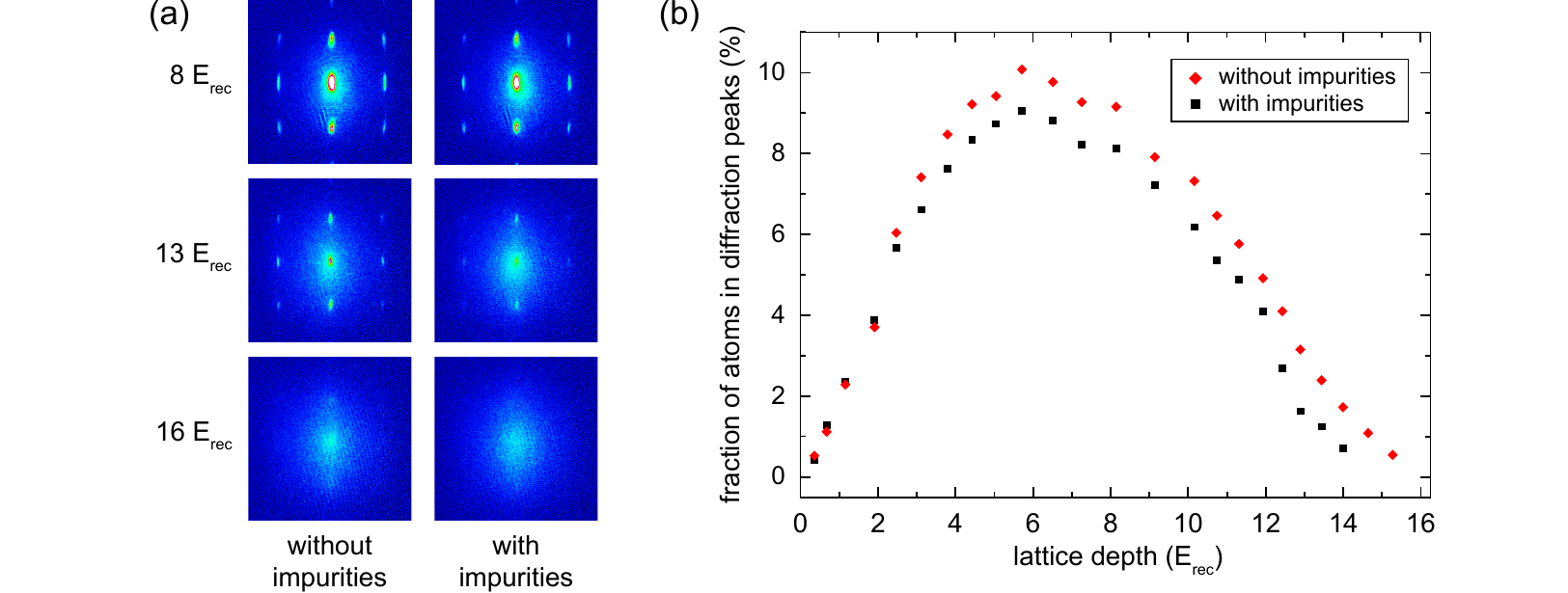}
\caption{Modification of the MI transition through fermionic impurity atoms. (a) Absorption images of the momentum distribution, taken for different lattice depths. (b) Fraction of atoms located in the diffraction peaks. For each step, we count the number of atoms in all eight interference peaks and normalize the sum by the overall atom number.}
\label{fig:Lattice3}
\end{figure}

The starting point for our experiment is a pure $^{84}$Sr BEC containing $6.0\times 10^5$ atoms with an average trap frequency of 80\,Hz. The 3D optical lattice with a lattice constant of 266\,nm is adiabatically ramped to a depth of $17\,E_{\rm rec}$. Absorption images of the momentum distribution are taken at various points throughout the adiabatic ramp; see Fig.~\ref{fig:Lattice3}(a).

In a second set of measurements, we perform the same experimental protocol, but add $2.1\times 10^5$ atoms of $^{87}$Sr to the bosons. The fermions are distributed among all ten $m_F$ states, and the degree of degeneracy is $T/T_F=0.15(2)$. The peak filling factor in the center of the trap is about 0.2 atoms of each spin component per lattice site, and ten times this value for all fermionic atoms. The interspecies scattering length between bosonic $^{84}$Sr and fermionic $^{87}$Sr is attractive and amounts to $-57\,a_0$. As can be seen in Fig.~\ref{fig:Lattice3}(a), fewer atoms are visible in the interference peaks for intermediate and large lattice depths.

To analyze the data more quantitatively, we determine the fraction of atoms in the diffraction peaks in dependence of lattice depth; see Fig.~\ref{fig:Lattice3}(b). We find that for lattice depths larger than a few $E_{\rm rec}$, a significantly smaller fraction of atoms is located in the the diffraction peaks if impurity atoms are present. The onset of the Mott-insulator state is shifted towards smaller lattice depths by about $1\,E_{\rm rec}$. We interpret this localization as an effect driven by the impurities. The total $^{84}$Sr atom number of the data set with impurities is systematically larger by 2.0\% compared to the case without impurities, where this value is comparable to the shot-to-shot fluctuation in the atom number, which is 1.1\% for both series. This small difference cannot account for the observed shifts, and we carefully normalize all values to the overall atom number.

Comparing now our experiment to the one of Ref.~\cite{Ospelkaus2006lob}, we find two main differences. At first, the shift observed in our experiment is substantially smaller, which can be explained by the fact that the interspecies scattering length in our case is a factor of 3.6 smaller than for the potassium-rubidium mixture. Second, we observe a difference in visibility already for intermediate lattice depths, where the alkali experiment observed deviations only for lattices deeper than $10\,E_{\rm rec}$. This might be explained by the fact that the filling factor in our case is larger, such that the presence of the fermions is more than just a small admixture of impurities and effects the system properties already in the superfluid phase.

\section{Sr$_2$ molecules}
\label{sec:Molecules}

The creation of ultracold molecular gases has made rapid progress over the last years. The rich internal structure of molecules combined with low translational energy enables precision measurements of fundamental constants, realizations of novel quantum phases, and applications for quantum computation \cite{Krems2009book}. A very successful route to large samples of ultracold molecules with complete control over the internal and external quantum state is association of molecules from ultracold atoms. Early experiments used magnetic Feshbach resonances to form weakly bound bi-alkali molecules, some of which have even been cooled to quantum degeneracy \cite{Ferlaino2009ufm}. Stimulated Raman adiabatic passage (STIRAP) \cite{Vitanov2001lip} has enabled the coherent transfer of these Feshbach molecules into the vibrational ground state \cite{Danzl2008qgo,Lang2008utm,Ni2008ahp}. In particular, heteronuclear molecules in the vibrational ground state have received a lot of attention, because they possess a strong electric dipole moment, leading to anisotropic, long-range dipole-dipole interactions, which will enable studies of fascinating many-body physics \cite{Pupillo2009cmp}. Efforts are underway to create samples of completely state-controlled molecules beyond bi-alkalis \cite{Nemitz2009poh,Hara2011qdm,Hansen2011qdm}, which will widen the range of applications that can be reached experimentally.

\begin{figure}
\centering
\includegraphics[width=165mm]{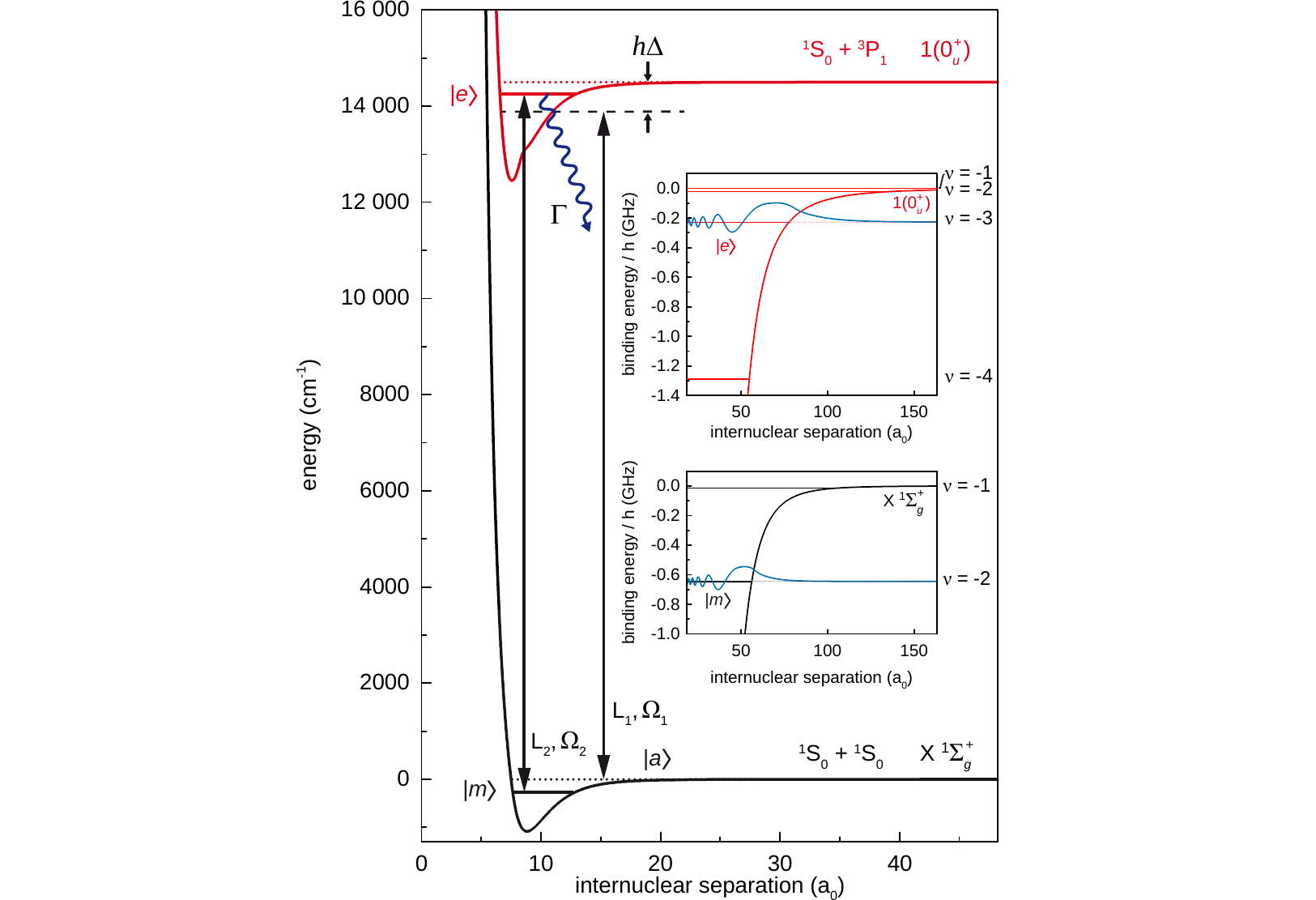}
\caption{Molecular potentials and vibrational levels of $^{84}$Sr$_2$ involved in STIRAP. The initial state $\left|a\right>$, an atom pair in the ground state of an optical lattice well, and the final molecular state $\left|m\right>$, are coupled by laser fields L$_1$ and L$_2$  to the excited state $\left|e\right>$ with Rabi frequencies $\Omega_1$ and $\Omega_2$, respectively. The parameter $\Delta$ is the detuning of L$_1$ from the ${^1S_0} - {^3P_1}$ transition and $\Gamma$ is the decay rate of $\left|e\right>$. The insets show the last vibrational levels of the molecular potentials and the wavefunctions of states $\left|m\right>$ and $\left|e\right>$. For comparison, the wavefunction of atomic state $\left|a\right>$ (not shown) has its classical turning point at a radius of $800\,a_0$, where $a_0$ is the Bohr radius. The potentials are taken from \cite{Stein2008fts,Skomorowski2012rdo} and the wavefunctions are calculated using the WKB approximation. The energies of states $\left|m\right>$ and $\left|e\right>$ are not to scale in the main figure. The figure is taken from Ref.~\cite{Stellmer2012cou}.}
\label{fig:molecules1}
\end{figure}

So far, the key step in the efficient creation of ultracold molecules has been molecule association using magnetic Feshbach resonances. This magnetoassociation technique cannot be used to form dimers of alkaline-earth atoms, because of the lack of magnetic Feshbach resonances in these nonmagnetic species. An example is Sr$_2$, which has been proposed as a sensitive and model-independent probe for time variations of the proton-to-electron mass ratio \cite{Zelevinsky2008pto,Kotochigova2009pfa,Beloy2011eoa}. Another class of molecules for which magnetoassociation is difficult, are dimers containing an alkali atom and a nonmagnetic atom, since in these cases magnetic Feshbach resonances are extremely narrow \cite{Zuchowski2010urm,Brue2012mtf}.

In the following, we show that ultracold Sr$_2$ molecules in the electronic ground state can be efficiently formed, despite the lack of a magnetic Feshbach resonance. Instead of magnetoassociation, we combine ideas from Refs.~ \cite{Mackie2000bsr,Jaksch2002coa,Drummond2002sra,Mackie2005cos,Drummond2005rtc,Tomza2011fou} and use optical transitions to transform pairs of atoms into molecules by STIRAP \cite{Vitanov2001lip}. The molecule conversion efficiency is enhanced by preparing pairs of atoms in a Mott insulator on the sites of an optical lattice \cite{Jaksch1998cba,Greiner2002qpt}. We use the isotope $^{84}$Sr for molecule creation, since it is ideally suited for the creation of a BEC \cite{Stellmer2009bec,Martinez2009bec}, and formation of a Mott insulator.

STIRAP coherently transfers an initial two-atom state $\left|a\right>$ into a molecule $\left|m\right>$ by optical transitions; see Fig.~\ref{fig:molecules1}. In our case, the initial state $\left|a\right>$ consists of two $^{84}$Sr atoms occupying the ground state of an optical lattice well. The final state $\left|m\right>$ is a Sr$_2$ molecule in the second-to-last ($\nu=-2$) vibrational level of the $X{^1\Sigma_g^+}$ ground-state molecular potential without rotational angular momentum. The molecules have a binding energy of 645\,MHz and are also confined to the ground state of the lattice well. States $\left|a\right>$ and $\left|m\right>$ are coupled by laser fields L$_1$ and L$_2$, respectively, to state $\left|e\right>$, the third-to-last ($\nu'=-3$) vibrational level of the metastable 1(0$_u^+$) state, dissociating to ${^1S_0} + {^3P_1}$.

\begin{figure}
\centering
\includegraphics[width=165mm]{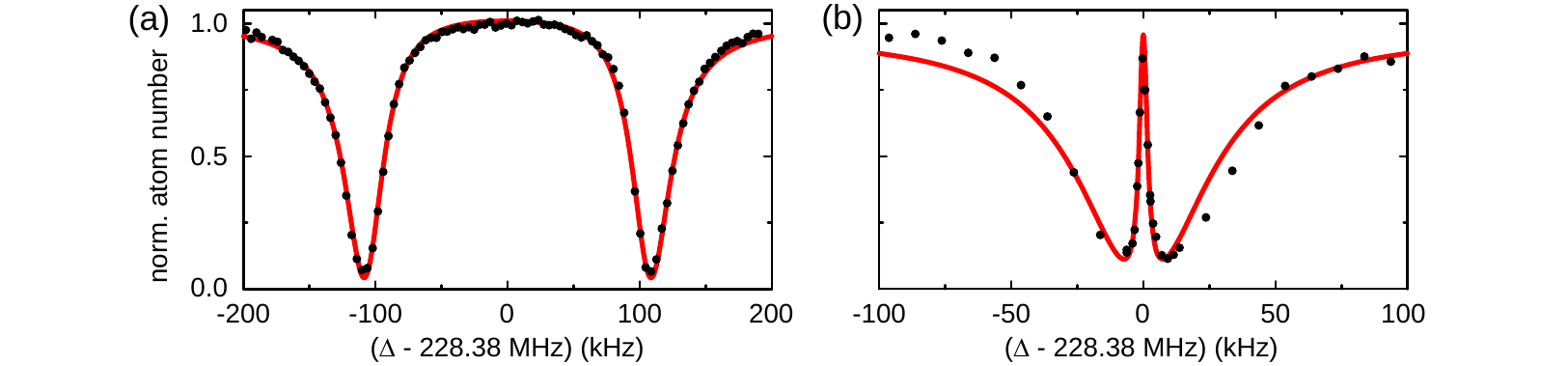}
\caption{Two-color PA spectra near state $\left|e\right>$ for two intensities of L$_2$. (a) For high intensity (20\,W/cm$^2$) the spectrum shows an Autler-Townes splitting. (b) For low intensity (80\,mW/cm$^2$) a narrow dark resonance is visible. For both spectra, the sample was illuminated by L$_1$ for 100\,ms with an intensity of 7\,mW/cm$^{2}$ at varying detuning $\Delta$ from the ${^1S_0}-{^3P_1}$ transition. The lines are fits according to a three-mode model \cite{Winkler2005amd}. The figure is adapted from Ref.~\cite{Stellmer2012cou}.}
\label{fig:Molecules2}
\end{figure}

The binding energies of the last bound vibrational states of the relevant potentials are determined using standard one- and two-color PA spectroscopy \cite{Stellmer2012cou}. These spectroscopy measurements are performed with a BEC of $^{84}$Sr, the binding energies are listed in Tabs.~\ref{tab:1ColorPA} and \ref{tab:2ColorPA}. The observation of dark resonances using two-color PA allows us to resolve the vibrational states of the $X{^1\Sigma_g^+}$ potential with high precision; see Fig.~\ref{fig:Molecules2}.

To enhance molecule formation, we create a Mott insulator by loading the BEC into an optical lattice. The local density increase on a lattice site leads to an increased free-bound Rabi frequency $\Omega_1$ compared to a pure BEC. Furthermore, molecules are localized on lattice sites and thereby protected from inelastic collisions with each other. The lattice is formed by three nearly orthogonal retroreflected laser beams with waists of 100\,$\mu$m on the atoms, derived from an 18-W single-mode laser operating at a wavelength of $\lambda=532\,$nm. Converting the BEC into a Mott insulator is done by increasing the lattice depth during 100\,ms to 16.5\,$E_{\rm rec}$. By inducing PA loss using L$_1$, we can show that half of these atoms occupy sites in pairs.

\begin{figure}[b]
\centering
\includegraphics[width=165mm]{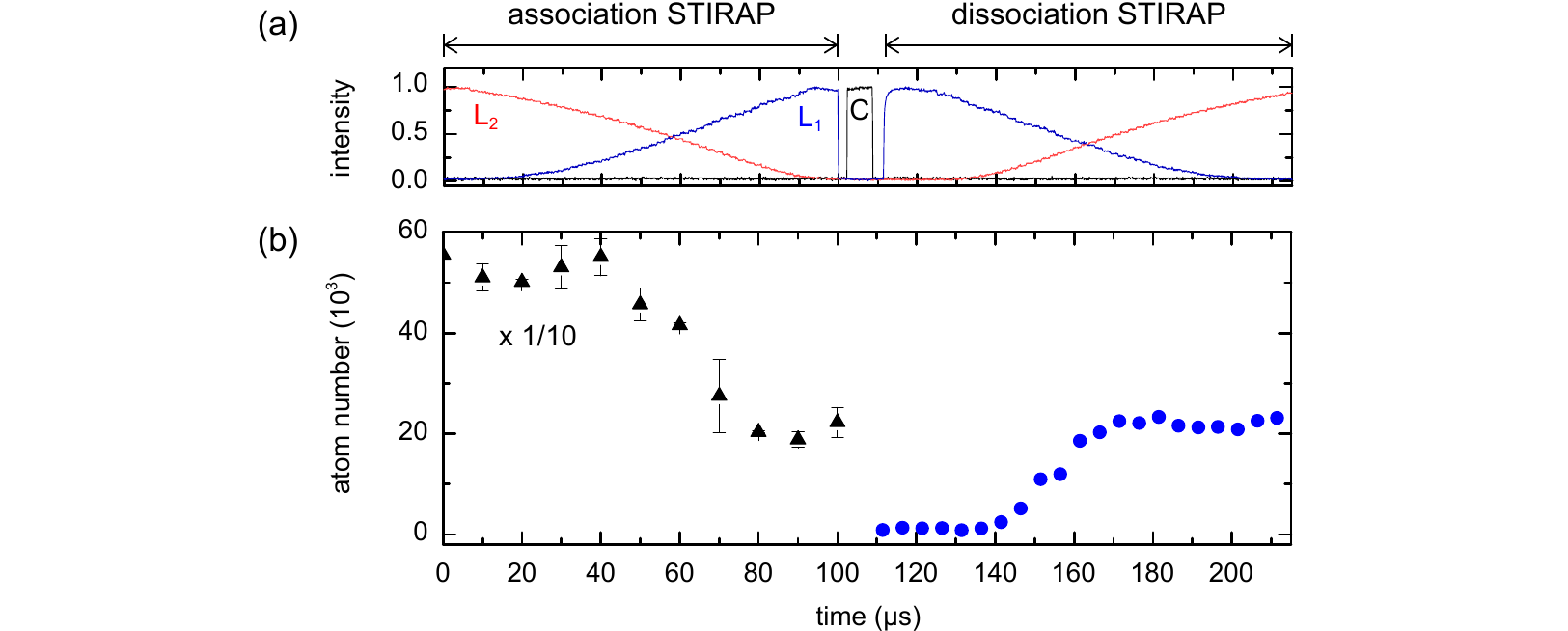}
\caption{Time evolution of STIRAP transfers from atom pairs to Sr$_2$ molecules and back. (a) Intensities of L$_1$, L$_2$, and cleaning laser C, normalized to one. (b) Atom number evolution. For these measurements, L$_1$ and L$_2$ are abruptly switched off at a given point in time and the atom number is recorded on an absorption image after 10\,ms free expansion. Note the scaling applied to data taken during the first 100\,$\mu$s (triangles). The starting condition for the time evolution shown is a Mott insulator. The figure is adapted from Ref.~\cite{Stellmer2012cou}.}
\label{fig:Molecules3}
\end{figure}

We are now ready to convert the atom pairs on doubly occupied sites into molecules by STIRAP. This method relies on a counterintuitive pulse sequence \cite{Vitanov2001lip}, during which L$_2$ is pulsed on before L$_1$. During this sequence, the atoms populate the dark state $\left|\Psi\right>=(\Omega_1 \left|m\right> + \Omega_2 \left|a\right> ) / (\Omega_1^2 + \Omega_2^2)^{1/2}$, where $\Omega_1$ and $\Omega_2$ are the time-dependent Rabi frequencies of the two coupling laser fields as defined in Ref.~\cite{Vitanov2001lip}, which can reach up to $\Omega_1^{\rm max}\approx2\pi \times 150$\,kHz and $\Omega_2^{\rm max}=2\pi \times 170(10)$\,kHz in our case. Initially the atoms are in state $\left|a\right>$, which is the dark state after L$_2$ is suddenly switched on, but L$_1$ kept off. During the pulse sequence, which takes $T=100\,\mu$s, L$_1$ is ramped on and L$_2$ off; see association STIRAP in Fig.~\ref{fig:Molecules3}(a). This adiabatically evolves the dark state into $\left|m\right>$ if $\Omega_{1,2}^{\rm max}\gg 1/T$, a condition, which we fulfill. To end the pulse sequence, L$_1$ is suddenly switched off. During the whole process, state $\left|e\right>$ is only weakly populated, which avoids loss of atoms by spontaneous emission if $\Omega_{1,2}^{\rm max}\gg \Gamma$. This condition is easily fulfilled with a narrow transition as the one used here. The association STIRAP transfer does not lead to molecules in excited lattice bands, since $T$ is long enough for the band structure to be spectrally resolved.

\begin{figure}
\centering
\includegraphics[width=165mm]{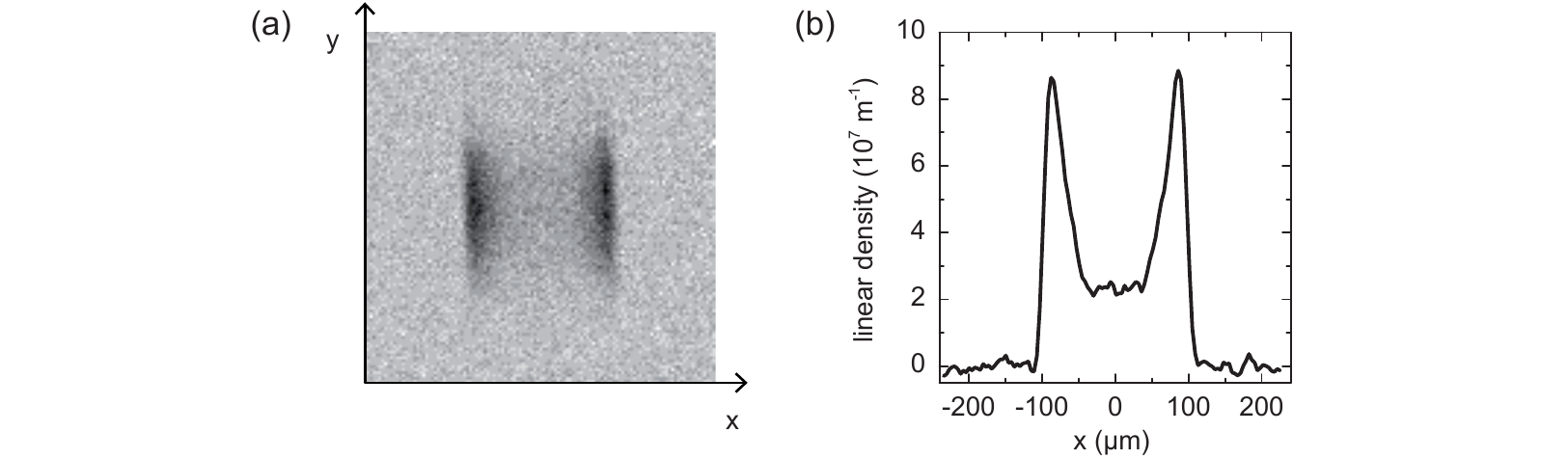}
\caption{Quasi-momentum distribution of repulsively bound pairs. (a) Average of 20 absorption images recorded 10\,ms after release of the atoms from the lattice. (b) Integral of the distribution along $y$. The figure is taken from Ref.~\cite{Stellmer2012cou}.}
\label{fig:Molecules4}
\end{figure}

We now characterize the molecule creation process. To detect molecules, we dissociate them using a time-mirrored pulse sequence (dissociation STIRAP in Fig.~\ref{fig:Molecules3}(a)) and take absorption images of the resulting atoms. The atom number evolution during molecule association and dissociation is shown in Fig.~\ref{fig:Molecules3}(b). After the molecule association pulse sequence, $2\times 10^5$ atoms remain, which we selectively remove by a pulse of light resonant to the ${^1S_0} - {^1P_1}$ atomic transition, out of resonance with any molecular transition; see ``cleaning'' laser C in Fig.~\ref{fig:Molecules3}(a). The recovery of $2 \times 10^4$ atoms by the dissociation STIRAP confirms that molecules have been formed. Further evidence that molecules are the origin of recovered atoms is that 80\% of these atoms occupy lattice sites in pairs. Quantitatively this is shown by removing atom pairs using PA and measuring the loss of atoms. Qualitatively we illustrate this fact by creating and detecting one-dimensional repulsively bound pairs along the $x$-direction \cite{Winkler2006rba}. The pairs were created by ramping the $x$-direction lattice beam to a value of 10\,$E_{\rm rec}$ before ramping all lattice beams off, which propels the pairs into free atoms with opposite momenta along $x$. Figure~\ref{fig:Molecules4} shows the characteristic momentum space distribution of these pairs.

To estimate the STIRAP efficiency and subsequently the number of molecules, we perform another round of molecule association and dissociation on such a sample of atoms with large fraction of doubly occupied sites. We recover $f=9$\% of the atoms, which corresponds to a single-pass efficiency of $\sqrt{f}=30$\%. The largest sample of atoms created by dissociating molecules contains $N_a=2.5 \times 10^4$ atoms, which corresponds to $N_m=N_a/(2\sqrt{f})=4 \times 10^4$ molecules.

We measure the lifetime of molecules in the lattice to be $\sim60\,\mu$s, nearly independent of the lattice depth. This time is surprisingly short and can neither be explained by scattering of lattice photons nor by tunneling of atoms or molecules confined to the lowest band of the lattice and subsequent inelastic collisions. By band mapping \cite{Greiner2001epc}, we observe that $3\times 10^4$ of the initial $6\times 10^5$ atoms are excited to the second band during the association STIRAP, and more atoms have possibly been excited to even higher bands. We speculate that these atoms, which move easily through the lattice, collide inelastically with the molecules, resulting in the observed short molecule lifetime. The short lifetime can explain the 30\%-limit of the molecule conversion efficiency. Without the loss, the high Rabi frequencies and the good coherence of the coupling lasers should result in a conversion efficiency close to 100\%. Related experiments using thermal samples of $^{88}$Sr \cite{Reinaudi2012opo} and $^{174}$Yb \cite{Kato2012ool} atoms observed molecule lifetimes of a few ms and 8\,s, respectively.

\section{Outlook}

In this review, we have touched on all relevant experiments performed with quantum-degenerate strontium until spring 2013. We have found that strontium combines very favorable electronic and collisional properties, which allow for the creation of large and robust BECs and deeply-degenerate Fermi gases. These results show that strontium is perfectly suited for research with degenerate quantum gases.

We envision a number of different experiments that could be performed in the near future. One such prospective experiment might investigate OFRs of $^{88}$Sr atoms confined to a lattice site. The attainment of BEC without evaporation might encourage new approaches to create a truly continuous atom laser. The implementation of techniques borrowed from optical clocks, such as an narrow-linewidth laser and a magic-wavelength lattice, constitute the next steps towards the coherent coupling of the $^1S_0$ and $^3P_0$ states, as required \textit{e.g.}~for the simulation of spin systems with two-orbital SU($N$) symmetry. A very similar technology is required for the creation of artificial gauge fields, which might be sufficiently strong to reach the quantum Hall regime, and for schemes of quantum information processing. Another promising topic, which is beyond the scope of this review, is the creation of alkali/alkaline-earth molecules such as RbSr \cite{Zuchowski2010urm,Aoki2013pli,Pasquiou2013rsd} in their ro-vibronic ground state, which offer additional degrees of tunability compared to bi-alkali molecules.

\section*{Acknowledgements}
We acknowledge the contributions of M.~K.~Tey, B.~Huang, B.~Pasquiou, and R.~Grimm to the Innsbruck experiment, and the contributions of Y.~N.~Martinez de Escobar, P.~G.~Mickelson, S.~B.~Nagel, A.~Traverso, M.~Yan, B.~J.~DeSalvo, B.~Ramachandhran, and H.~Pu to the Rice experiment. The Innsbruck group gratefully acknowledges support from the Austrian Ministry of Science and Research (BMWF) and the Austrian Science Fund (FWF) through a START grant under Project No.~Y507-N20. As member of the project iSense, financial support of the Future and Emerging Technologies (FET) programme within the Seventh Framework Programme for Research of the European Commission is also acknowledged under FET-Open Grant No.~250072. The Rice group acknowledges support from the Welch Foundation (C-1579 and C-1669) and the National Science Foundation (PHY-1205946 and PHY-1205973).

\bibliographystyle{apsrev}

\bibliography{SrBECReview}

\end{document}